\documentclass[]{aastex631}

\newcommand\ha{H$\alpha$}
\newcommand\stwo{[S~II]}

\shorttitle{Herbig-Haro Content and Star Formation in Circinus West and East}
\shortauthors{Rector et al.}
\graphicspath{{./}{figures/}}

\begin{document}

\title{The Herbig-Haro Outflow Content and Star-Forming Environment of Circinus West and East}

\correspondingauthor{Travis A. Rector}
\email{tarector@alaska.edu}

\author[0000-0001-8164-653X]{T.~A. Rector}
\affil{Department of Physics and Astronomy, University of Alaska Anchorage, Anchorage, AK 99508, USA}

\author[0000-0001-7998-226X]{L. Prato}
\affil{Lowell Observatory, 1400 West Mars Hill Road, Flagstaff, AZ 86001, USA}

\author[0000-0002-6549-9792]{R. Kerr}
\affiliation{Dunlap Institute for Astronomy \& Astrophysics, University of Toronto\\
Toronto, ON M5S 3H4, Canada\\}

\author{E. Papraniku}
\affil{Department of Physics and Astronomy, University of Alaska Anchorage, Anchorage, AK 99508, USA}

\author{K. Fisk}
\affil{Department of Physics and Astronomy, University of Alaska Anchorage, Anchorage, AK 99508, USA}



\begin{abstract}

We report the results of a spatially compete, high-sensitivity survey for Herbig-Haro (HH) outflows in the Western and Eastern Circinus molecular clouds. We have detected 28 new HH objects in Circinus West, doubling the number known in this dark nebula.  We have also discovered 9  outflows in Circinus East, the first to be identified here. Although both Circinus West and East appear to be located at $\sim$800 pc, their morphologies are distinct. Circinus West shows filamentary structure, while Circinus East is dominated by amorphous dark clouds. North-east of Circinus East, an extended distribution of young stars is centered on the $\sim$6 Myr-old open cluster \object{ASCC 79}, which may have triggered the sequential formation of younger surrounding populations. New transverse velocities from \textit{Gaia} show two dynamically distinct stellar populations in Circinus East; their velocity distribution is consistent with an active cloud-cloud collision between material ejected by the formation of O and B stars in ASCC~79 and a dynamically similar interloping cloud. Given the similar distances to Circinus West and East, and the presence in both of HH objects -- a phenomenon associated with stellar ages of $\sim$1 Myr --  it is likely that these clouds are nominally related, but only Circinus East is subject to substantial feedback from the central cluster in the parent complex. This feedback appears to guide the morphology and evolution of Circinus East, resulting in a complex and possibly disruptive dynamical environment rich in star formation potential that contrasts with the relatively quiescent environment in Circinus West.

\end{abstract}



\section{Introduction} \label{sec:intro}

Herbig-Haro (HH) objects \citep{1951ApJ...113..697H,1952ApJ...115..572H} result from the energetic, bipolar jets driven by the youngest stars and protostars, often members of binary systems \citep{2010AJ....140..699R,2012ApJ...753..143W}. The jet-launching mechanism is active in young stars of a range of masses, from brown dwarfs \citep{2007Ap&SS.311...63S} to proto-Herbig Ae stars \citep{2023AJ....165..209R}. The impact between collimated, high-velocity (40-100 km s$^-1$) gas and ambient molecular cloud material or wider outflows in which jets are entrained \citep{1982ApJ...263L..73D} produce shocks that collisionally excite \ha\ and \stwo\ emission \citep[e.g.,][]{2001ApJ...546..299B}. The shock heating of external molecular gas can also result in $\rm H_2$ emission at $2.12 \mu$m\citep[e.g.,][]{1989ApJ...342..337Z}. Because the timescale for the production of sufficiently energetic outflows and jets is limited to the first $\sim$1 Myr of a star's lifetime, HH objects are definitive markers of active star formation. HH objects also provide a vehicle for the injection of small-scale stellar feedback into the molecular cloud environment, disrupting and/or propagating successive waves of star formation \citep{Reipurth97, Grudic21} and shaping the ambient medium \citep{1999AJ....117..410B}.

For HH objects, electron densities range from 10$^2$--10$^3$ cm$^{-3}$ with temperatures of 8,000--12,000K \citep{1999A&A...342..717B}. In spite of the relatively low mass (1--20 M$_{\earth}$) entrained in these outflows, the associated morphologies are highly varied and may present as bright bow shocks, faint amorphous nebulosity, collimated flows, or clusters of small knots. Given their wide range of brightness and structure, HH objects can be challenging to identify in young, complex, dust-enshrouded star forming regions, as the very young progenitor stars are often deeply embedded in cloud cores. The obscuration present in the youngest regions traces inhomogeneous densities in local cloud gas structure and likely prevents the detection of all the HH objects present.

Mapping the distribution and frequency of HH objects in diverse star forming environments provides a rough chronometer of the region's evolutionary state. How many HH objects we can see in a given cloud complex is dictated by the balance between the region's age and morphology. The latter may be modulated in part by the presence of high-mass stars that carve out cavities in the cloud structure, yielding clues to the IMF of the embedded stellar population; in the NGC~1333 molecular cloud complex, the high abundance of HH objects may be indicative of a relatively coeval microburst of star formation \citep{1996ApJ...473L..49B}, i.e., a short-lived ($<$1 Myr) epoch of star formation within a small area ($<$1 pc). The ubiquity of the resulting shocked gas in this region must have a significant impact on the star formation rate and environment \citep{1996swhs.conf..491B}.

In \citet{2020AJ....160..189R}, we published a preliminary study of new HH objects in a portion of Circinus West. Here we report the results of the first spatially complete, uniform sensitivity survey for HH objects in the clouds of both Circinus West and Circinus East (hereafter referred to as Cir-W and Cir-E respectively). This investigation is augmented with a new characterization of the distances and ages of the local young stellar population, extending to the north-east of Cir-E, which may have shaped the dynamical state of the Circinus clouds through stellar feedback.

The relatively isolated Circinus molecular clouds are described in more detail by \citet{2008hsf2.book..285R}. They lie at low galactic latitudes with a distance of $\sim700$~pc \citep[e.g.,][but see \S~\ref{app:distance} of this paper for a revised estimate]{2020A&A...633A..51Z}. The $^{12}$CO mass of $\sim5\times 10^4 M_{\odot}$ for the entire Circinus cloud complex as determined by \citet{1987ApJ...322..706D} is about 10$\times$ greater than either of the $^{13}$CO mass estimates of $\sim5\times 10^3 M_{\odot}$ each for Cir-W and Cir-E alone \citep{2011ApJ...731...23S}.

Cir-W consists of a central dense core of gas surrounded by sinuous, filamentary structures (Figure~\ref{fig:fovs}). Within Cir-W, the core Cir-MMS cluster \citep{1996A&A...314..258R} contains several low-mass YSOs \citep{2011ApJ...733L...2L}. Near-infrared polarization indicates a strong magnetic field that likely regulates the low-mass cluster formation in this region \citep{2018ApJS..234...42K}. \citet{1999AJ....117..410B} suggested that, on small scales, the last few Myr of local star formation have yielded many complex CO outflows, carving out cavities in the dust and gas and revealing numerous HH objects. \citet{2011ApJ...731...23S} found that the core of $^{13}$CO gas in Cir-W is composed of a single component with V$_{LSR}$ of $\sim-6.5$ km s$^{-1}$, indicating that the star formation dynamics show little evidence for external perturbations.


The amorphous morphology of Cir-E contrasts starkly with the filamentary structures in Cir-W.  
\citet{2011ApJ...731...23S} found Cir-E to be more compact, with a higher velocity dispersion and at least two distinct $^{13}$CO components. The distribution of IRAS and H$\alpha$ sources along the $^{13}$CO ridges correspond to the different velocity components, with V$_{LSR}$ of $\sim-6$ km s$^{-1}$ and $\sim-4$ km s$^{-1}$ for Cir-Ea and Cir-Eb, respectively \citep{2011ApJ...731...23S}. It suggests a complex dynamical past in this region. Collisions between the cloud components may have triggered the star formation events leading to the concentrated grouping of young sources. The relative paucity of HH objects in Cir-E may be the result of the region's morphology, a slightly younger age than Cir-W, and/or the inhibiting impact of relatively recent feedback from star formation in the extended environment.



In \S~\ref{sec:obs} we summarize the data, observations, analysis, and identification of HH objects. In \S~\ref{sec:sources}, we discuss the individual HH objects discovered and potential YSO progenitors. \S \ref{sec:context} provides an analysis of the distance and age characteristics of the distributed young star population in the proximity of Cir-E. In \S~\ref{sec:disc} we provide a discussion of the distribution of HH objects and of the implications for the interconnections between the broader star formation environment. In \S~\ref{sec:concl}, we present a summary and conclusions.

\section{Data} \label{sec:obs}




\subsection{Observations}

Observations were completed with DECam on the {\it Blanco} 4-meter telescope at Cerro Tololo Interamerican Observatory.  DECam is a wide-field CCD imager that consists of sixty-two 2048x4096 pixel red-sensitive CCDs, yielding 520 megapixels in total.  The field of view is about 3 square degrees with a scale of $0\farcs26$ pixel$^{-1}$, facilitating spatially complete surveys for HH outflows in both of these molecular clouds.  The observations are centered on $\alpha = 14:59:55.3$, $\delta = -63:22:13$ for Cir-W and $\alpha = 15:15:08.8$, $\delta = -62:45:32$ for Cir-E.  The two fields slightly overlap.

We used the broadband DES {\it g,i} filters, which are similar to the SDSS survey filters of the same name.  To isolate \ha\ and \stwo\ emission, observations were also made with the DECam narrowband N662 and N673 filters, which are centered at 6620 and 6730\AA, with FWHM values of 160 and 100\AA~respectively. Exposures were taken with a 5-pointing dither pattern to fill in chip gaps, with exposure times of 5x240s in {\it g}, 5x120s in {\it i}, and 5x600s each in N662 and N673. These images are significantly deeper than those obtained with the Swope~1m telescope as reported in \citet{2020AJ....160..189R}. Cir-W and Cir-E were observed with the {\it g}, {\it i}, and N662 filters on March 10th, 2021.  Follow-up observations in N673 were completed on April 23rd, 2022, more than a year later.  The data were reduced with the DECam Community Pipeline \citep{2014ASPC..485..379V}.  

\subsection{Analysis}

\begin{figure}[ht]
\plottwo{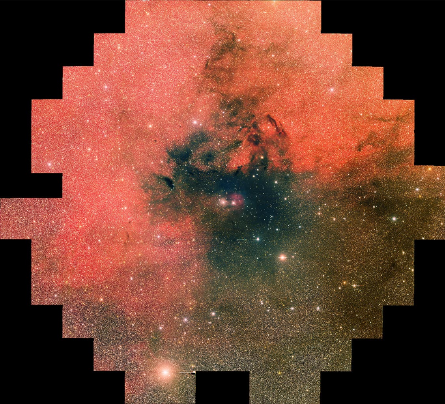}{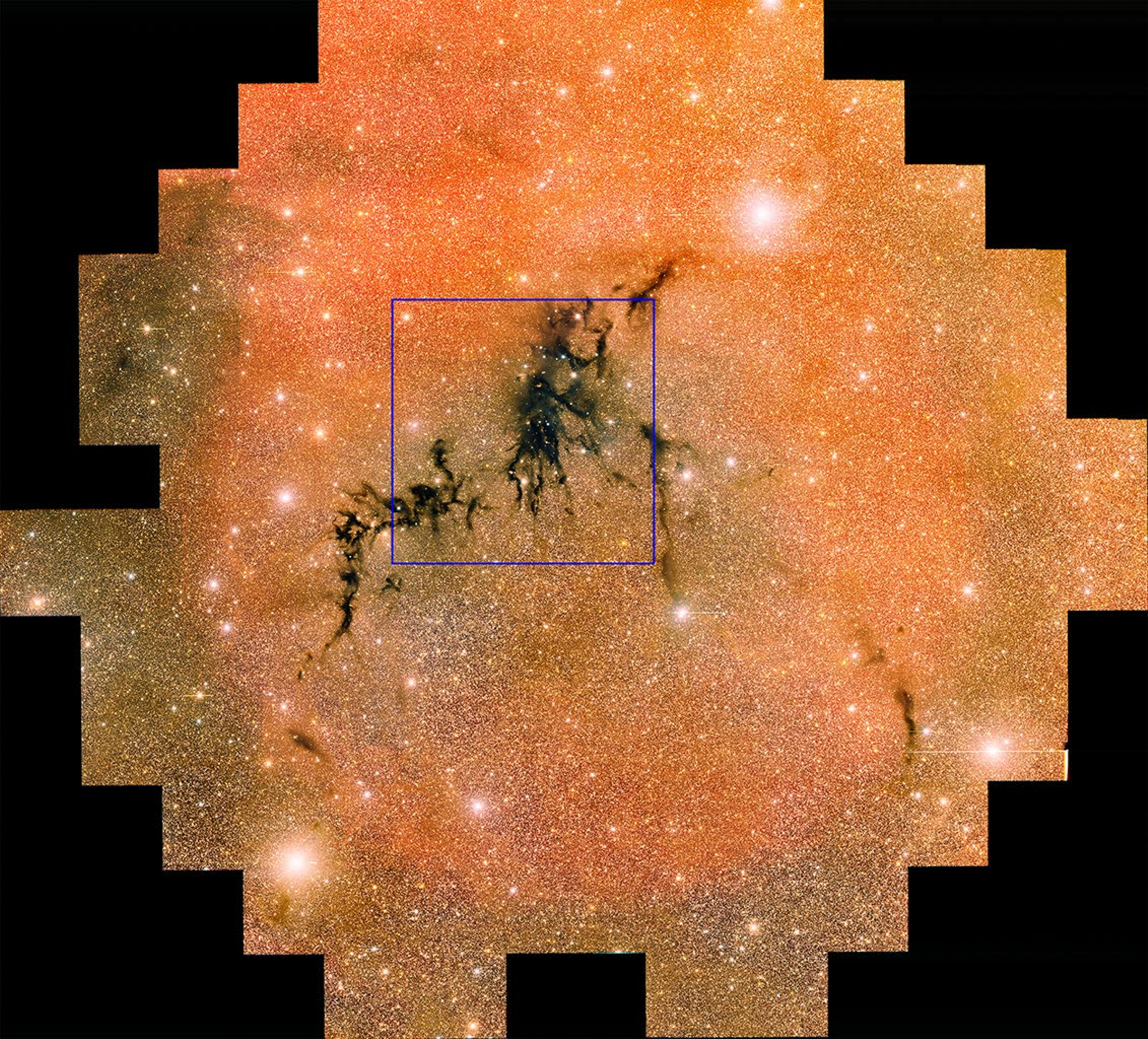}
\caption{The fields of view for the Cir-E (left) and Cir-W (right) DECam pointings.  The two fields slightly overlap.  The field of view of the Swope~1m observations from \citet{2020AJ....160..189R} is shown in the Cir-W image for comparison.  Note that each image is composited slightly differently to maximize sensitivity to faint outflows within the dynamic range of the nebulosity–– the color differences between the two images therefore are not meaningful. 
\label{fig:fovs}}
\end{figure}

To better see faint HH objects, the {\it g, i}, and \ha\ data were combined to form a color-composite image with the methodology described in \citet{2007AJ....133..598R}.  Specifically, in the color images the broadband filters $g$ and $i$ are assigned the colors cyan and yellow respectively, while \ha\ is assigned to red.  The resultant color composites are shown in Figure~\ref{fig:fovs}.  An advantage of searching for HH objects in this manner is that the red \ha\ emission is chromatically distinct from structure visible in the other filters.  Objects that emit only \ha\ will appear as a deep red.  Further, the broadband filters reveal the relative amounts of obscuration from dust and gas.  Thus, faint outflows can be found more easily in the complex environments typical of star-forming regions.  The \stwo\ observations are not included in the color-composite image but are instead used for confirmation. All of the HH objects are visible in the \ha\ and \stwo\ filters, but are not detected in the broadband $i$ filter.  Thus we are confident they are sources of line emission only.  We were also kindly given access to $2.12 \mu$m $\rm H_2$ narrowband images of this region obtained by B. Reipurth and J. Bally.  All but two of our newly discovered HH objects are detected in these infrared images as well, further solidifying our confidence in their identifications as such.

\begin{figure}[ht]
\plotone{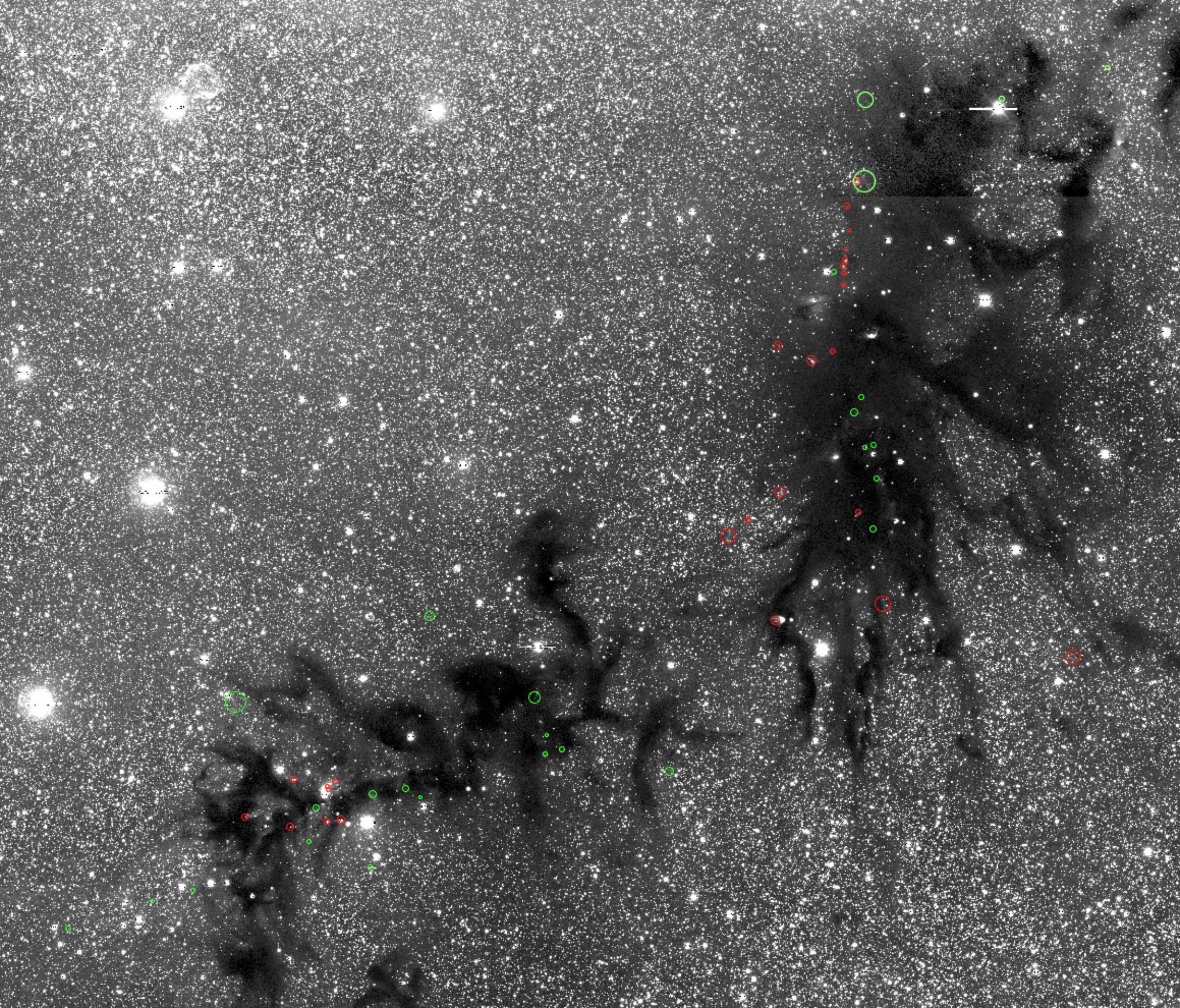}
\caption{This region of Cir-W, shown here in \ha, contains all of the known HH objects in this nebula except for HH~1259, which lies at the southeastern tip of the cloud.  Red circles show the locations of previously known HH objects; and green circles show new ones reported in this paper.  The size of each circle roughly reflects the size of the HH object. North is up, east is left.
\label{fig:circ_w_ds9}}
\end{figure}

The coordinates of newly discovered HH objects in Cir-W and Cir-E are given in Tables~\ref{tbl:newHH_w} and \ref{tbl:newHH_e} respectively.  The positions correspond to the center of the brightest knot in each object.  Because the sources are extended, the coordinates are only given to a precision of 1\arcsec.    


\begin{deluxetable}{lll}
\tablecaption{Newly Discovered HH objects in Cir-W\label{tbl:newHH_w}}
\tablewidth{0pt}
\tablehead{\colhead{HH} & \colhead{RA(2000)} & \colhead{DEC}}
\startdata
1236 & 14:59:17.2 & -62:57:26 \\ 
1237 & 14:59:50.1 & -62:58:32 \\ 
1238 & 15:00:29.1 & -63:11:58 \\ 
1239 & 15:00:30.3 & -63:10:45 \\
1240 & 15:00:30.4 & -63:13:44 \\ 
76c2 & 15:00:31.7 & -63:01:39 \\ 
1241 & 15:00:32.6 & -63:10:50 \\ 
76i & 15:00:33.4 & -62:58:26 \\ 
1242 & 15:00:33.6 & -63:09:05 \\ 
1243 & 15:00:35.8 & -63:09:38 \\ 
1244 & 15:00:42.2 & -63:04:40 \\
1245 & 15:01:34.6 & -63:22:16 \\
1246 & 15:02:08.3 & -63:21:29 \\ 
1247 & 15:02:13.3 & -63:20:58 \\ 
1248 & 15:02:13.9 & -63:21:39 \\ 
1249 & 15:02:17.1 & -63:19:38 \\ 
1250 & 15:02:49.7 & -63:16:43 \\ 
1251 & 15:02:53.4 & -63:23:08 \\ 
1252 & 15:02:58.0 & -63:22:48 \\ 
1253 & 15:03:08.3 & -63:22:57 \\ 
1254 & 15:03:09.0 & -63:25:35 \\
1255 & 15:03:26.4 & -63:23:25 \\ 
1256 & 15:03:28.5 & -63:24:38 \\ 
1257 & 15:03:51.3 & -63:19:37 \\ 
1258a & 15:04:05.5 & -63:26:19 \\ 
1258b & 15:04:18.1 & -63:26:40 \\ 
1258c & 15:04:45.4 & -63:27:35 \\ 
1259 & 15:04:31.0 & -63:38:03 \\ 
\enddata
\end{deluxetable}

\subsection{Extended Stellar Population} \label{sec:gaiadata}

While the literature view of the Circinus Star-forming Region has focused primarily on its gas, dust, and protostars \citep{2008hsf2.book..285R}, recent work has indicated that older, largely gas-free populations also reside in the area. In \citet{Kerr23}, protostars associated with the Circinus Molecular Cloud were grouped together as SCYA-6 (hereafter the Circinus Complex), which was defined by clustering, photometrically young \textit{Gaia} stars using the HDBSCAN algorithm \citep{McInnes2017}. The sample of 1756 photometrically young stars used to define the Circinus Complex far surpasses the tens of protostars known in the Circinus Molecular Cloud, while the mean age of 5 Myr is incompatible with the presence of HH objects. This discrepancy with the properties of the Circinus Molecular Cloud is explained by the dominance of stellar populations largely located to the galactic east and northeast, centered on the open cluster ASCC~79. \citet{KerrFarias25} estimated that approximately 3100 stars exist across the entire complex, and only 19\% reside in populations associated with the Circinus Molecular Cloud, compared to 69\% in populations contiguous with ASCC~79. The broader stellar population defined by \citet{KerrFarias25} in Circinus therefore provides information on adjacent populations that may have contributed to the star formation history of Circinus, while expanding our stellar sample in the Circinus region to include stars without strong H-alpha signatures or outflows. 

We therefore include \citet{KerrFarias25}'s high-quality stellar sample in our analysis. We use a subset of the stars presented in Table 1 of that paper, including only objects that pass the quality and membership probability flags described in \citet{KerrFarias25} Section 3, to define the broader stellar population in Circinus. For a more detailed analysis of the Circinus Molecular Cloud, we also define a separate sample that consists only of stars included in the high-quality stellar sample with a probability $P_{Age<50 Myr} > 0.95$ that \citet{KerrFarias25} assigns to subgroups CIR-3 and CIR-4. These two subgroups correspond to Cir-W and Cir-E, respectively. The stringent cut to youth probability $P_{Age<50 Myr}$ was shown in \citet{KerrFarias25} to repress a component of older ejecta from ASCC~79 and heavily-reddened subgiants that detract from the dynamical substructure in the region. This dataset therefore represents a maximally pure sample of \textit{Gaia} stars connected to the Circinus Molecular Cloud, and we refer to it as the \textit{Gaia} Circinus Molecular Cloud sample.

\section{Discussion of Individual Sources} \label{sec:sources}

Here we discuss the nature of each newly discovered HH object, including their locations relative to other HH objects as well as possible progenitor(s).  Sources are discussed collectively where appropriate.

\subsection{Circinus West}

\begin{figure}
\gridline{\fig{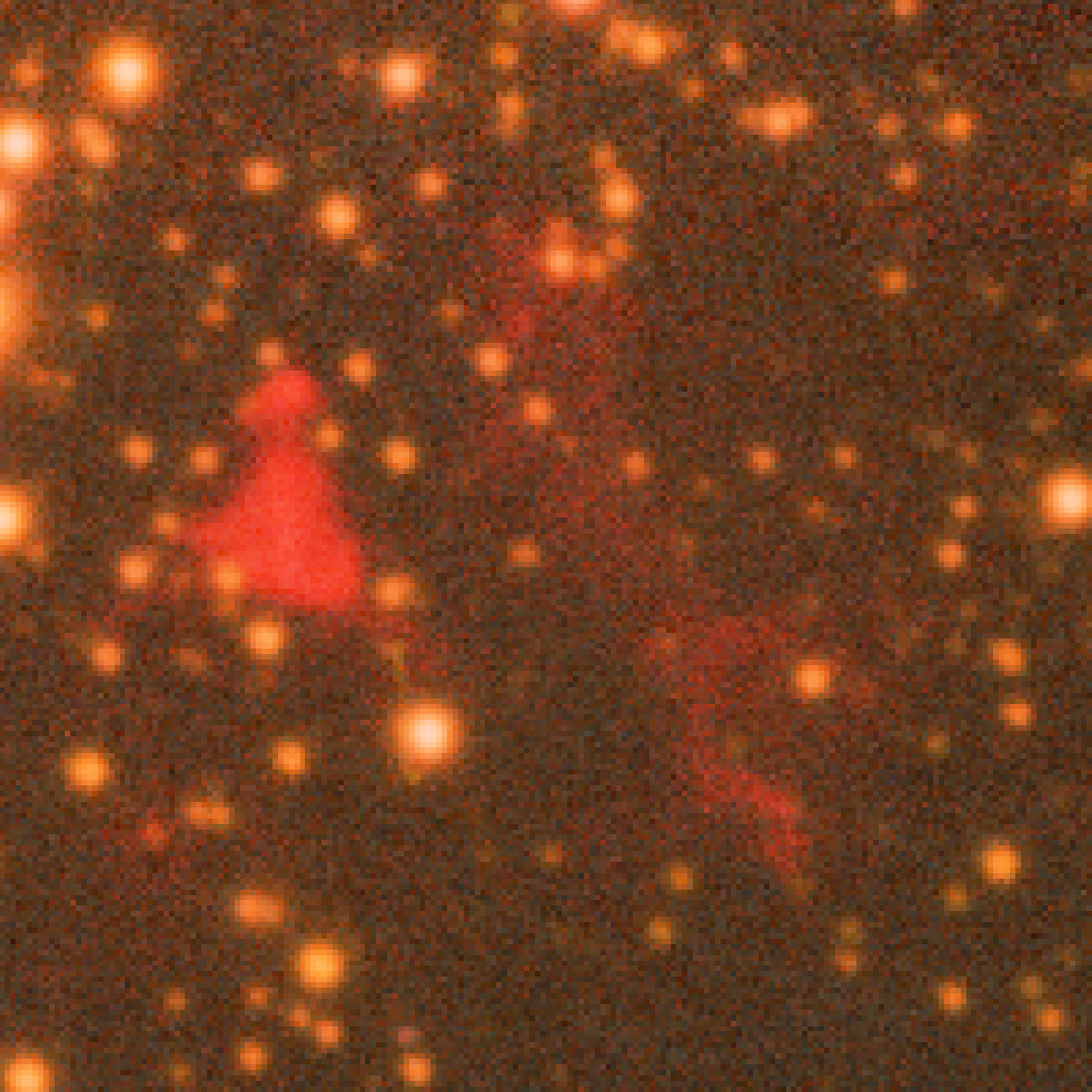}{0.25\textwidth}{HH~76c \& 76c2}
          \fig{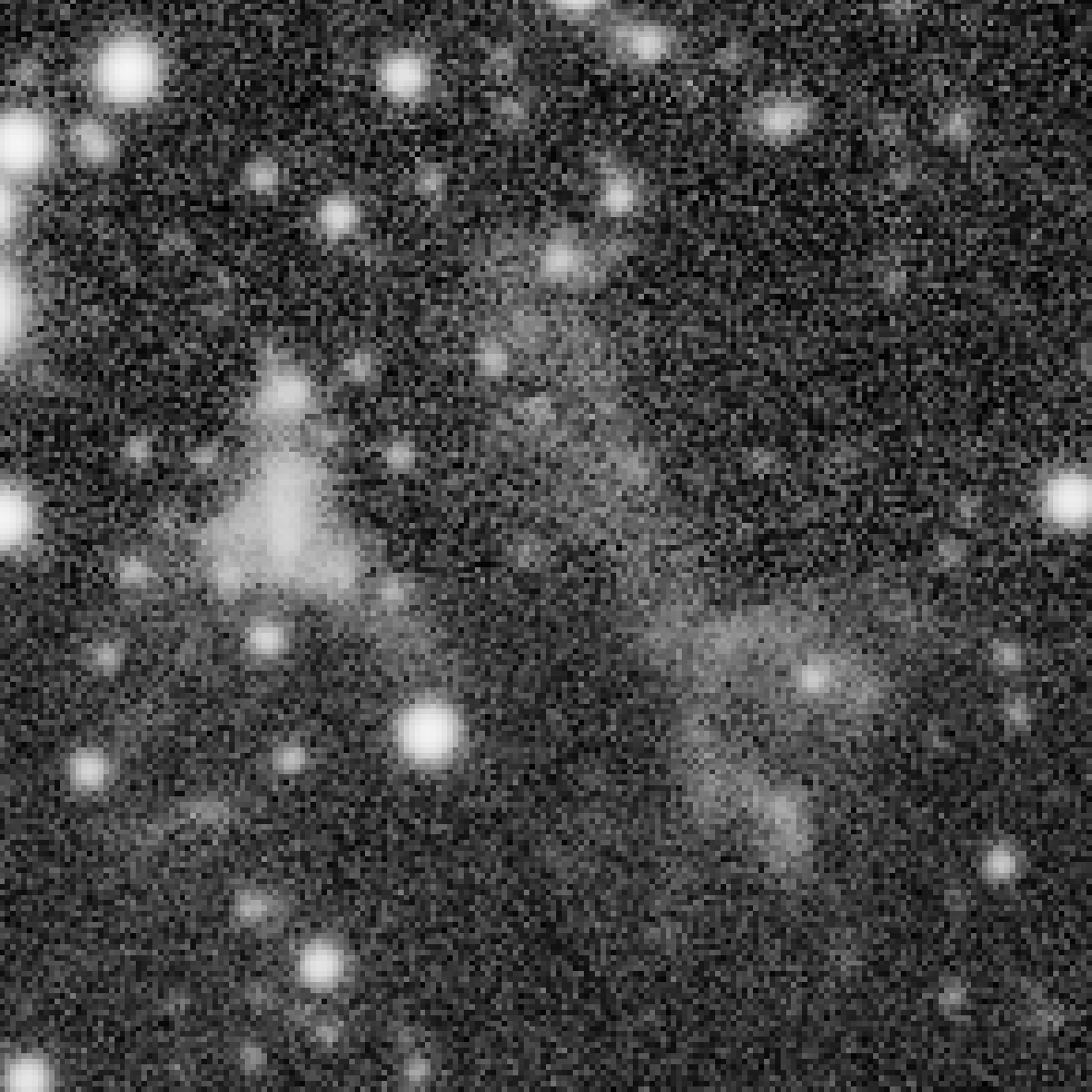}{0.25\textwidth}{HH~76c \& 76c2}
          \fig{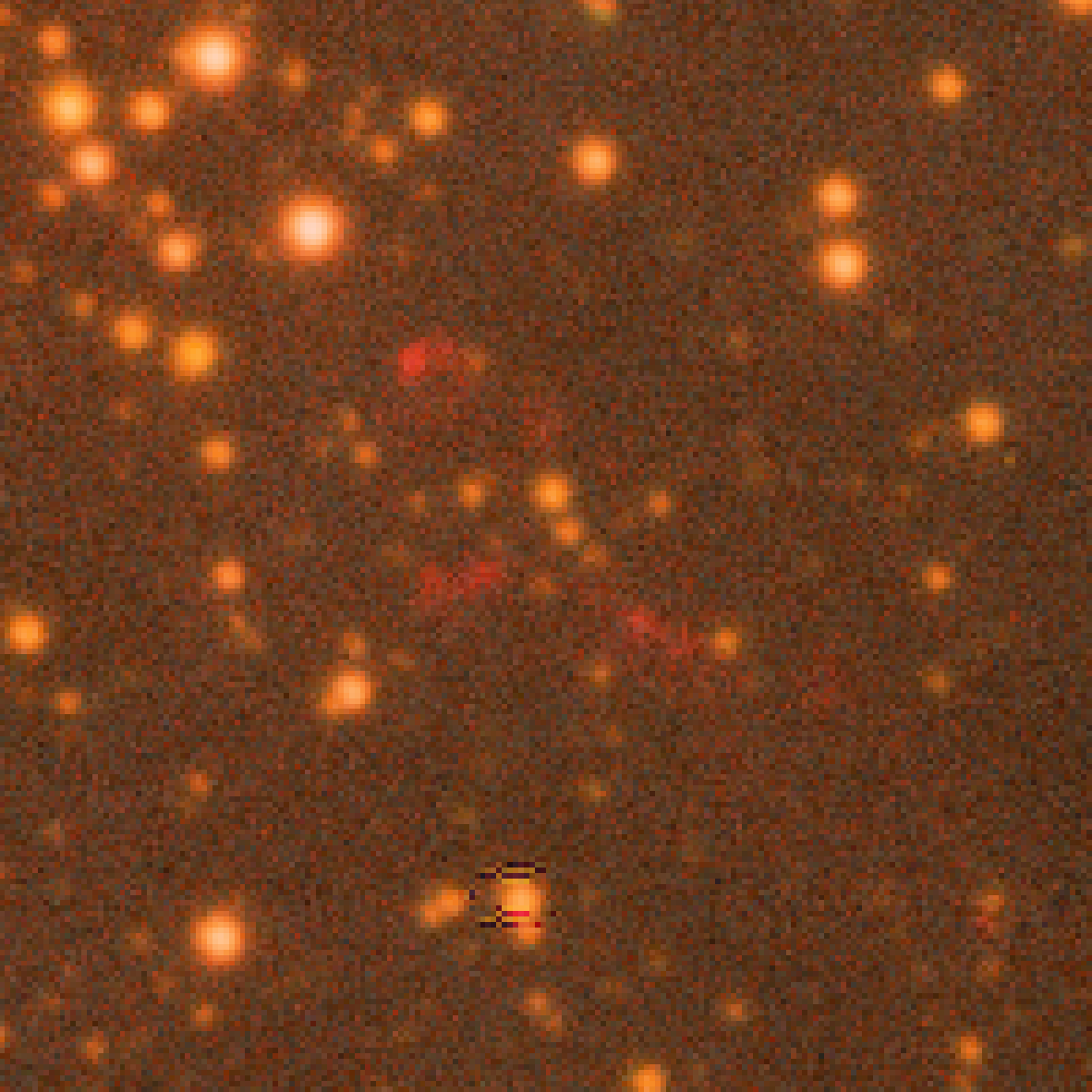}{0.25\textwidth}{HH~76i}
          \fig{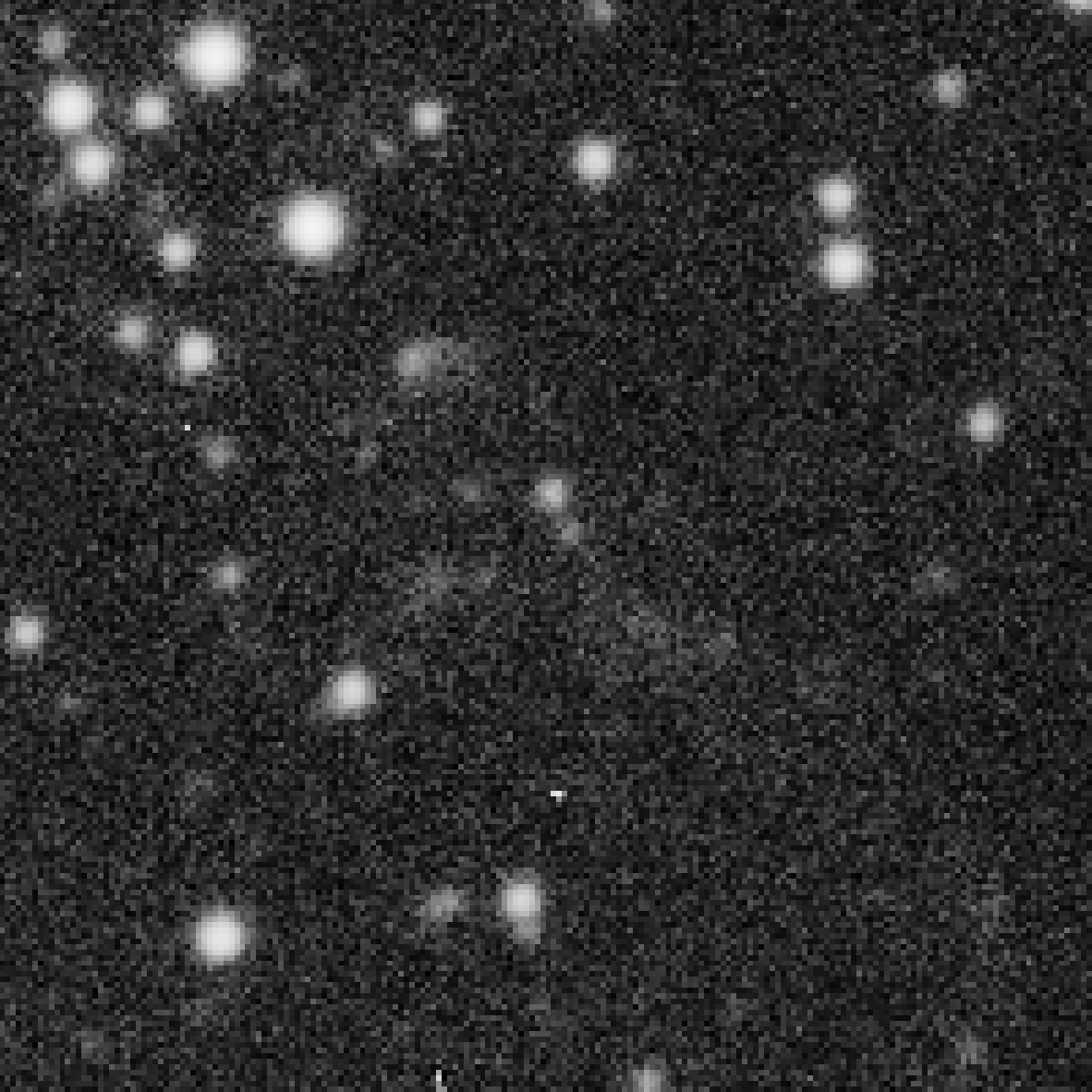}{0.25\textwidth}{HH~76i}
          }
\caption{Cutouts of newly detected components of the HH~76 outflow. Each cutout is 200 pixels (52\arcsec) on side.  Color composite images were made with the {\it g} (cyan), {\it i} (yellow), and N662~\ha\ (red) filters so that the \ha\ emission from the HH objects is distinctly red.  The grayscale images are from follow-up N673~\stwo\ observations.  HH objects are therefore deep red in the color images and are also detected in \stwo.
\label{fig:cutouts_AABB}}
\end{figure}

HH~76c2 \& 76i:  These objects (Figure~\ref{fig:cutouts_AABB}) appear to be a northern continuation of the \object{HH76} outflow, for which \object{2MASS J15004103-6306381} is thought to be the progenitor. HH~76c2 is directly west -- and assumed to be associated with --  HH~76c, which was discovered by \citet{2020AJ....160..189R}. HH~76i expands the known length of this outflow from 5.2\arcmin\ to 8.3\arcmin.  Assuming a distance of 777~pc to Cir-W (see \S~\ref{app:distance}), this is a projected length of 1.8~pc.

\begin{figure}
\gridline{\fig{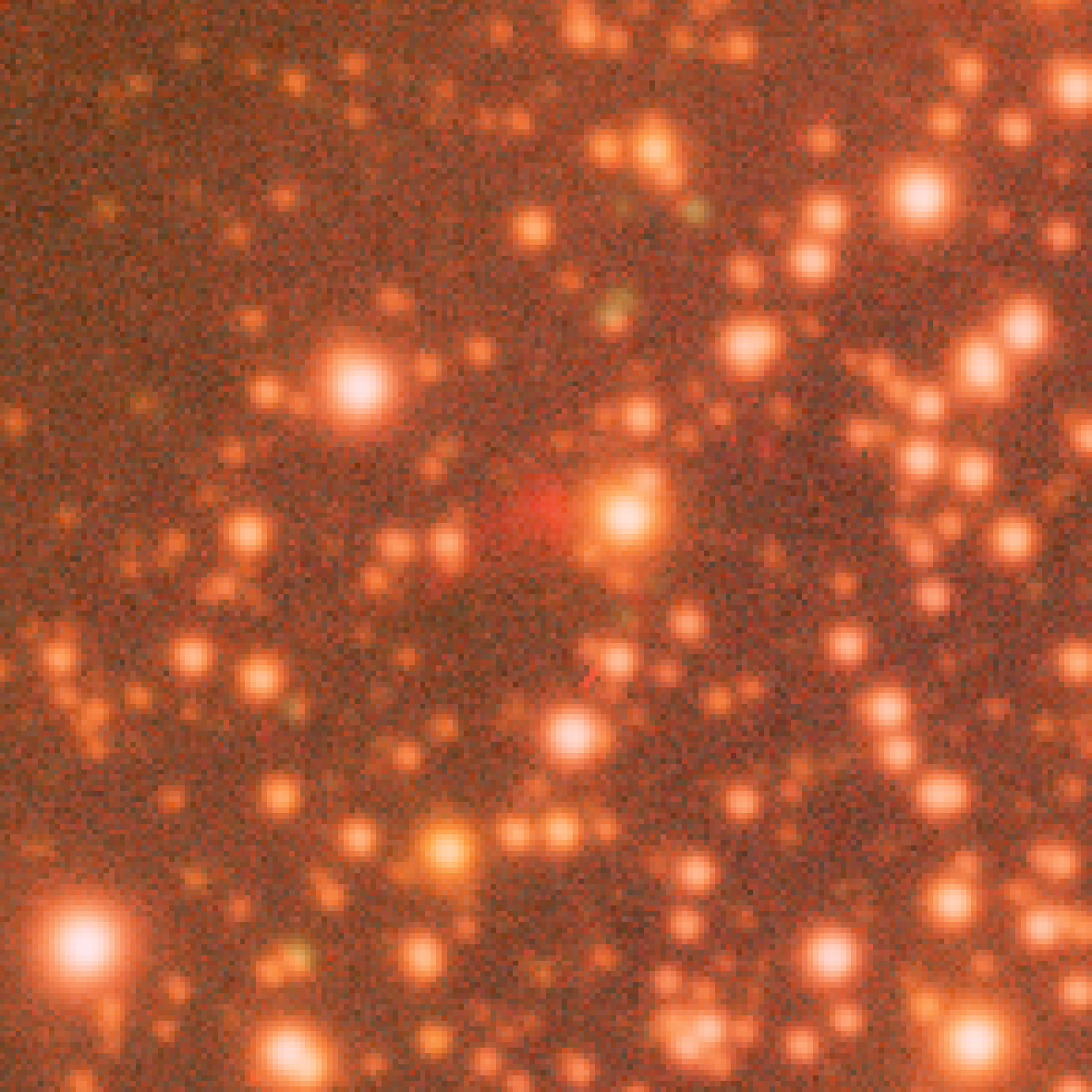}{0.25\textwidth}{HH~1236}
          \fig{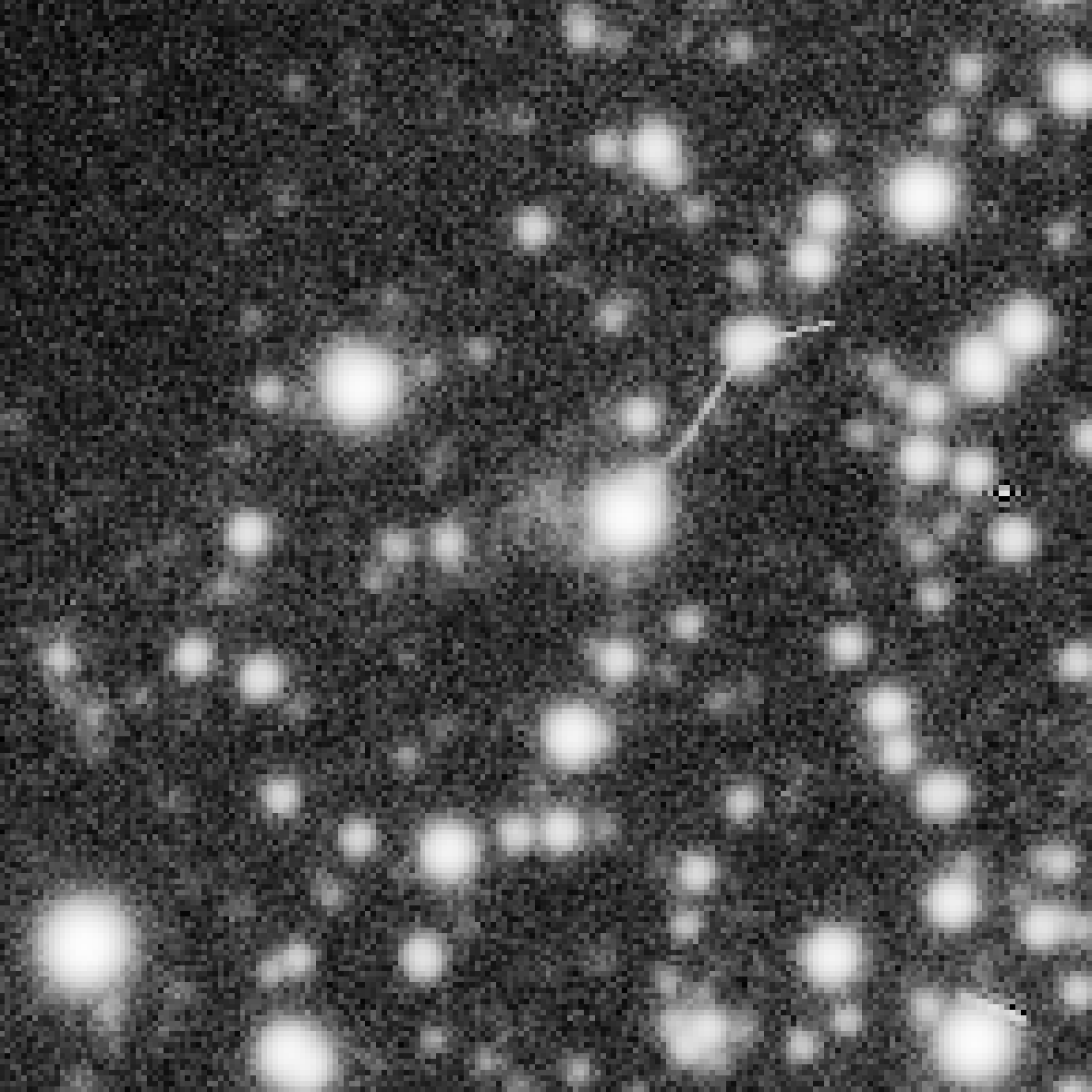}{0.25\textwidth}{HH~1236}
          \fig{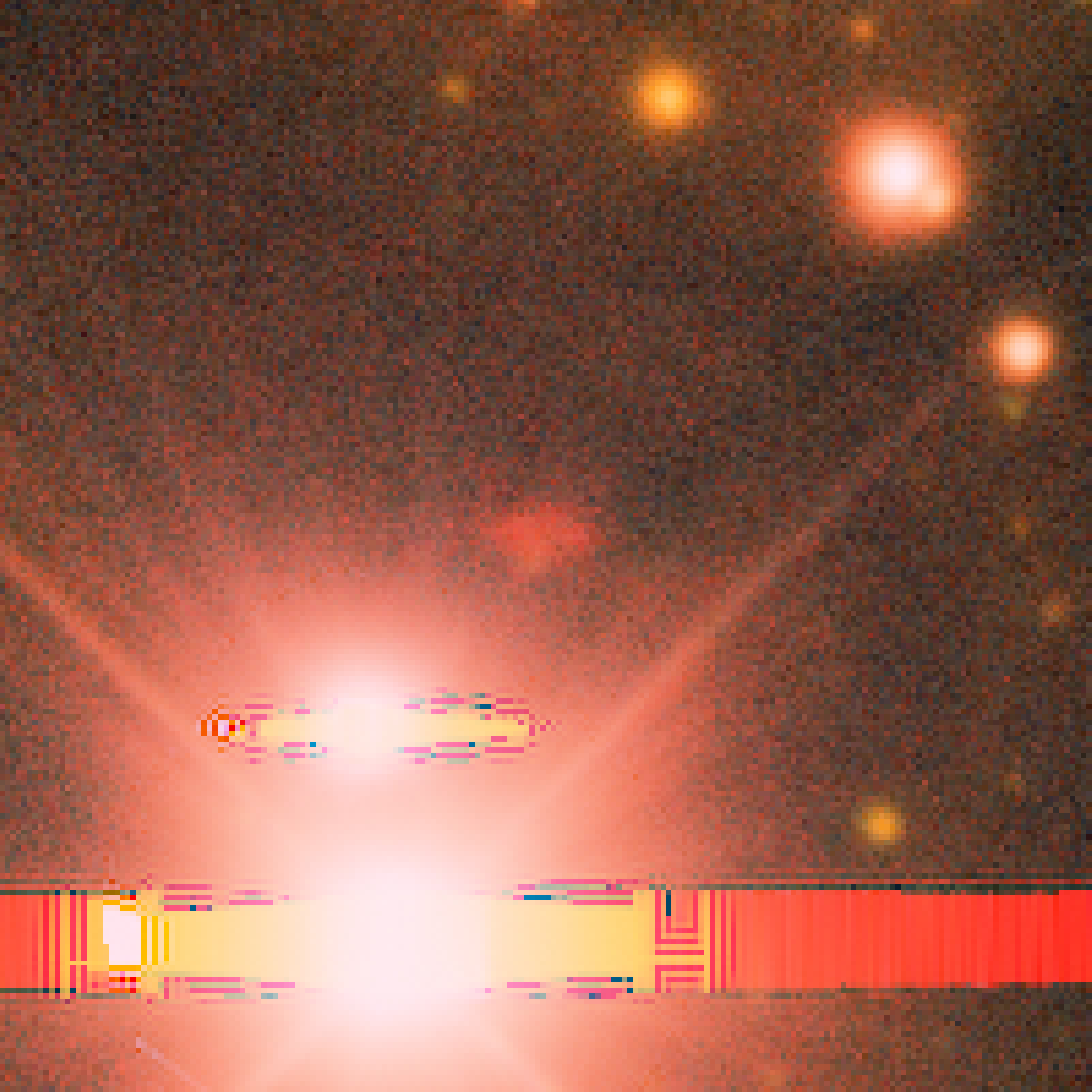}{0.25\textwidth}{HH~1237}
          \fig{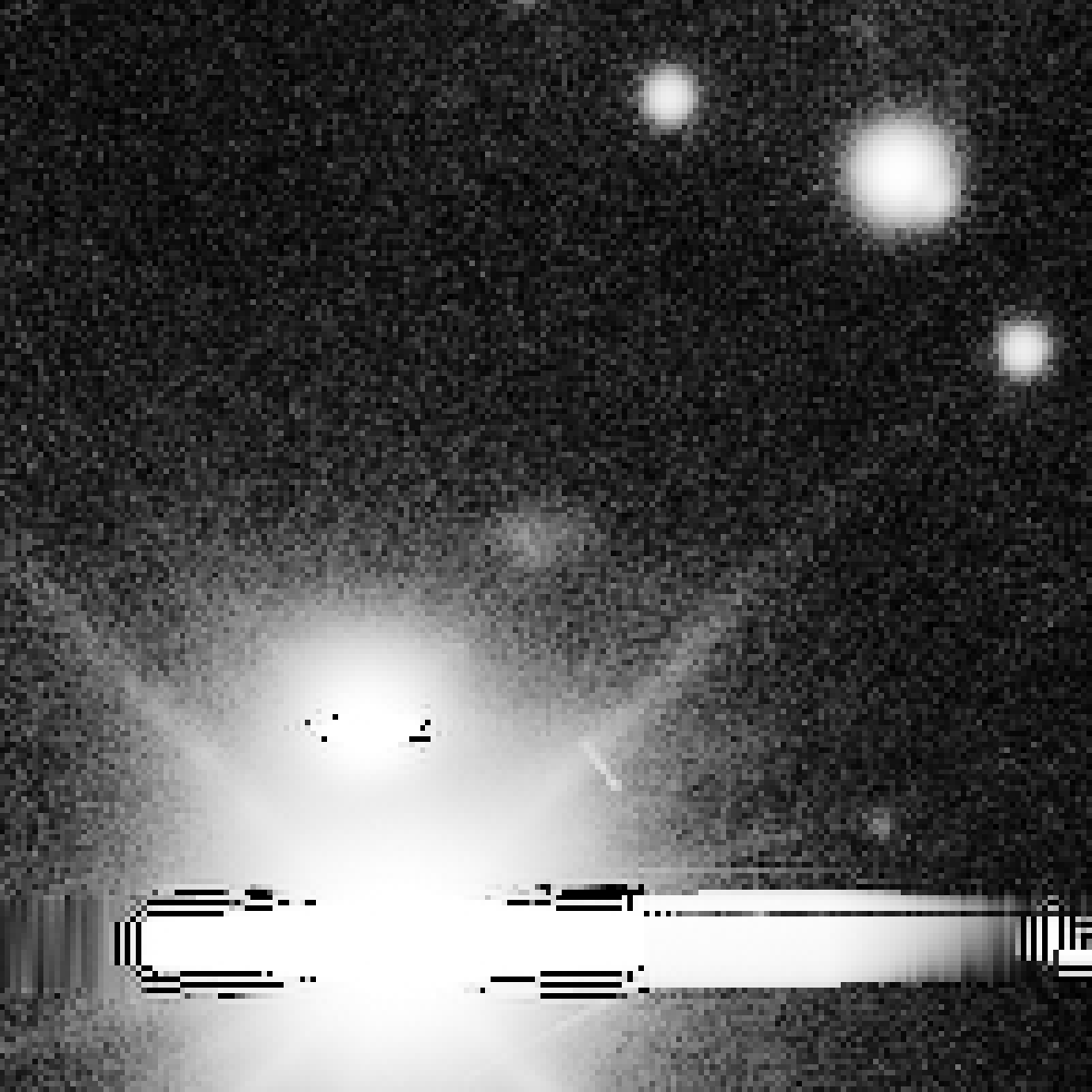}{0.25\textwidth}{HH~1237}
          }
\caption{Cutouts of newly discovered HH objects within Cir-W, to the west of the HH~76 outflow. Cutout size and color schemes are the same as for Figure~\ref{fig:cutouts_AABB}.
\label{fig:cutouts_CCDD}}
\end{figure}

HH~1236:  This HH object (Figure~\ref{fig:cutouts_CCDD}) is relatively isolated.  The nearest YSO candidate, \object{WISEP J150024.04-630155.5}, is about 1.5\arcmin\ to the south.

HH~1237:  Located at the northern tip of the Cir-W dust cloud (Figure~\ref{fig:cutouts_CCDD}), HH~1237 is near the YSO candidate \object{WISEP J145950.07-625833.1}; however, the nearby YSO candidates \object{WISEP J145959.01-625936.7} and \object{WISEP J150024.04-630155.5} are also possible progenitors.

\begin{figure}
\gridline{\fig{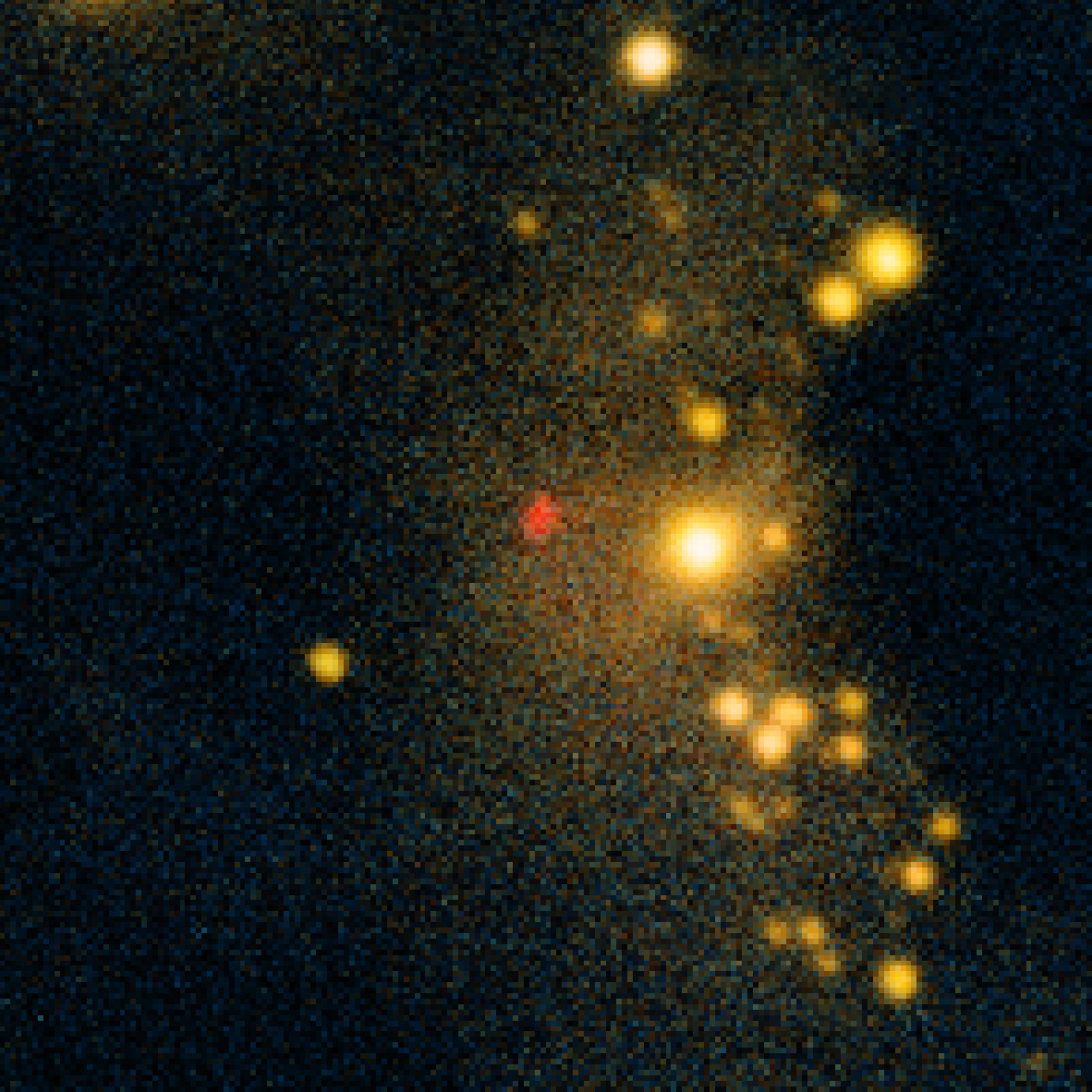}{0.25\textwidth}{HH~1238}
          \fig{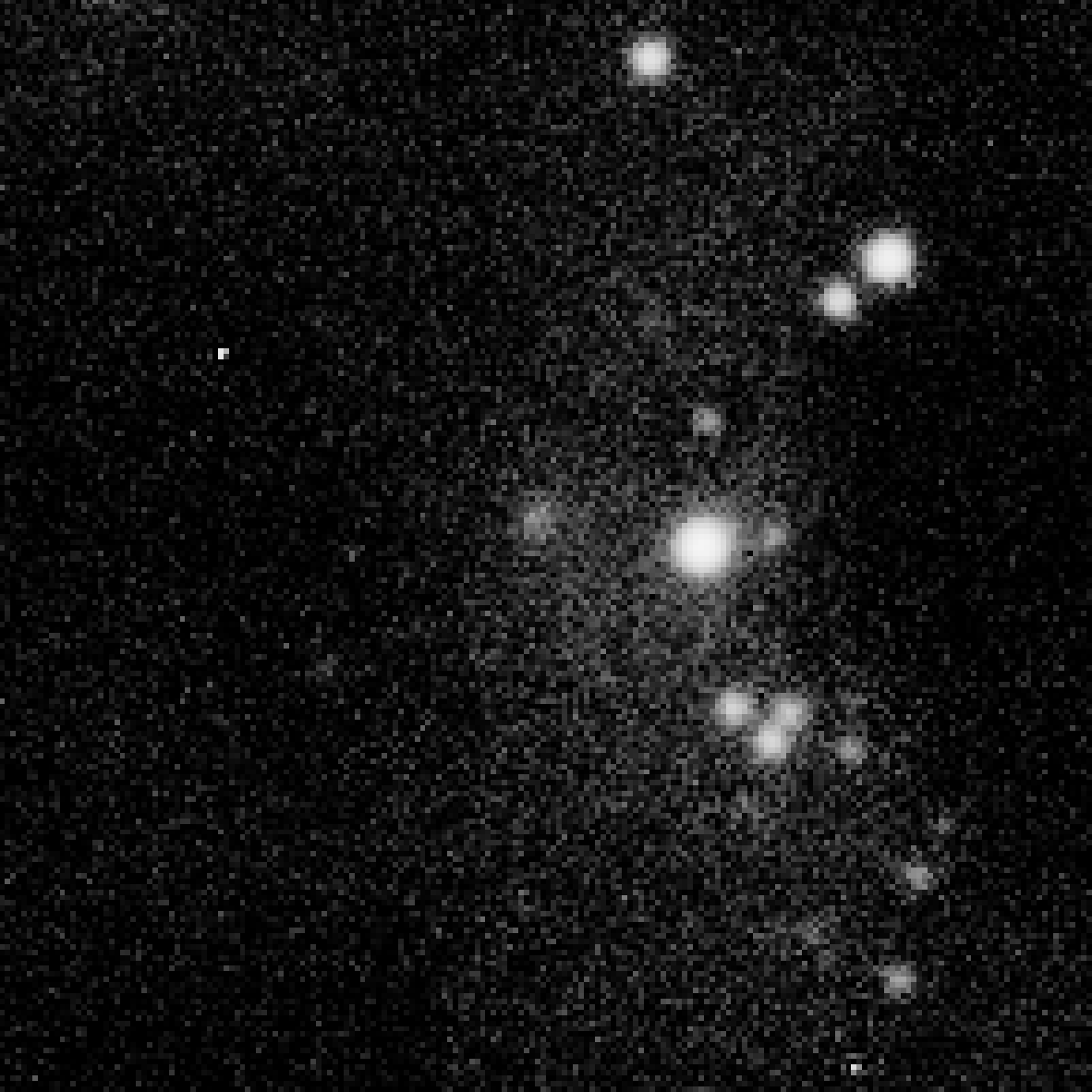}{0.25\textwidth}{HH~1238}
          \fig{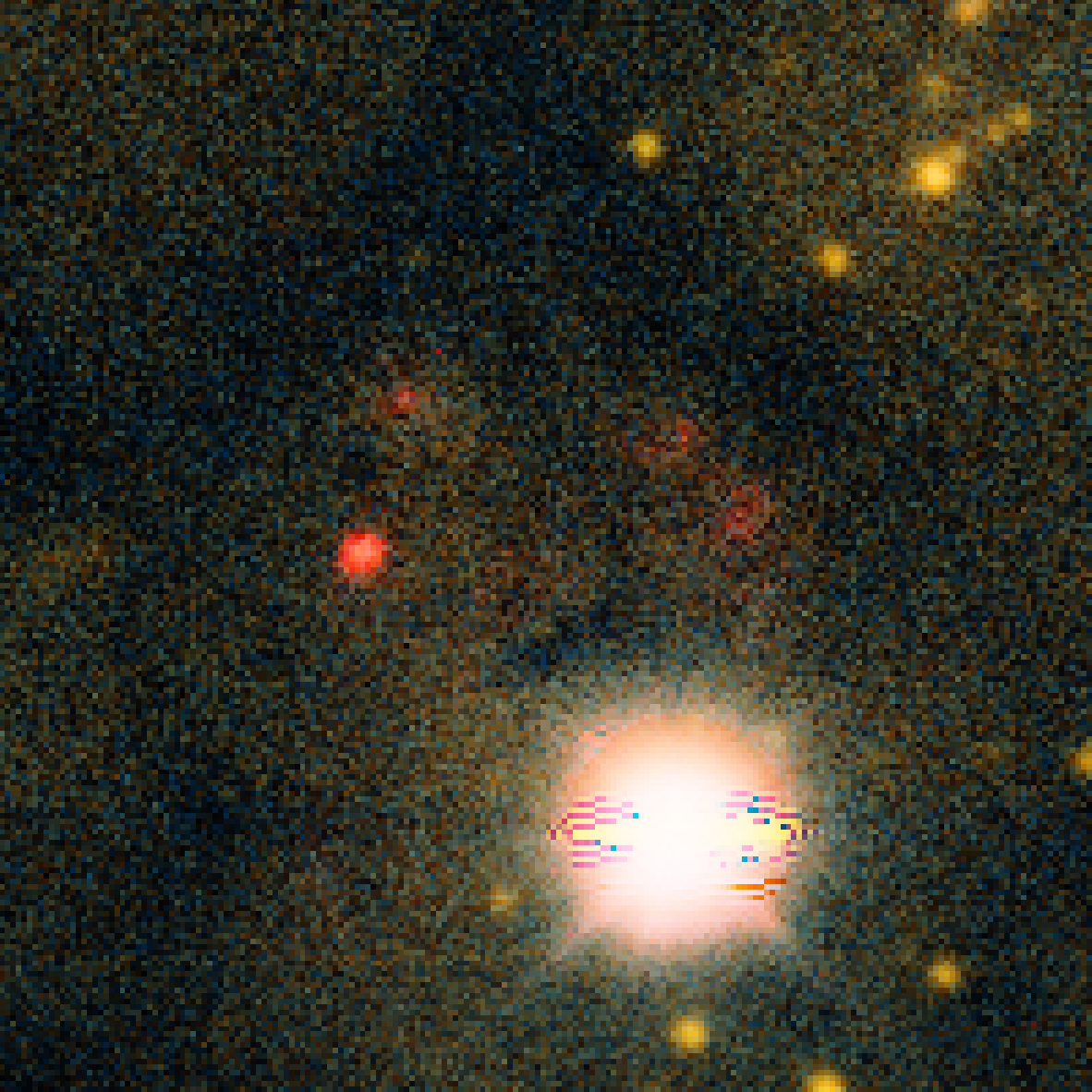}{0.25\textwidth}{HH~1241 \& 1239}
          \fig{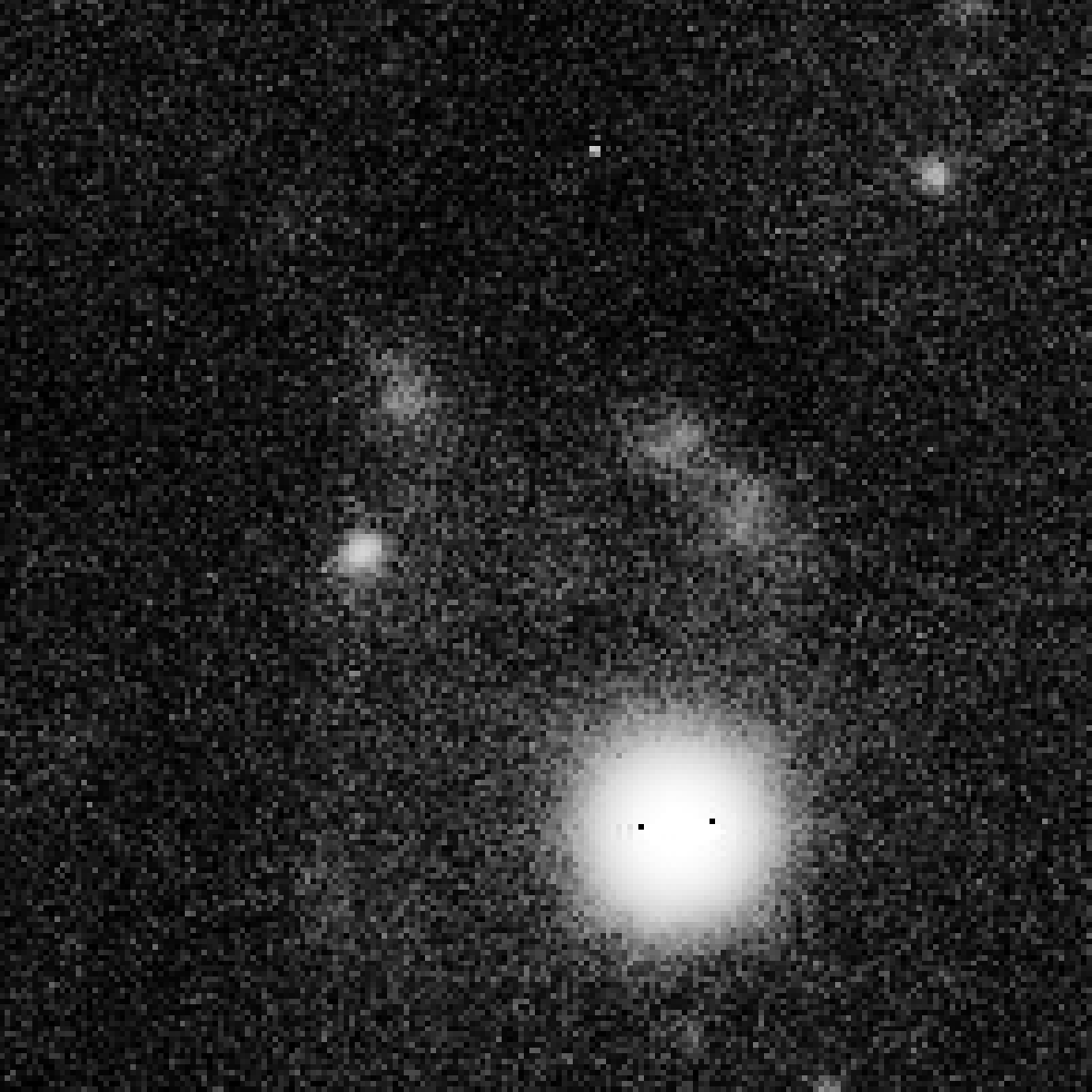}{0.25\textwidth}{HH~HH~1241 \& 1239}
          }
\gridline{\fig{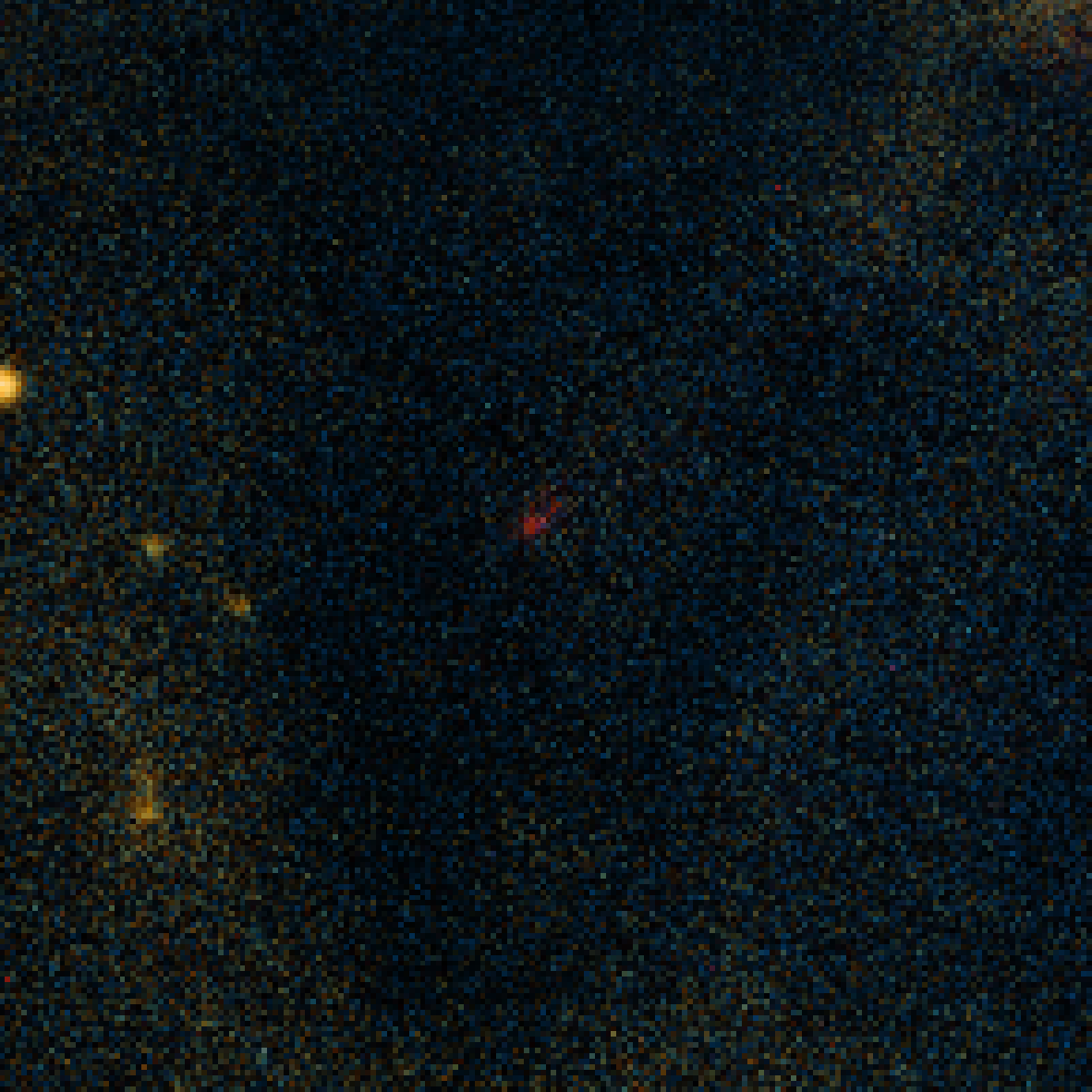}{0.25\textwidth}{HH~1240}
          \fig{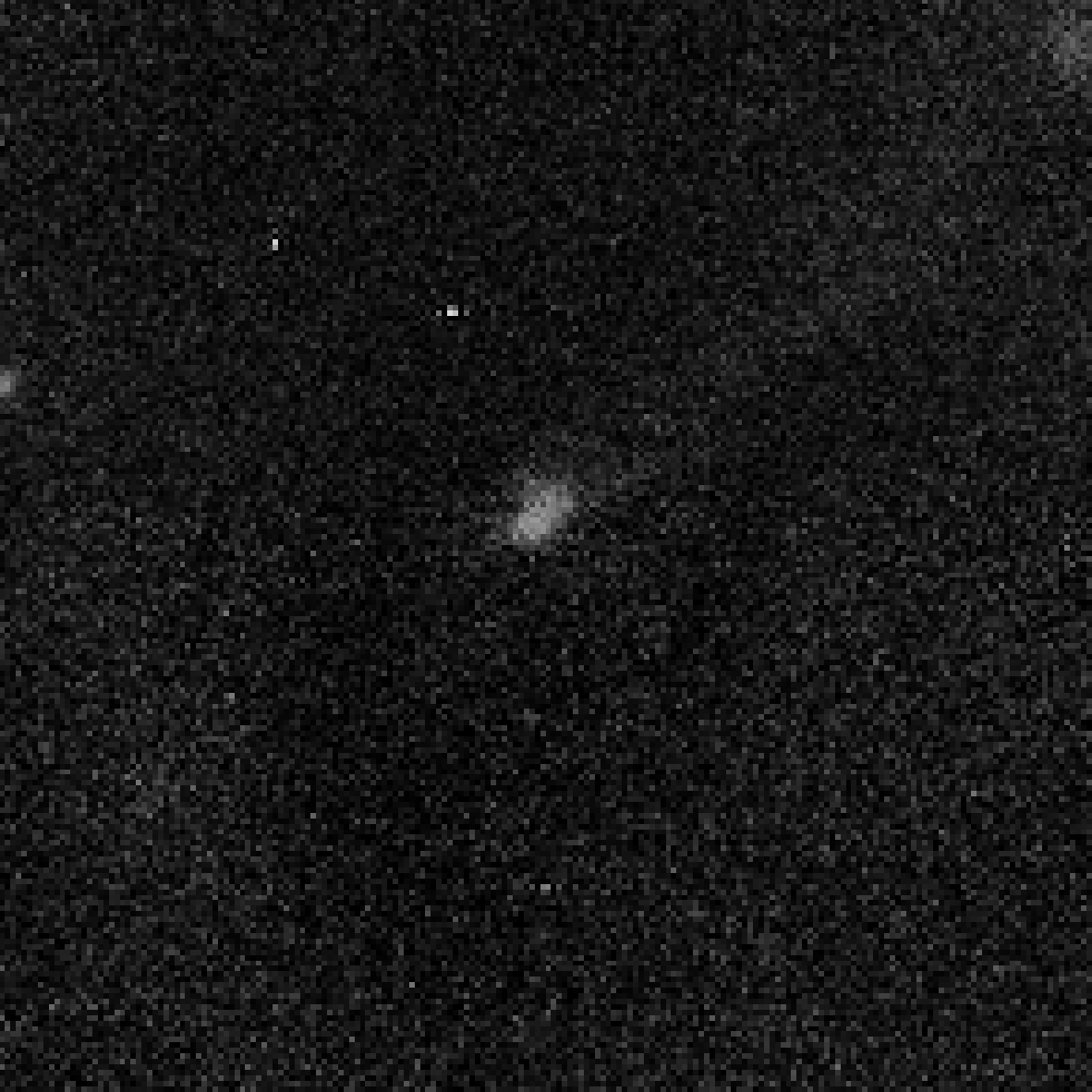}{0.25\textwidth}{HH~1240}
          \fig{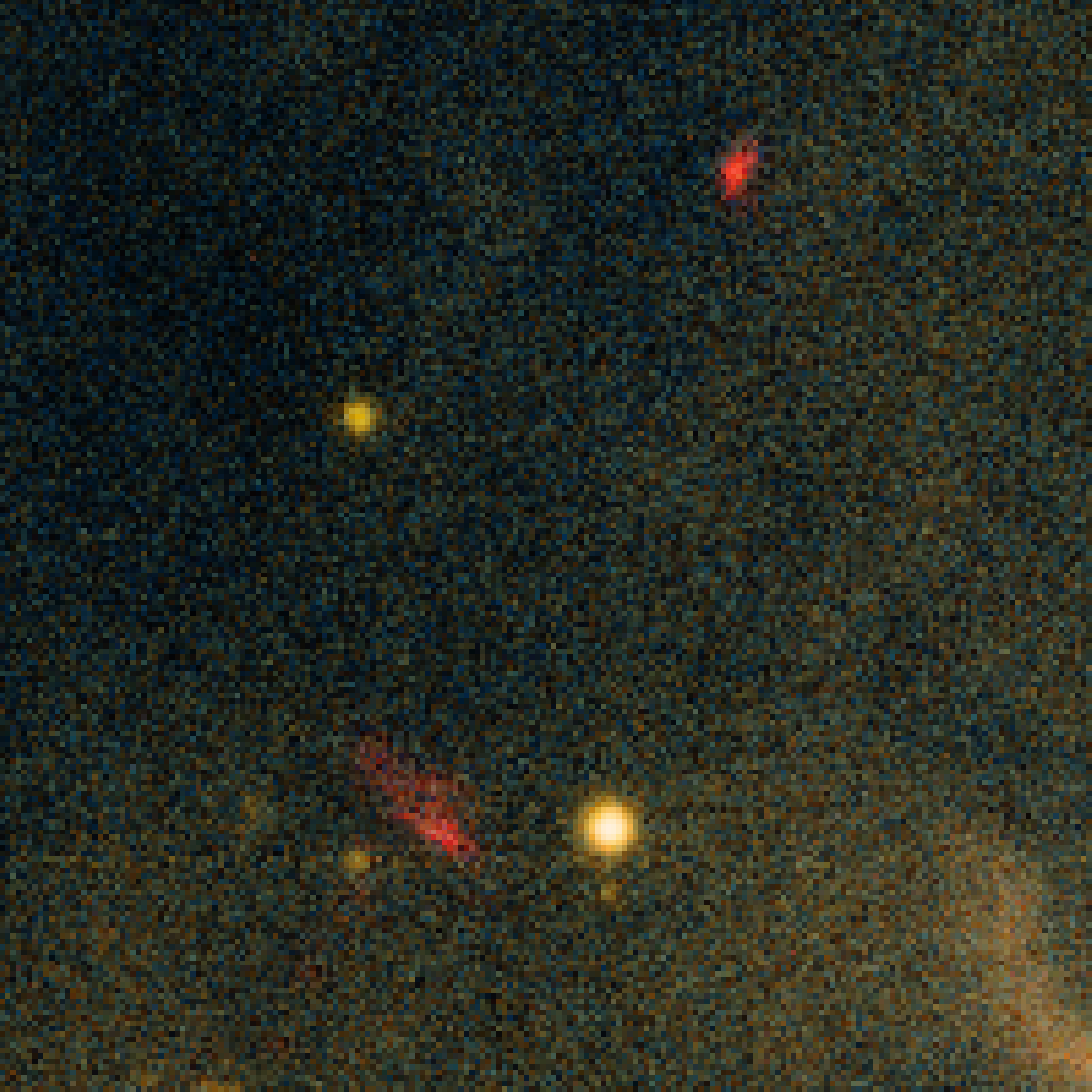}{0.25\textwidth}{HH~1243 \& 1242}
          \fig{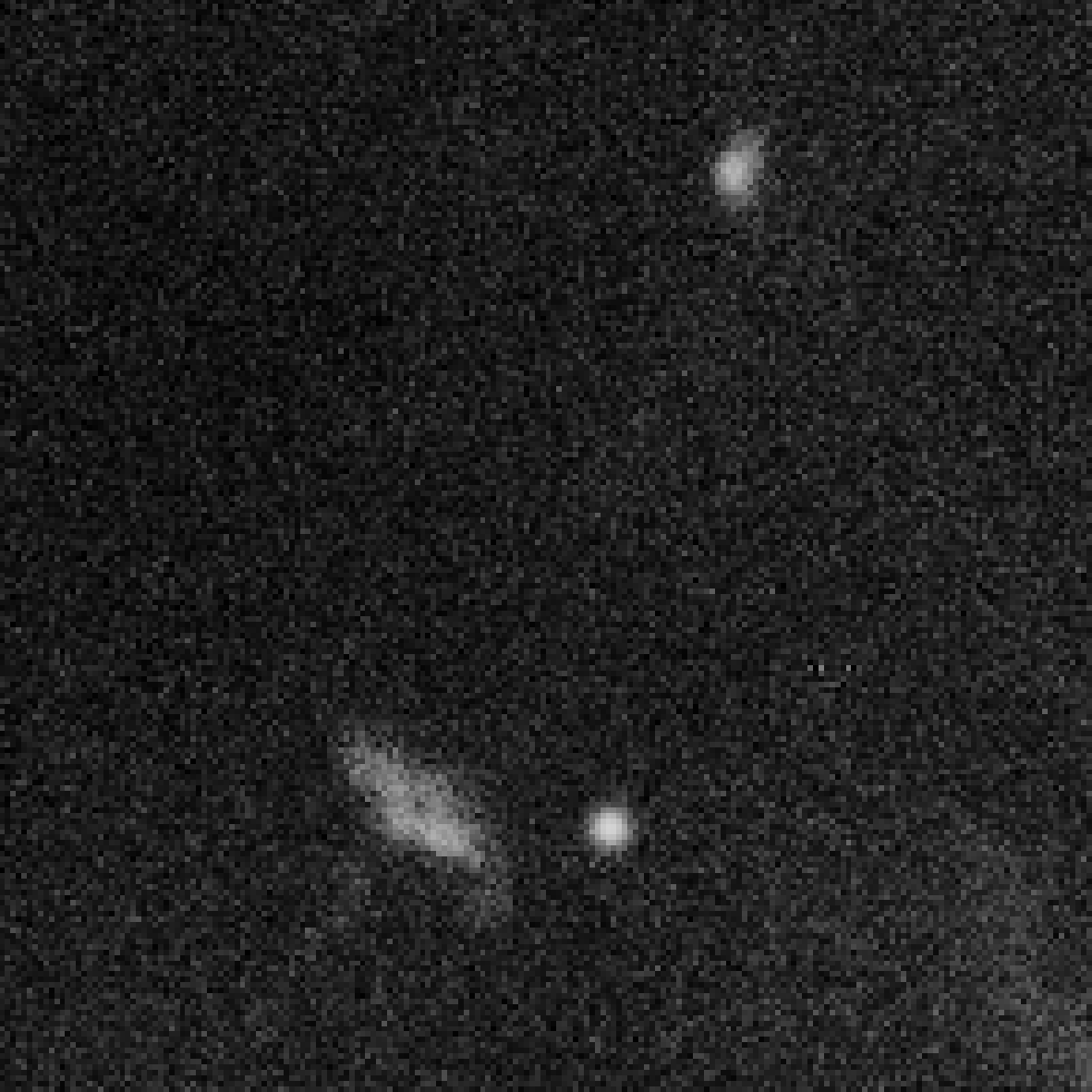}{0.25\textwidth}{HH~1243 \& 1242}
          }
\caption{Cutouts of six new HH objects in the highly obscured area south of the Cir-MMS complex.  Cutout size and color schemes are the same as for Figure~\ref{fig:cutouts_AABB}.
\label{fig:cutouts_MMWW}}
\end{figure}

HH~1238-1243:  These faint HH objects (Figure~\ref{fig:cutouts_MMWW}) are embedded in the dusty region south of the Cir-MMS complex.  None appear to align with previously known outflows.  About a dozen YSOs and YSO candidates lie within this region, making it nearly impossible to identify the progenitor for each HH object.

\begin{figure}
\gridline{\fig{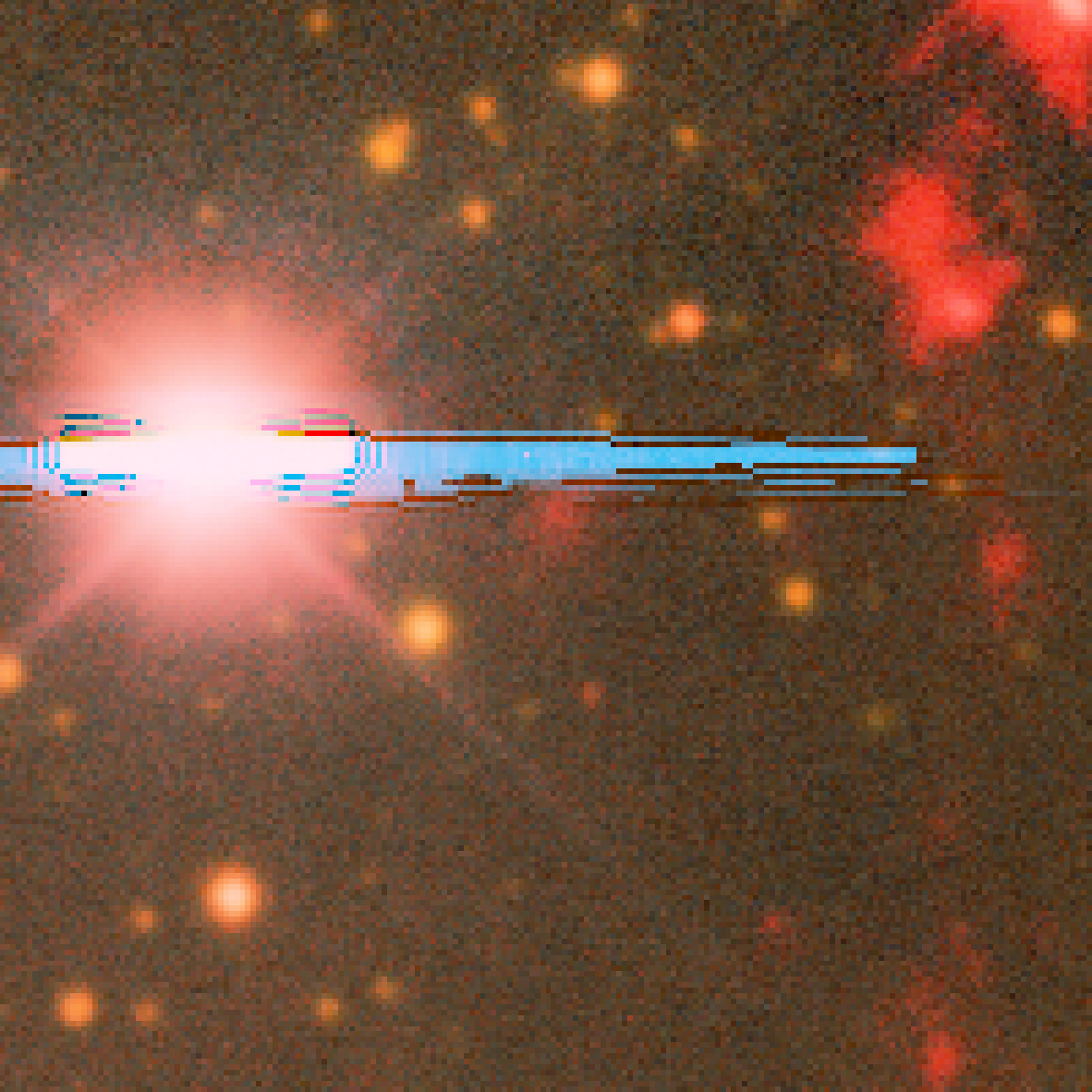}{0.25\textwidth}{HH~1244 \& HH~76}
          \fig{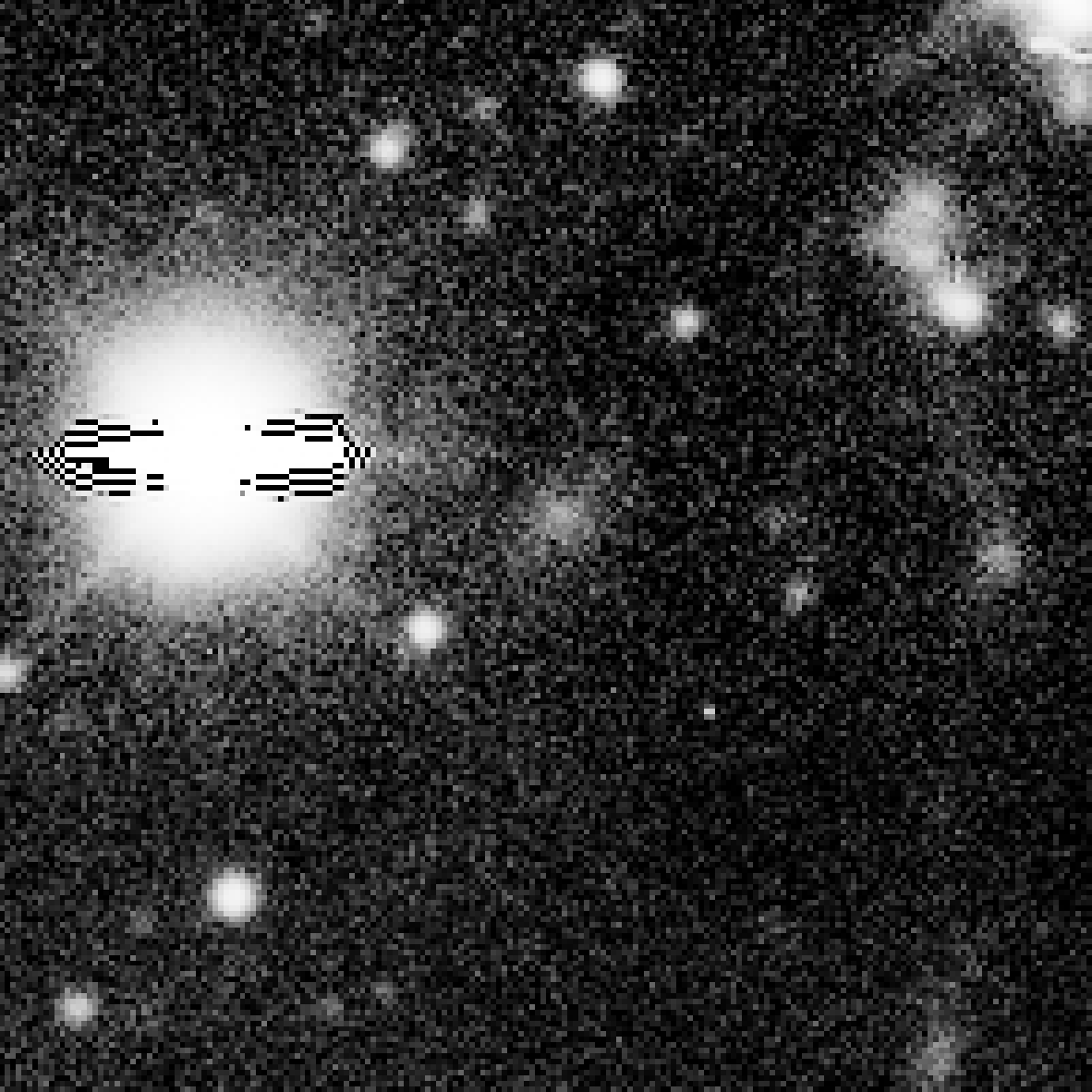}{0.25\textwidth}{HH~1244 \& HH~76}
          \fig{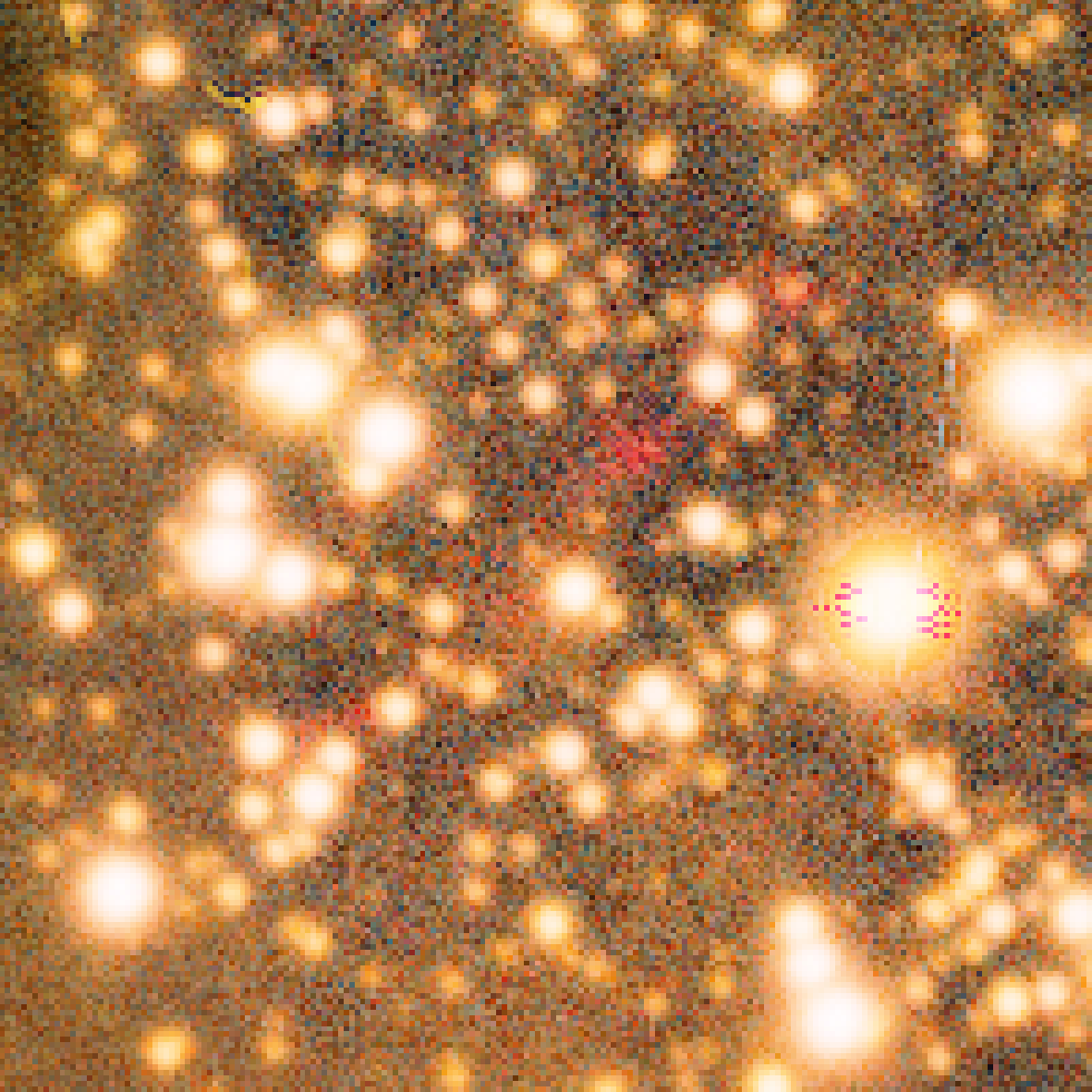}{0.25\textwidth}{HH~1245}
          \fig{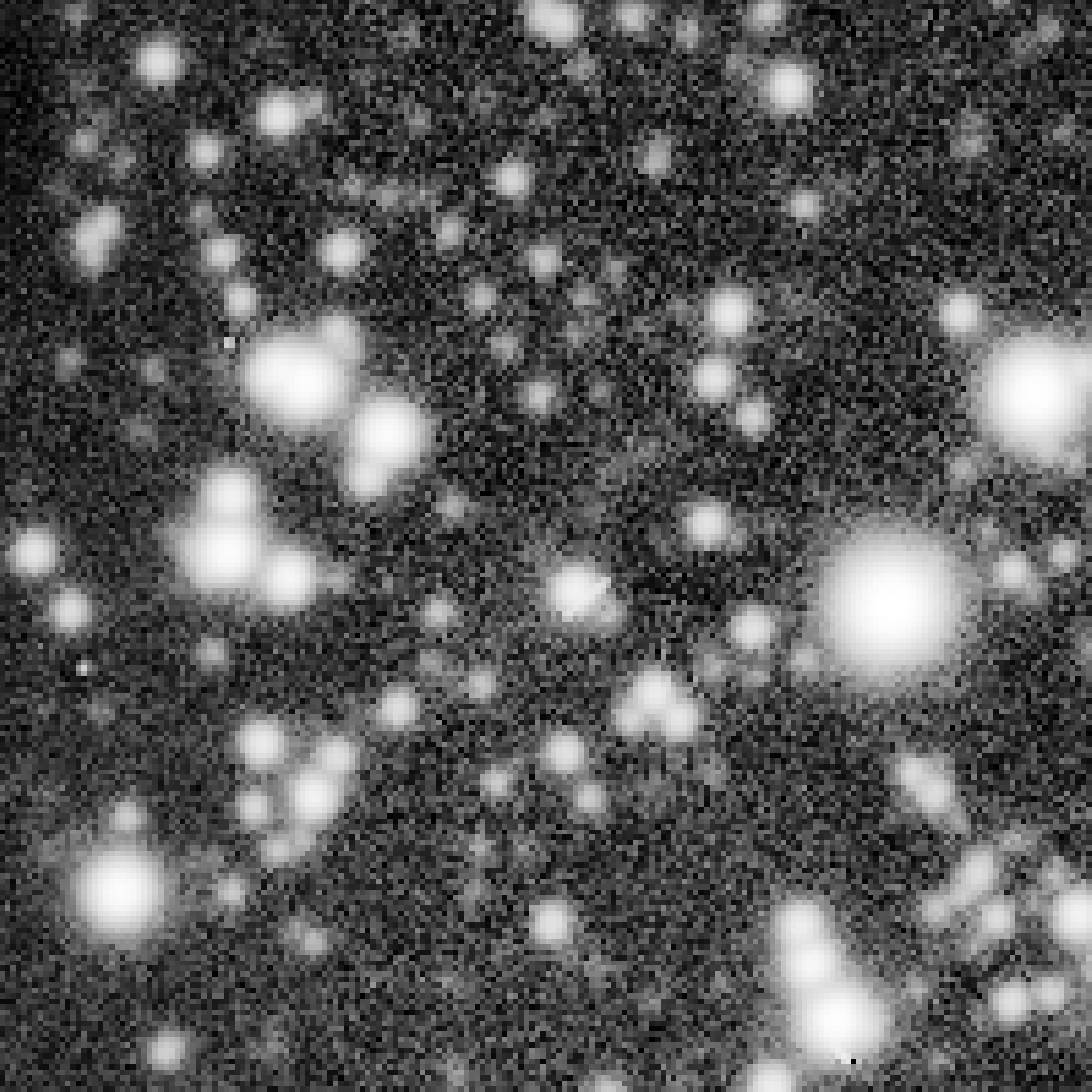}{0.25\textwidth}{HH~1245}
          }
\caption{Cutouts of new HH objects within Cir-W.  HH~1244 is just east of the HH~76 outflow (seen just below the blue charge bleed in the color image above).  And HH~1245 is west of the HH~1246-1249 group.  Cutout size and color schemes are the same as for Figure~\ref{fig:cutouts_AABB}.
\label{fig:cutouts_FFJJ}}
\end{figure}

HH~1244:  This object (Figure~\ref{fig:cutouts_FFJJ}) is just to the east of HH~76g; however, it is not clear if it is part of the HH~76 outflow.

HH~1245:  This object (Figure~\ref{fig:cutouts_FFJJ}) is faint but detected in \ha\ and \stwo.  It is isolated from other HH objects.  The nearest YSO candidate, \object{WISEP J150206.96-632047.7}, is 4.1\arcmin\ to the east.

\begin{figure}
\gridline{\fig{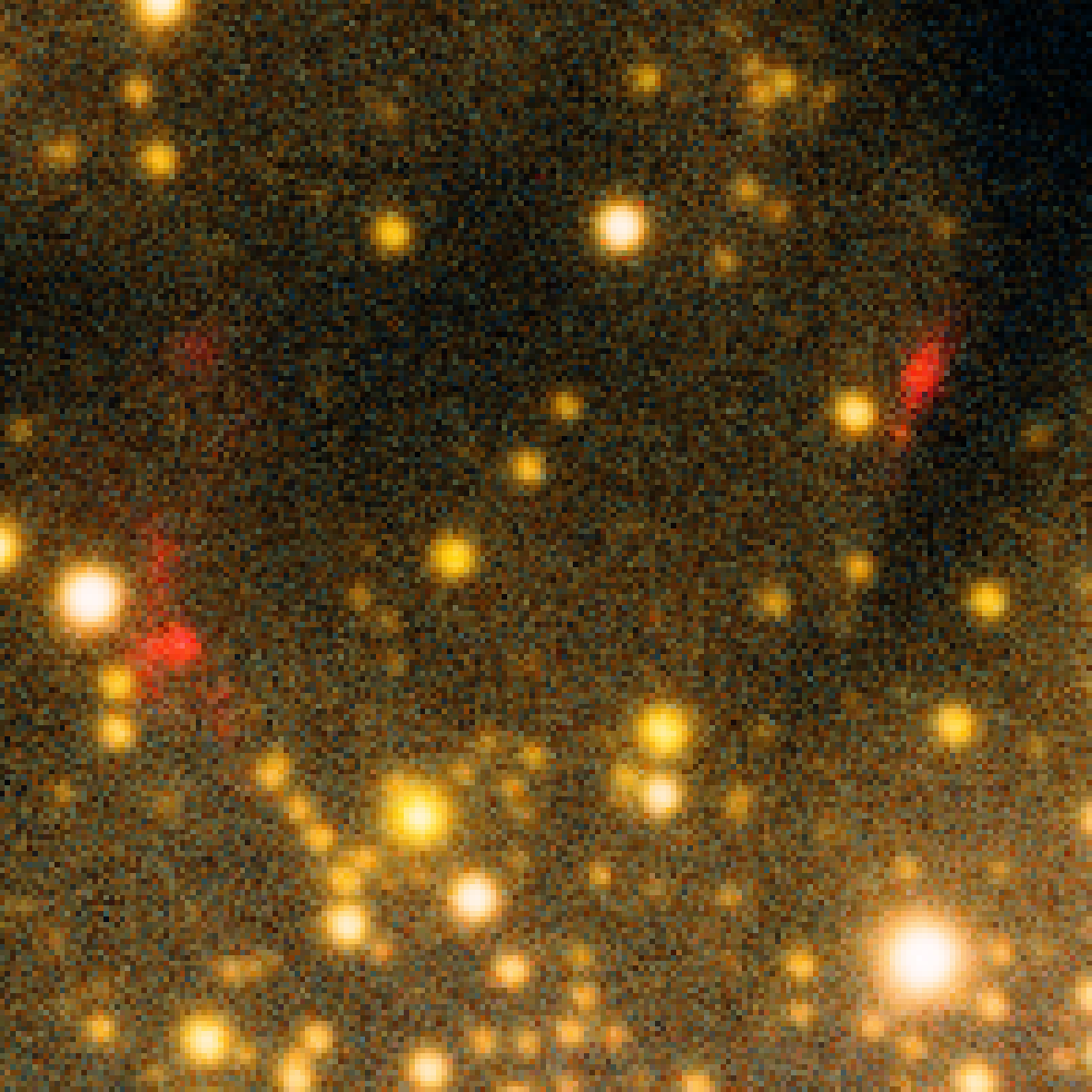}{0.25\textwidth}{HH~1248 \& 1246}
          \fig{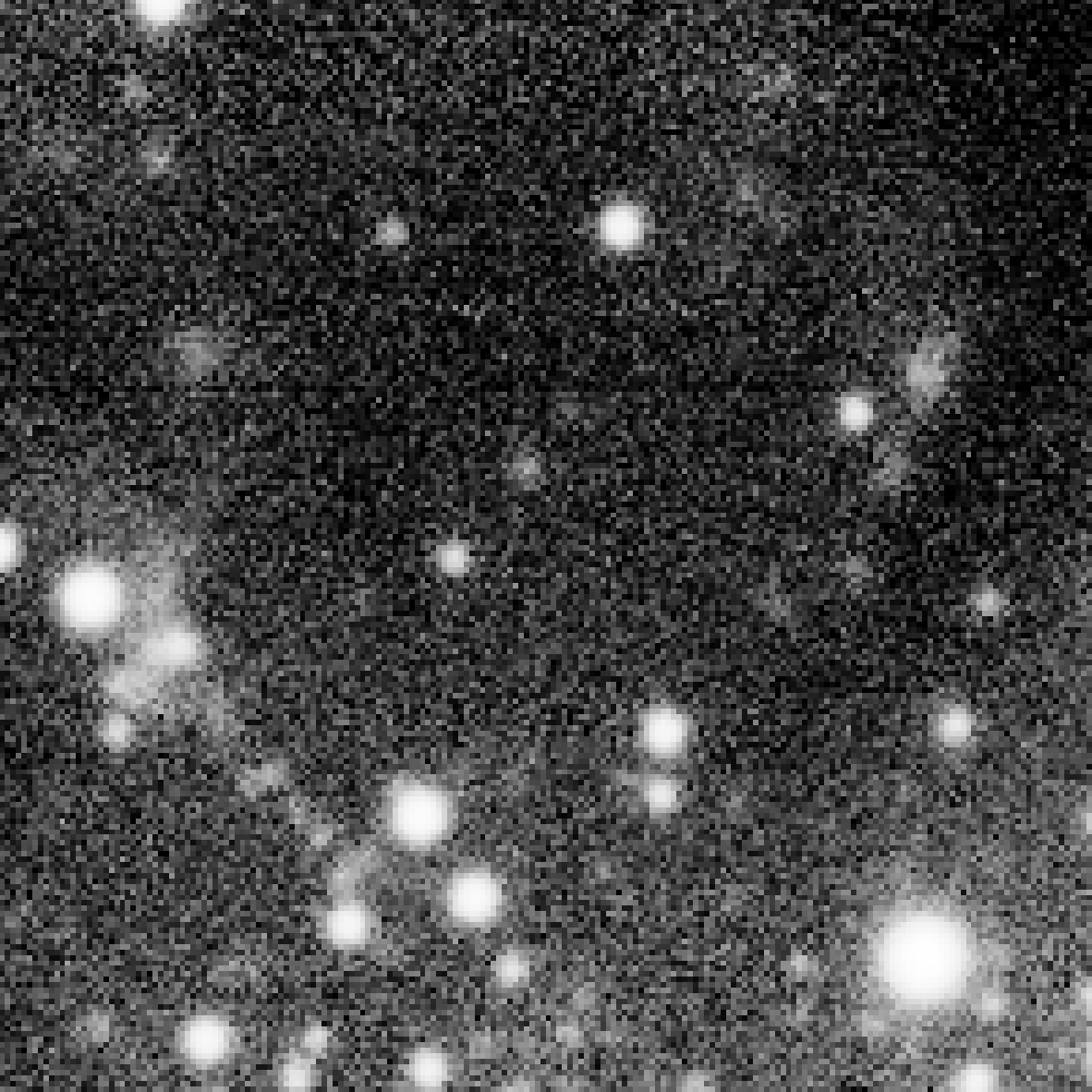}{0.25\textwidth}{HH~1248 \& 1246}
          \fig{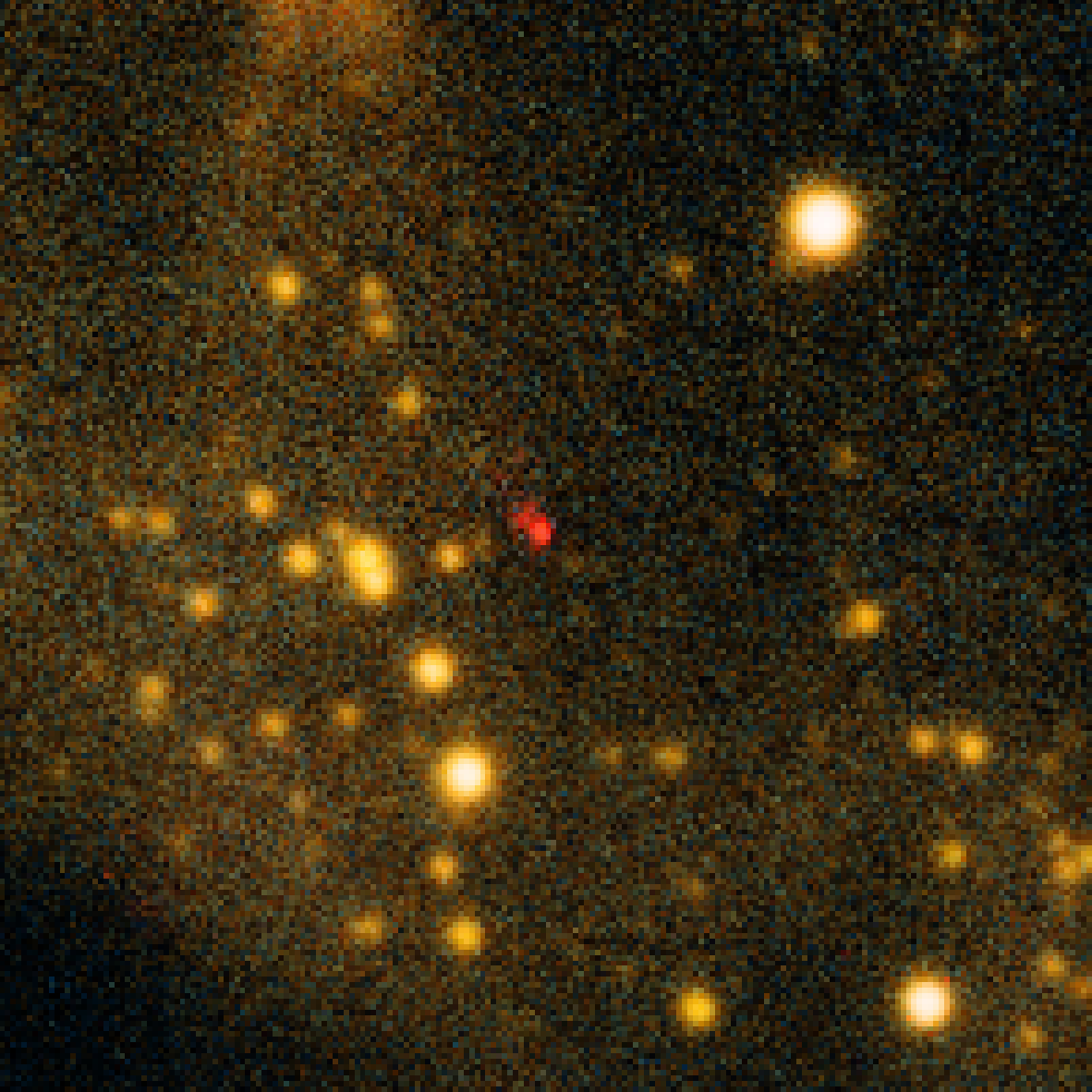}{0.25\textwidth}{HH~1247}
          \fig{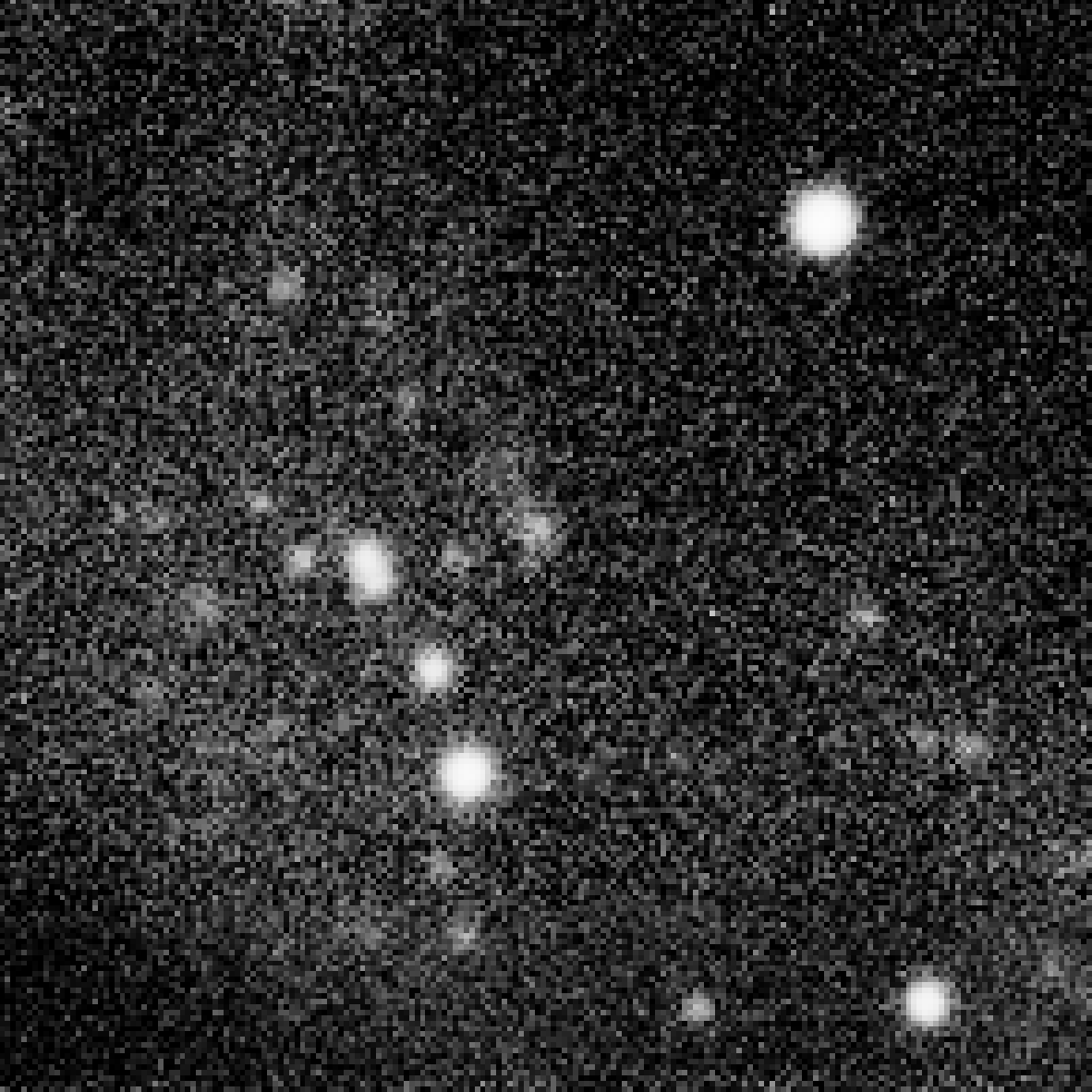}{0.25\textwidth}{HH~1247}
          }
\gridline{\fig{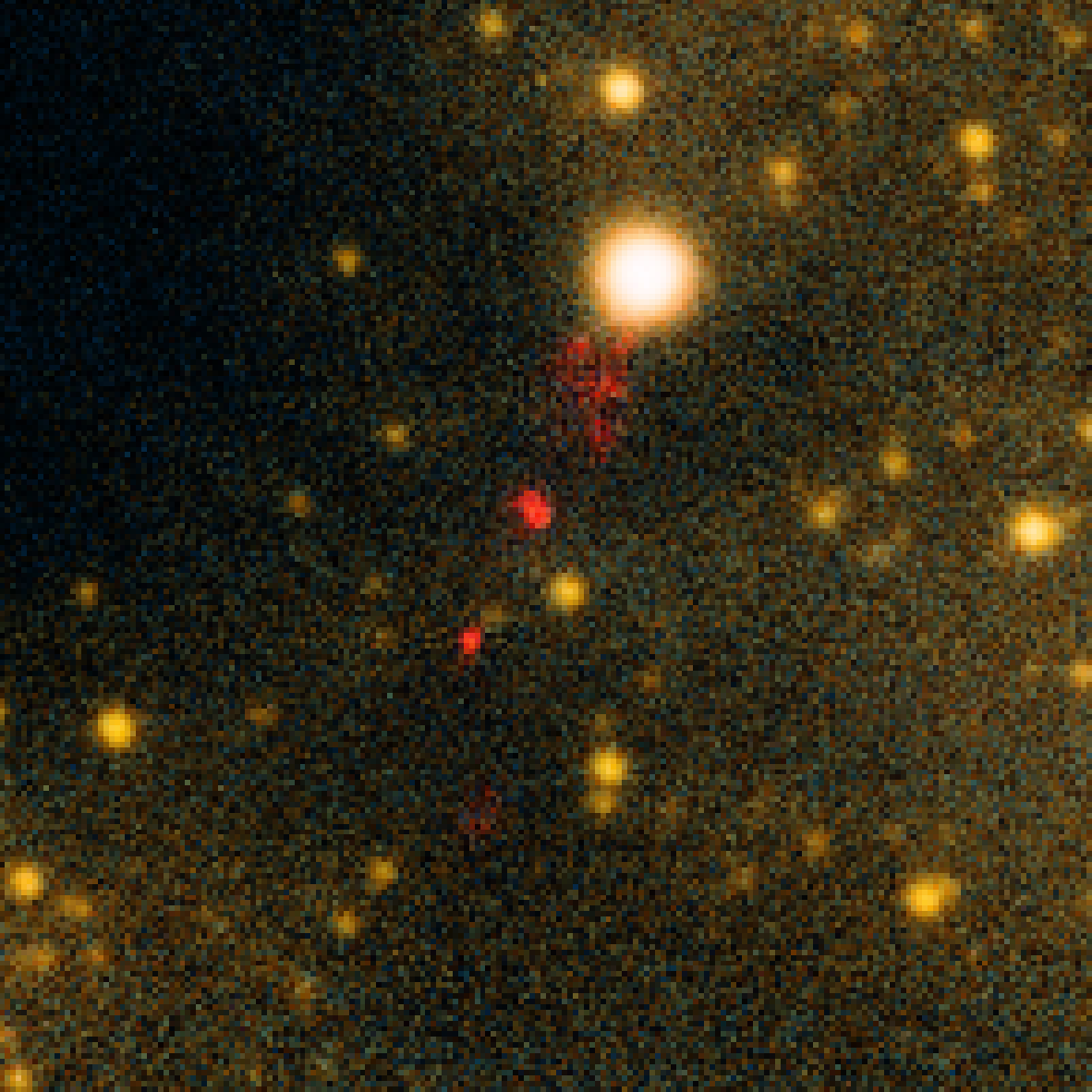}{0.25\textwidth}{HH~1249}
          \fig{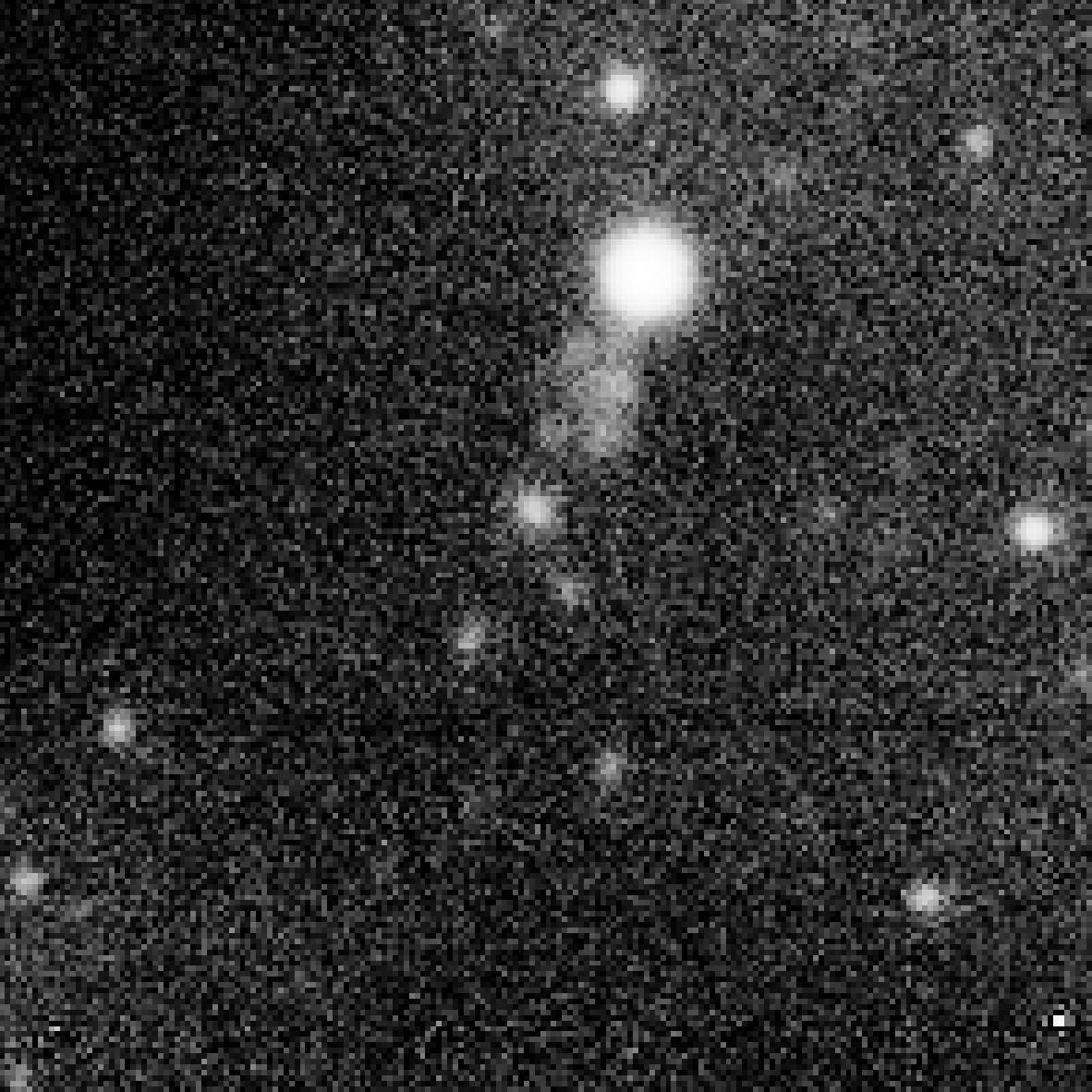}{0.25\textwidth}{HH~1249}
          \fig{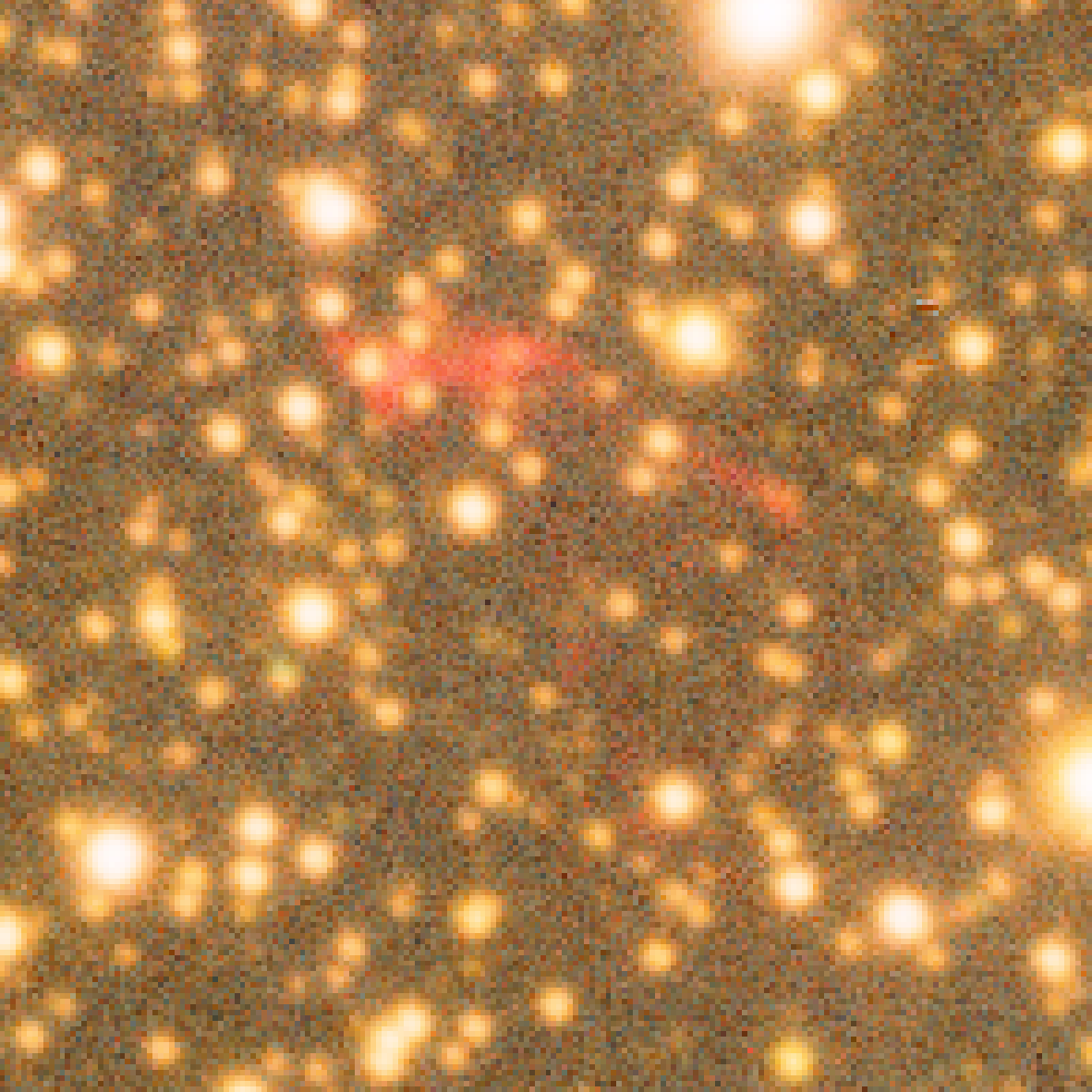}{0.25\textwidth}{HH~1250}
          \fig{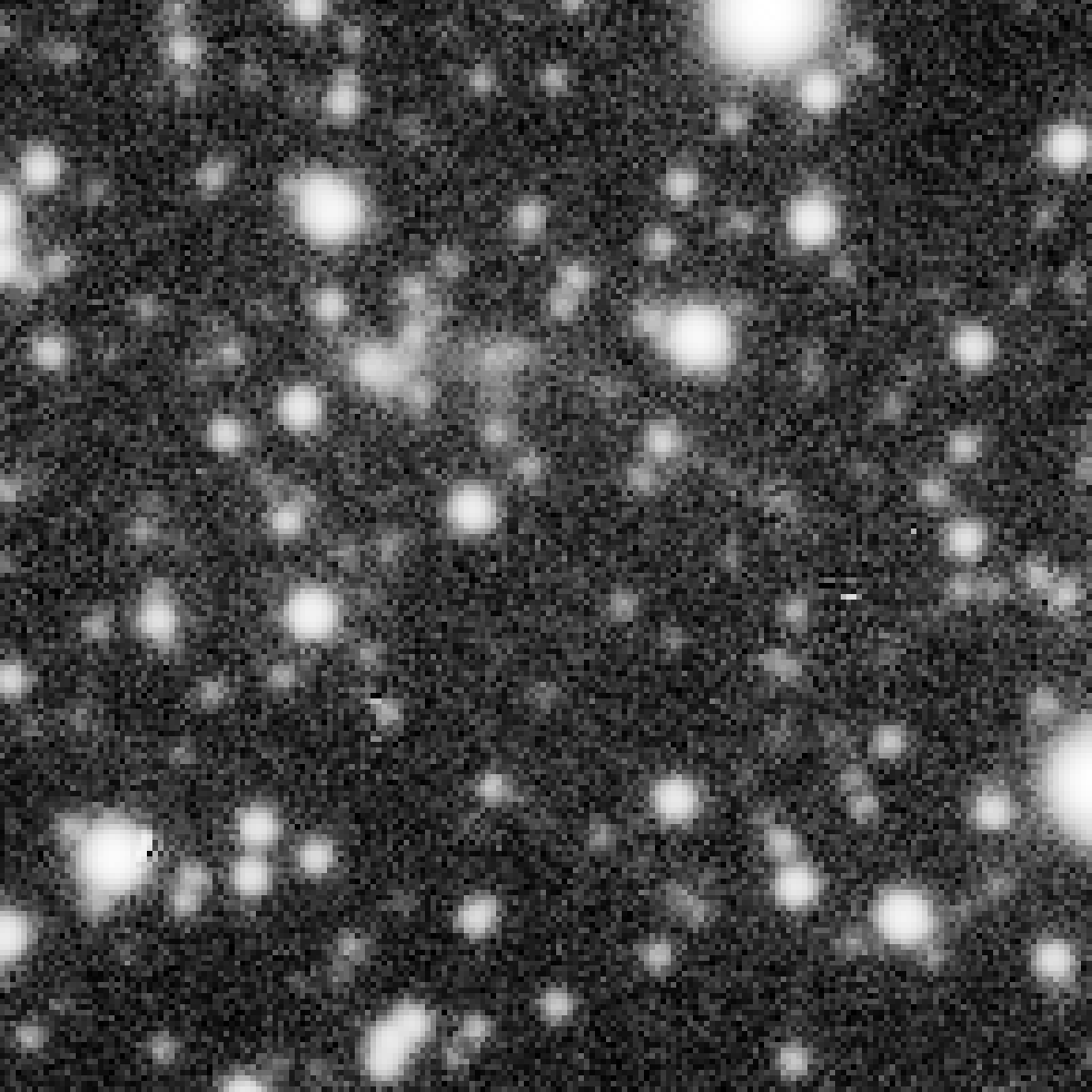}{0.25\textwidth}{HH~1250}
          }
\caption{Cutouts of five new HH objects between the VdBH~65b and Cir-MMS regions.  Cutout size and color schemes are the same as for Figure~\ref{fig:cutouts_AABB}.
\label{fig:cutouts_TY}}
\end{figure}

HH~1246-1249:  These outflows (Figure~\ref{fig:cutouts_TY}) are clustered around the YSO candidates \object{2MASS J15021535-6320284} and \object{WISEP J150206.96-632047.7}.  

HH~1250:  This HH object (Figure~\ref{fig:cutouts_TY}) is relatively isolated.  The nearest IR sources are YSO candidate \object{WISEP J150238.02-631900.3} and \object{IRAS 14580-6303}, which are 2.5\arcmin\ and 4.4\arcmin\ away respectively.

\begin{figure}
\gridline{\fig{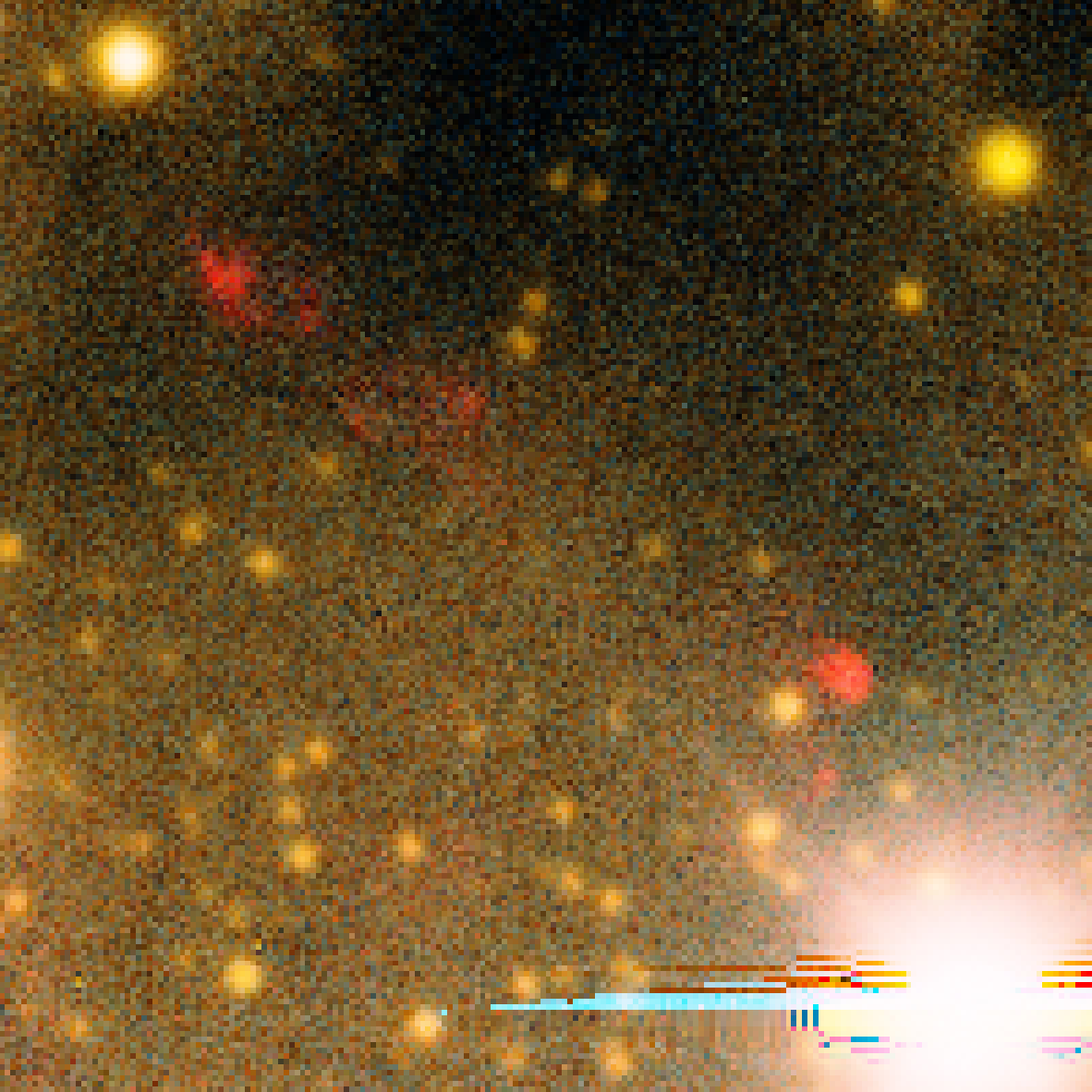}{0.25\textwidth}{HH~1252 \& 1251}
          \fig{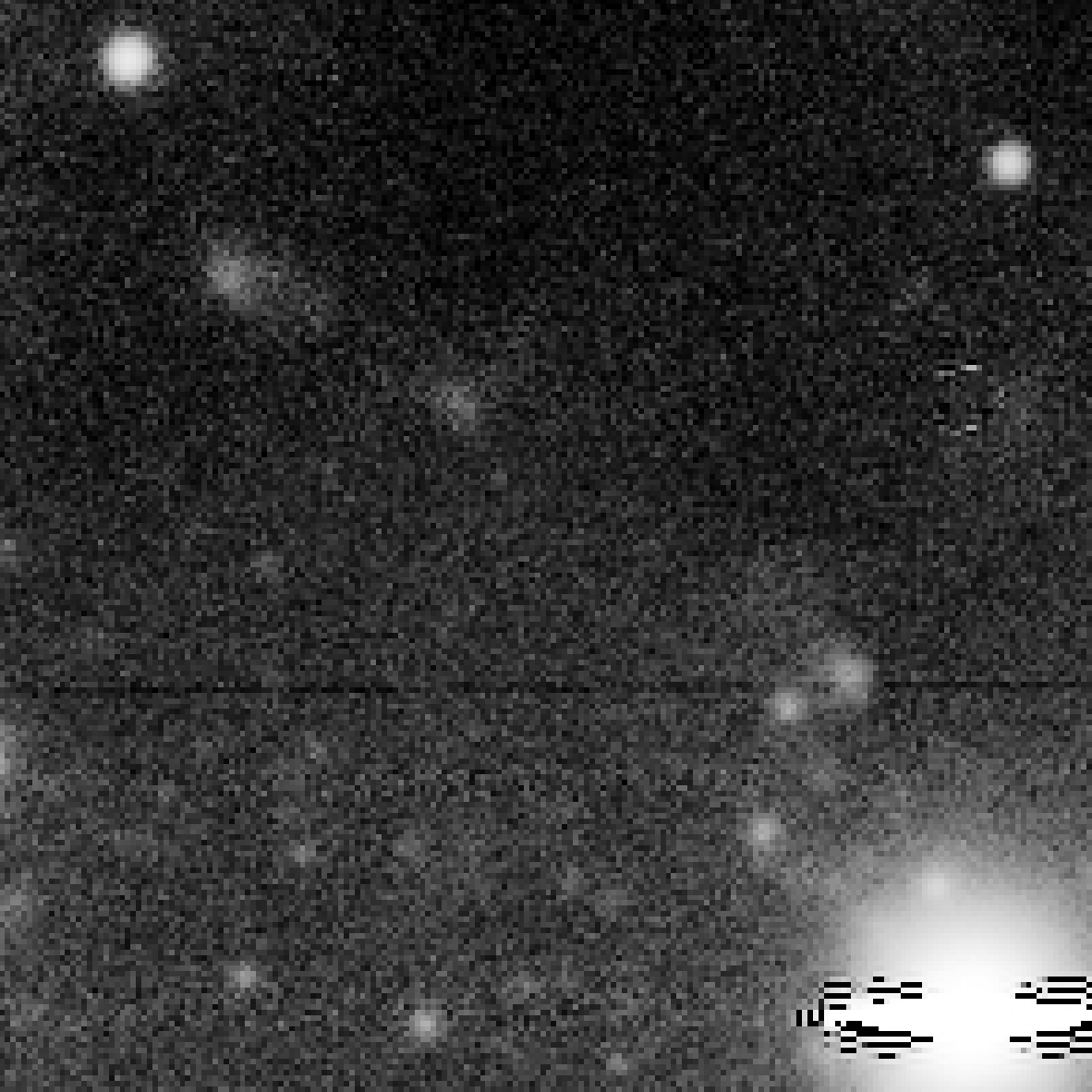}{0.25\textwidth}{HH~1252 \& 1251}
          \fig{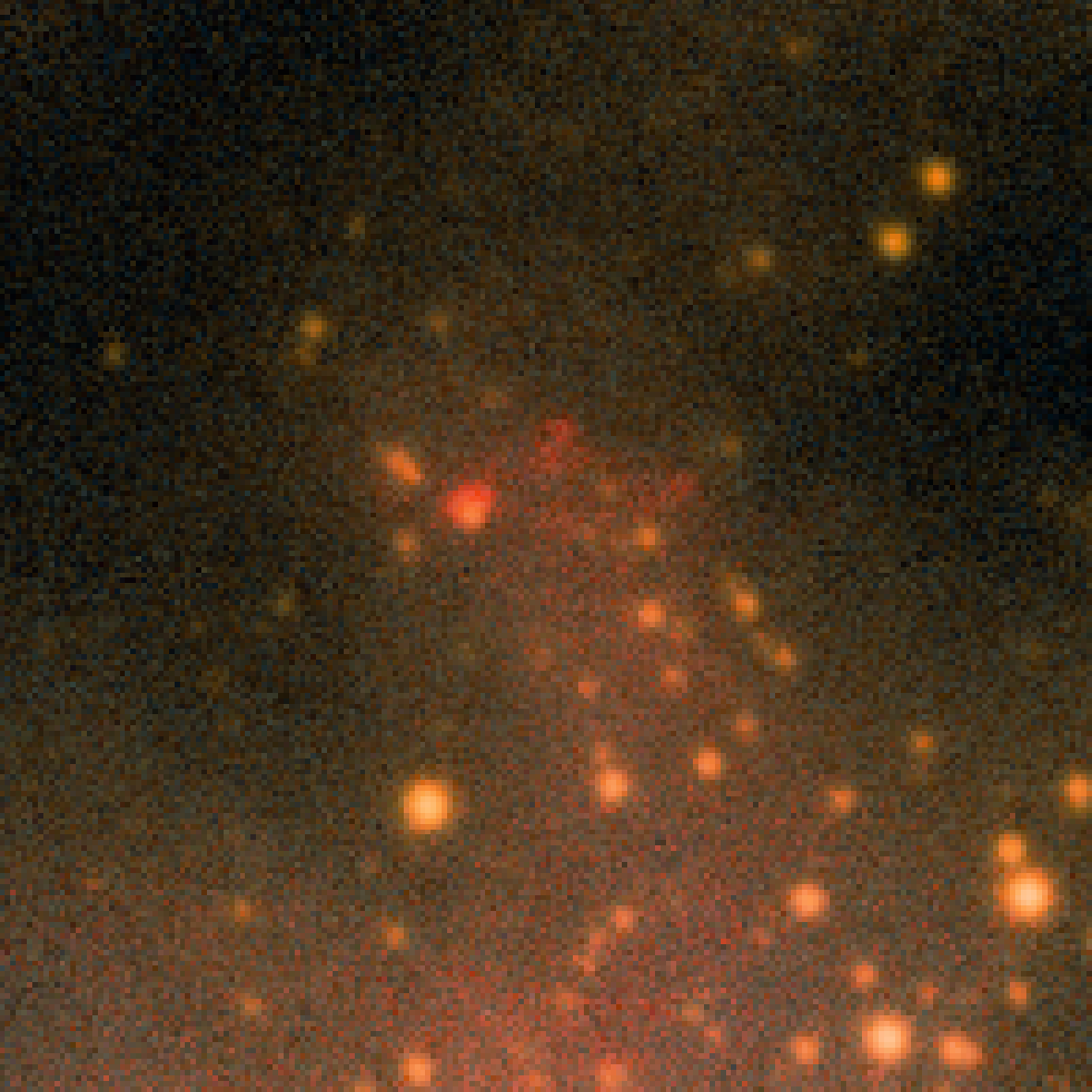}{0.25\textwidth}{HH~1253}
          \fig{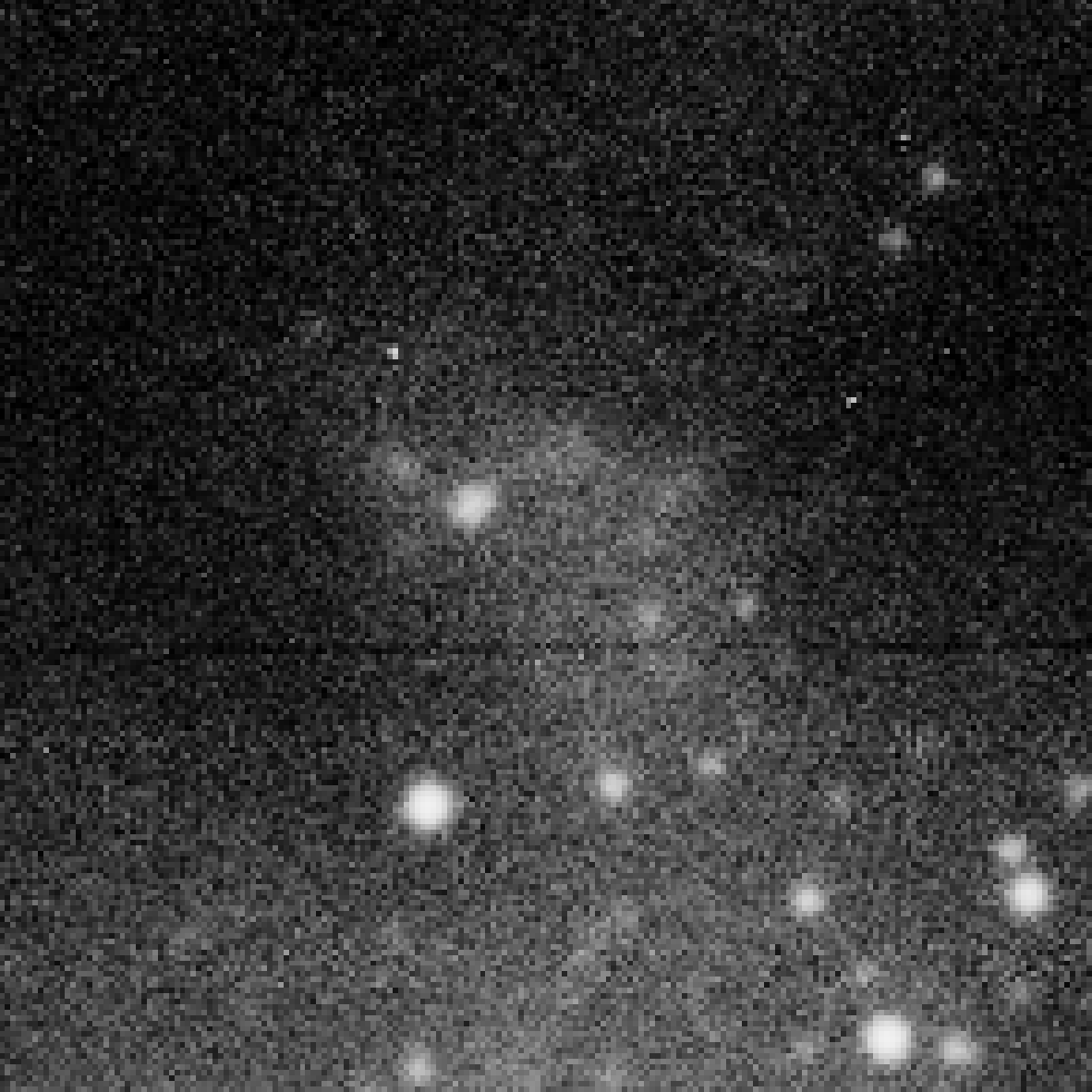}{0.25\textwidth}{HH~1253}
          }
\gridline{\fig{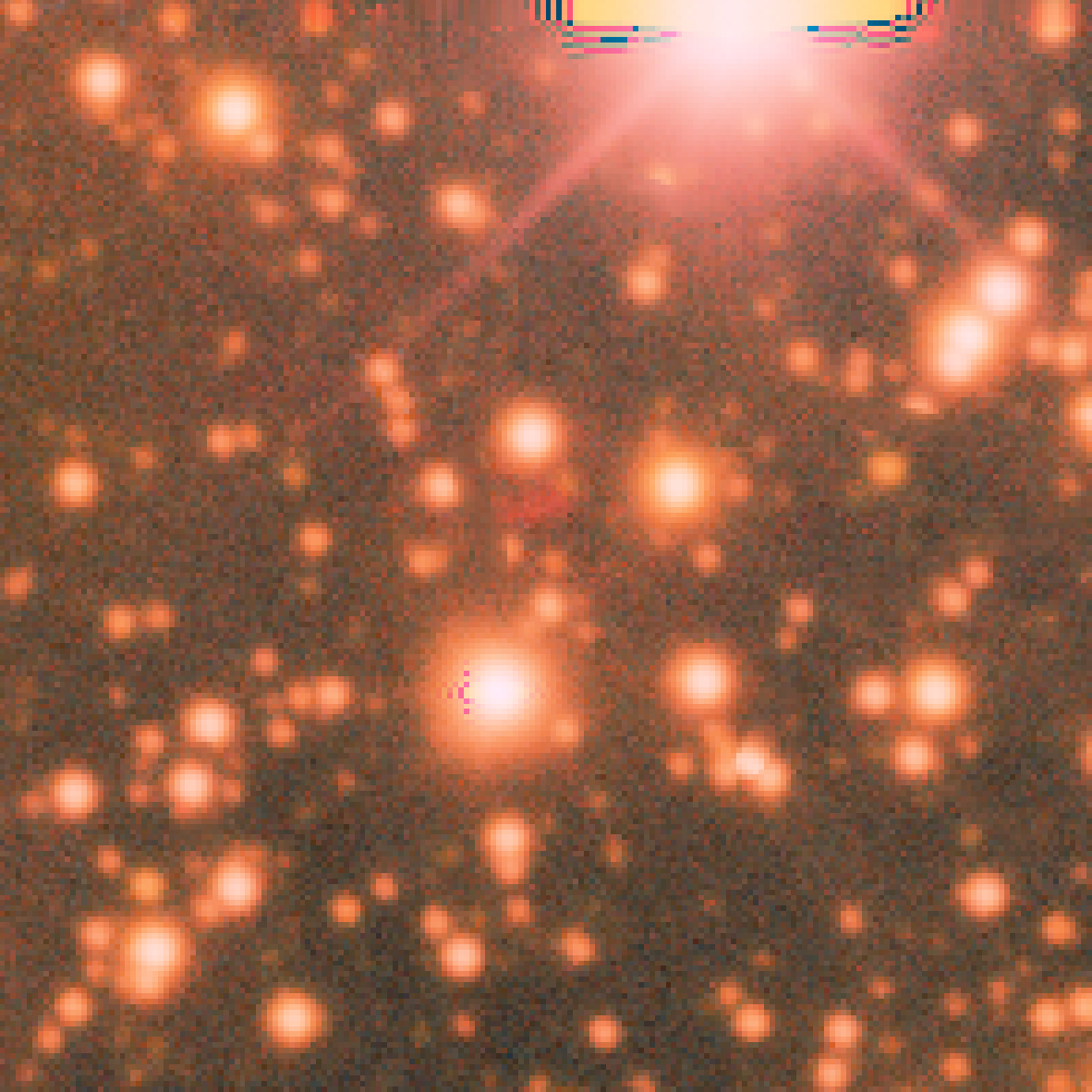}{0.25\textwidth}{HH~1254}
          \fig{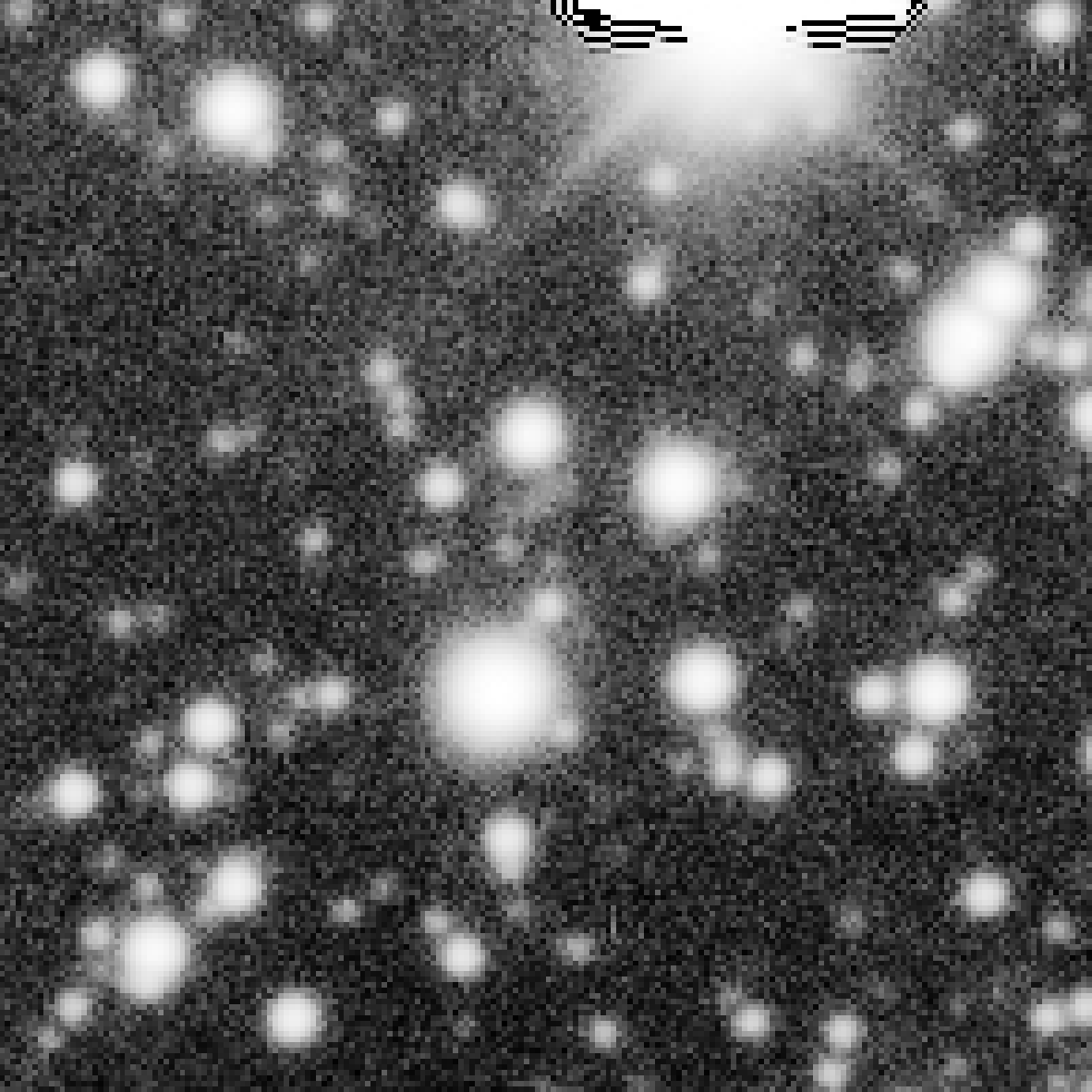}{0.25\textwidth}{HH~1254}
          \fig{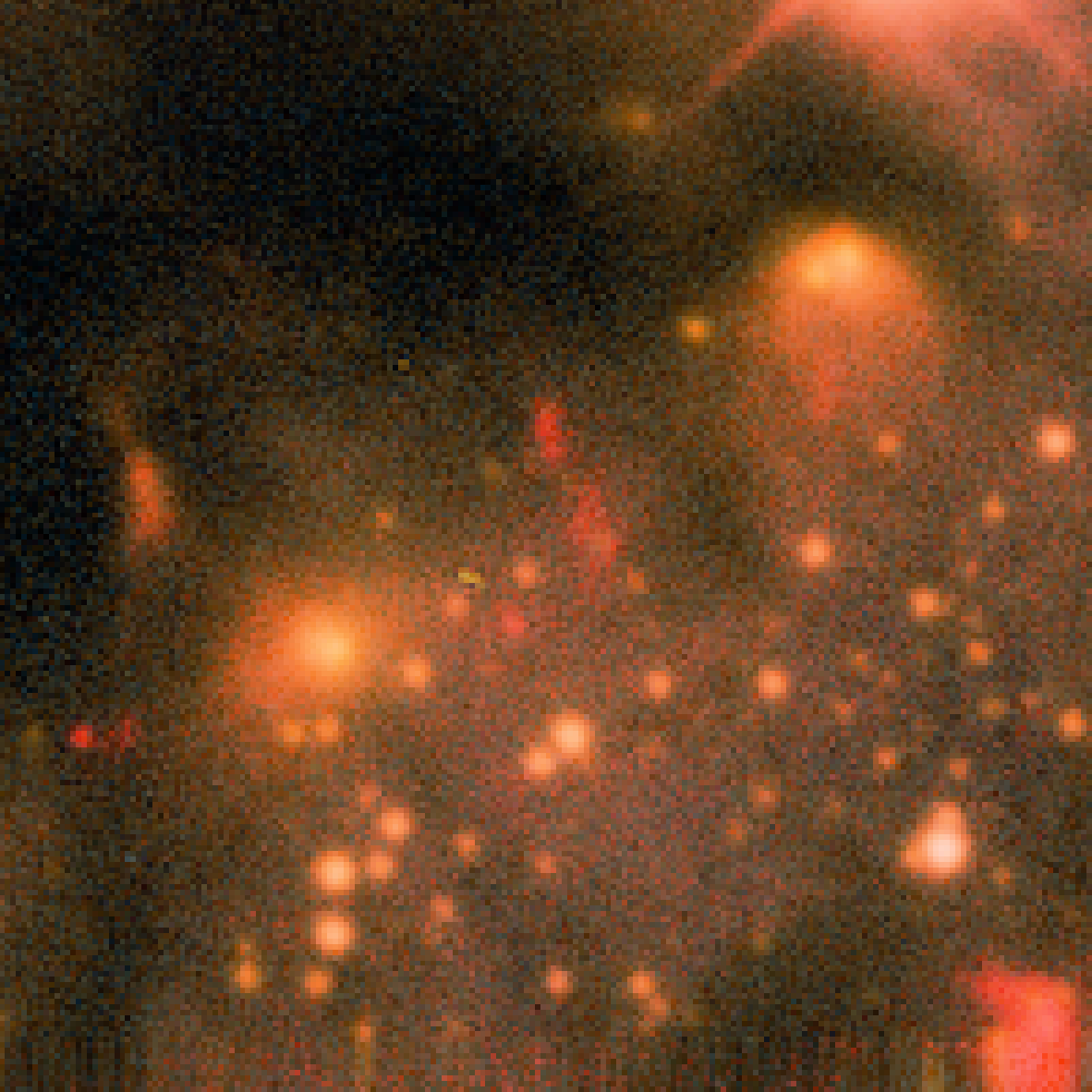}{0.25\textwidth}{HH~1255}
          \fig{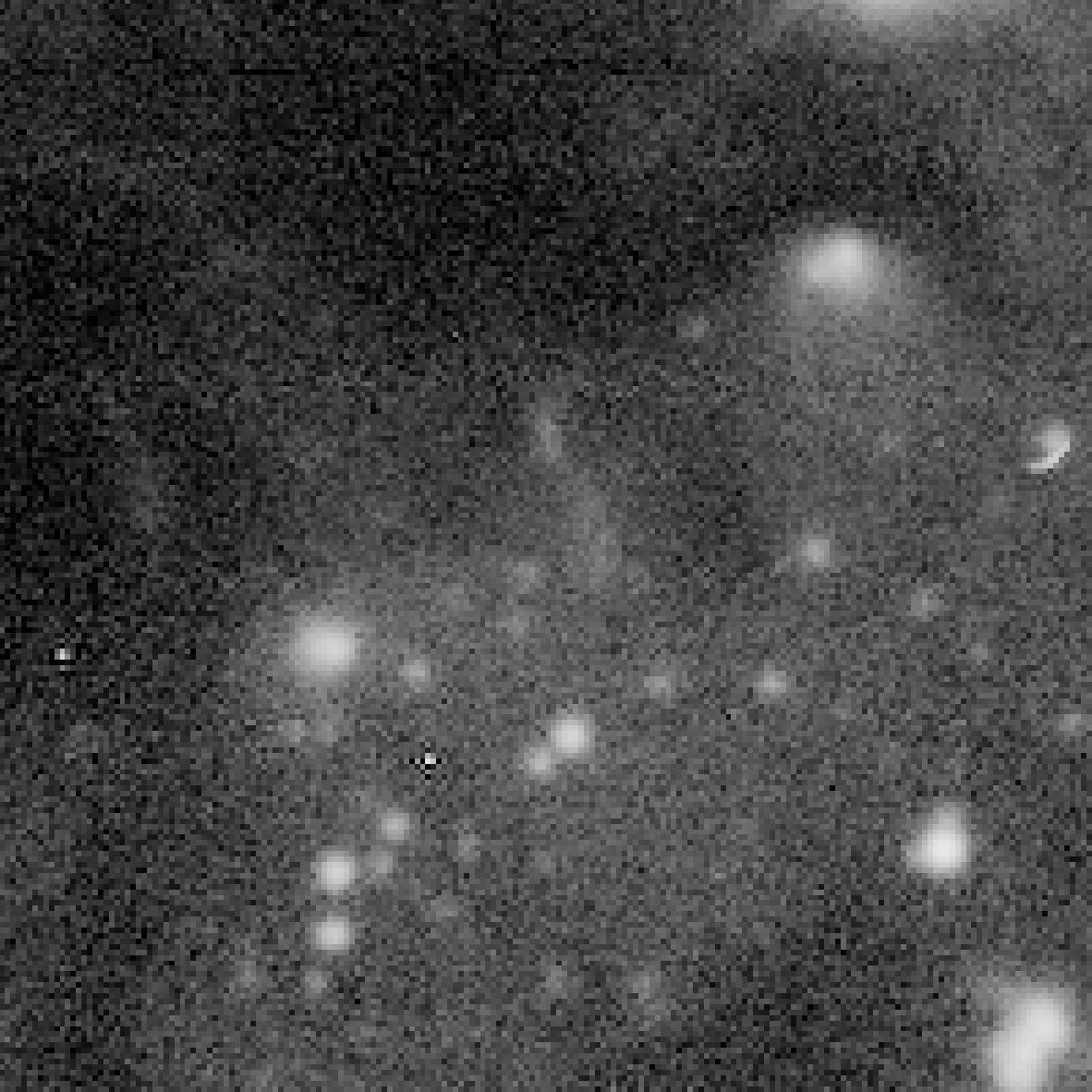}{0.25\textwidth}{HH~1255}
          }
\gridline{\fig{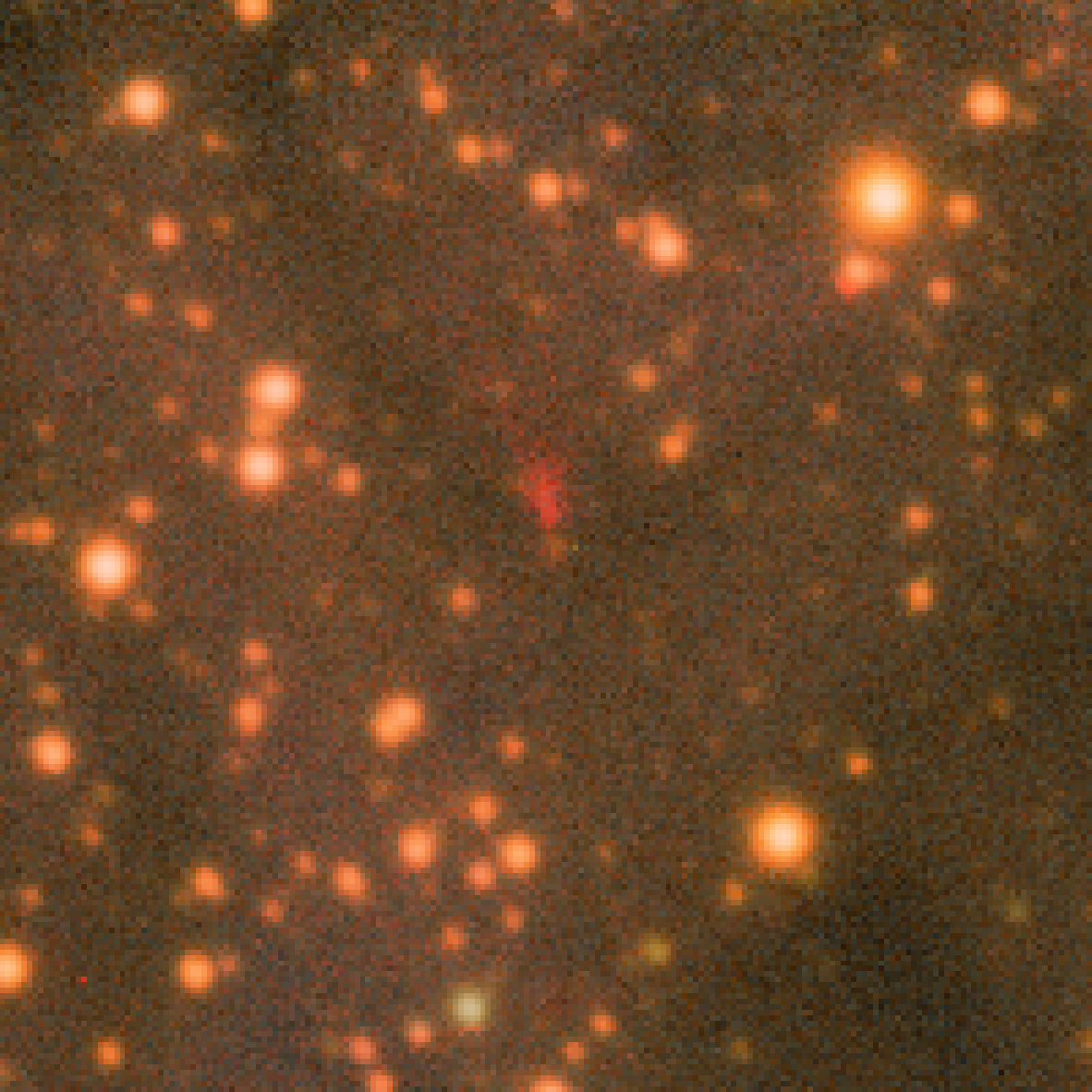}{0.25\textwidth}{HH~1256}
          \fig{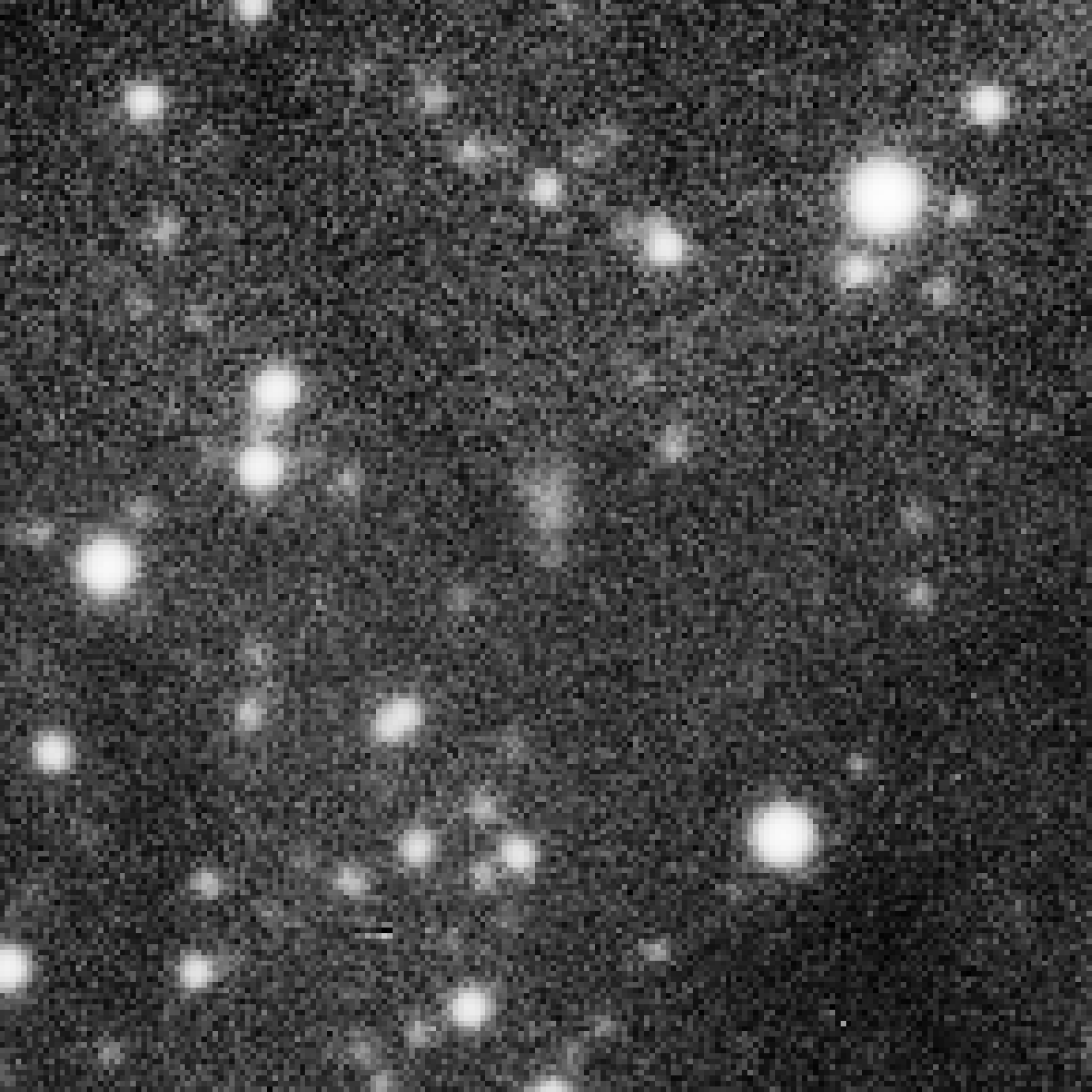}{0.25\textwidth}{HH~1256}
          \fig{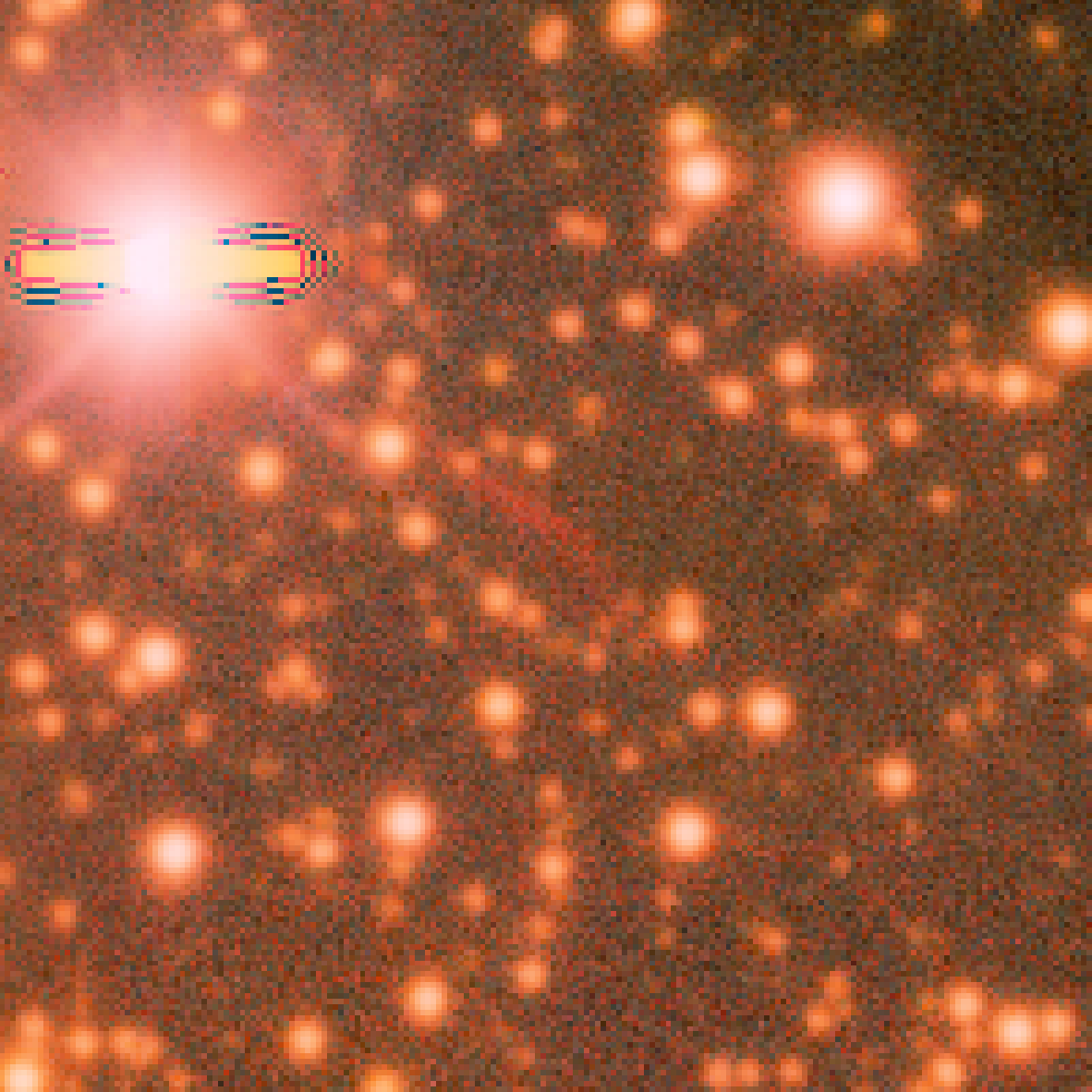}{0.25\textwidth}{HH~1257}
          \fig{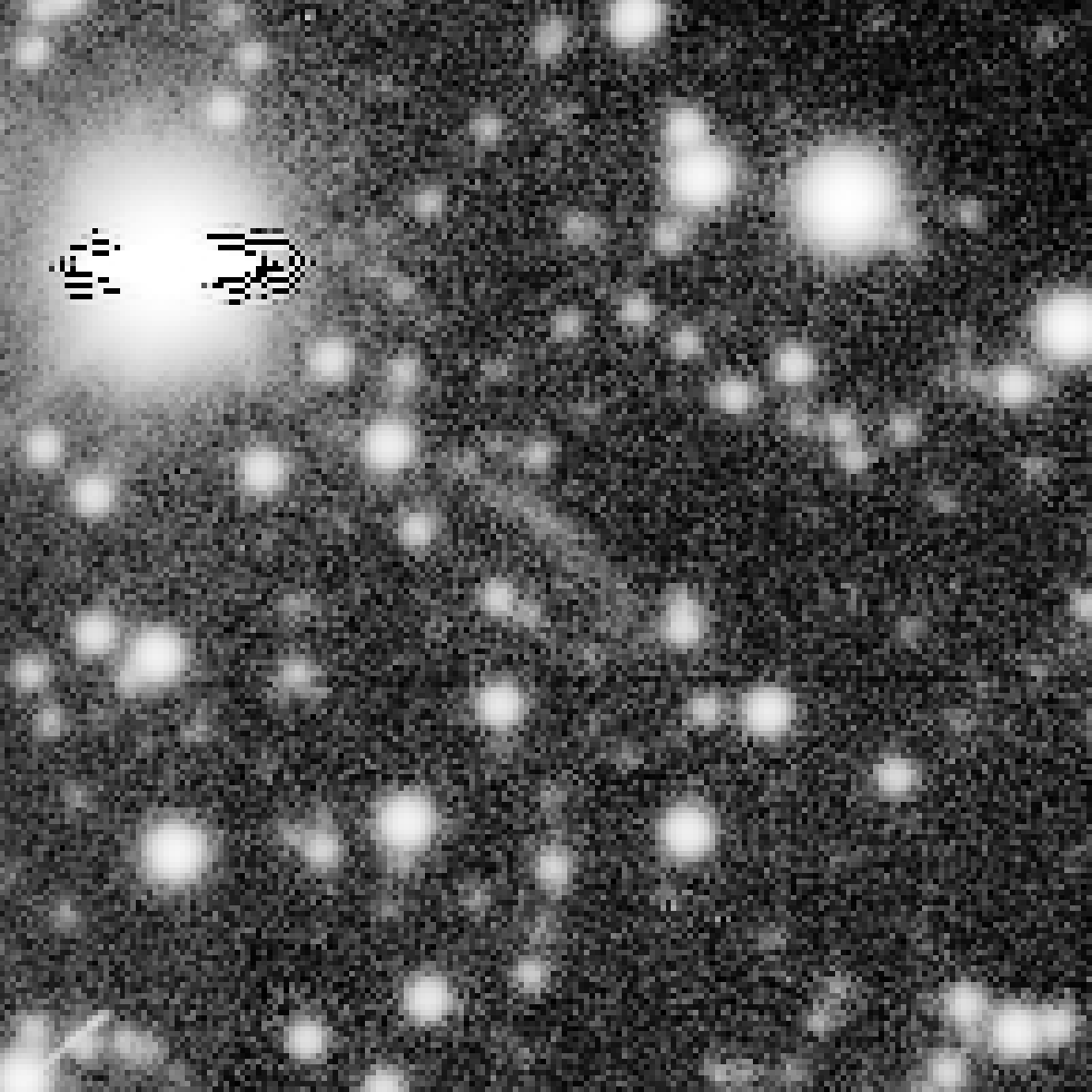}{0.25\textwidth}{HH~1257}
          }
\caption{Cutouts of seven new HH objects in the VdBH~65b region.  Cutout size and color schemes are the same as for Figure~\ref{fig:cutouts_AABB}.
\label{fig:cutouts_EVV}}
\end{figure}

HH~1251-1257:  These HH objects (Figure~\ref{fig:cutouts_EVV}) are clustered around Herbig Ae/Be star \object{VdBH 65b}, as well as the YSO candidates \object{2MASS J15032868-6323164} and \object{2MASS J15033229-6323565}.  Two additional YSO candidates, \object{2MASS J15034569-6323413} and \object{2MASS J15034898-6323436}, are about 3\arcmin\ to the east. This region also contains HH~140-143, previously discovered by \citet{1994A&A...290..605R}.  HH~1254 and 1257 also likely originate from sources within this group, even though they are about 3-5\arcmin\ distant.

\begin{figure}
\gridline{\fig{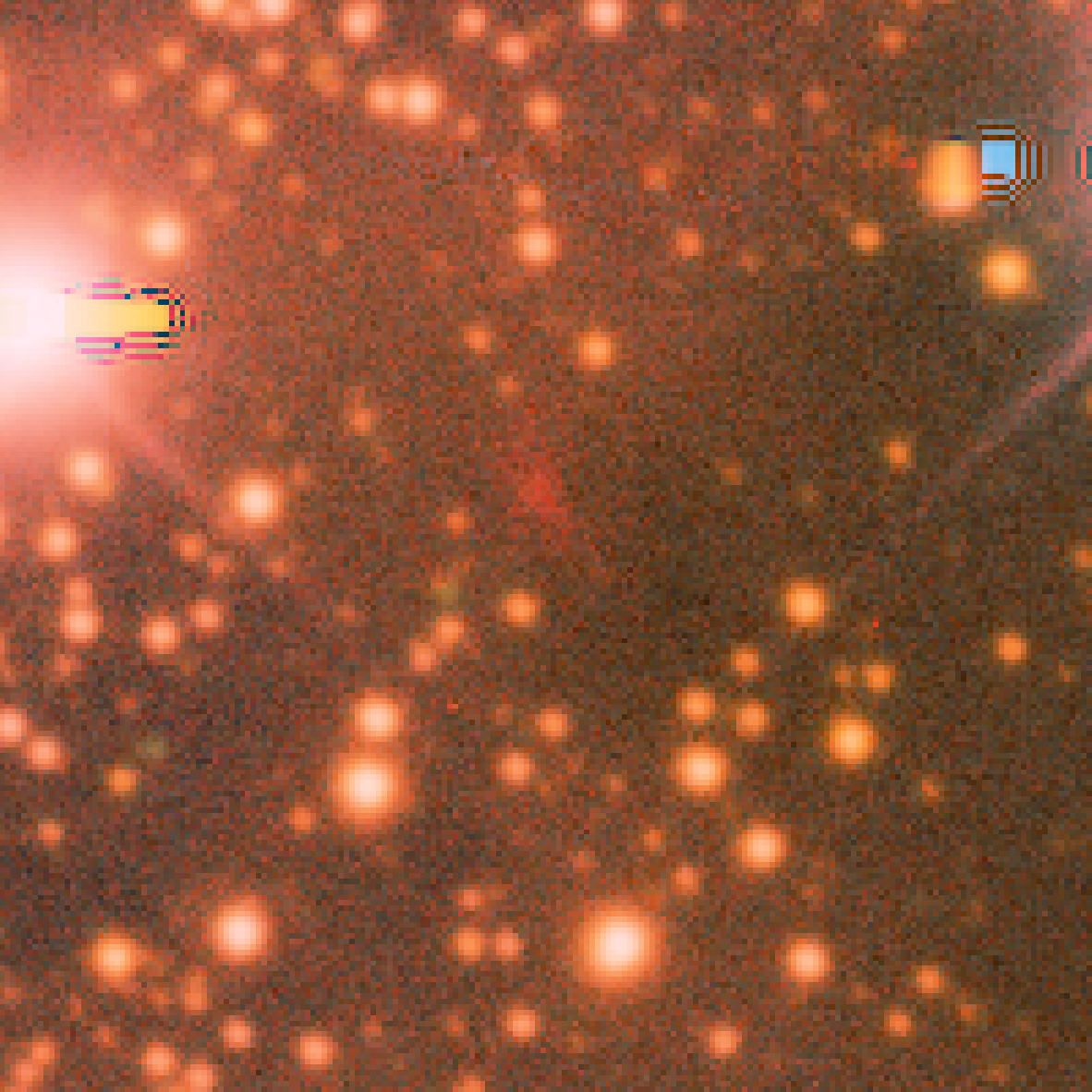}{0.25\textwidth}{HH~1258a}
          \fig{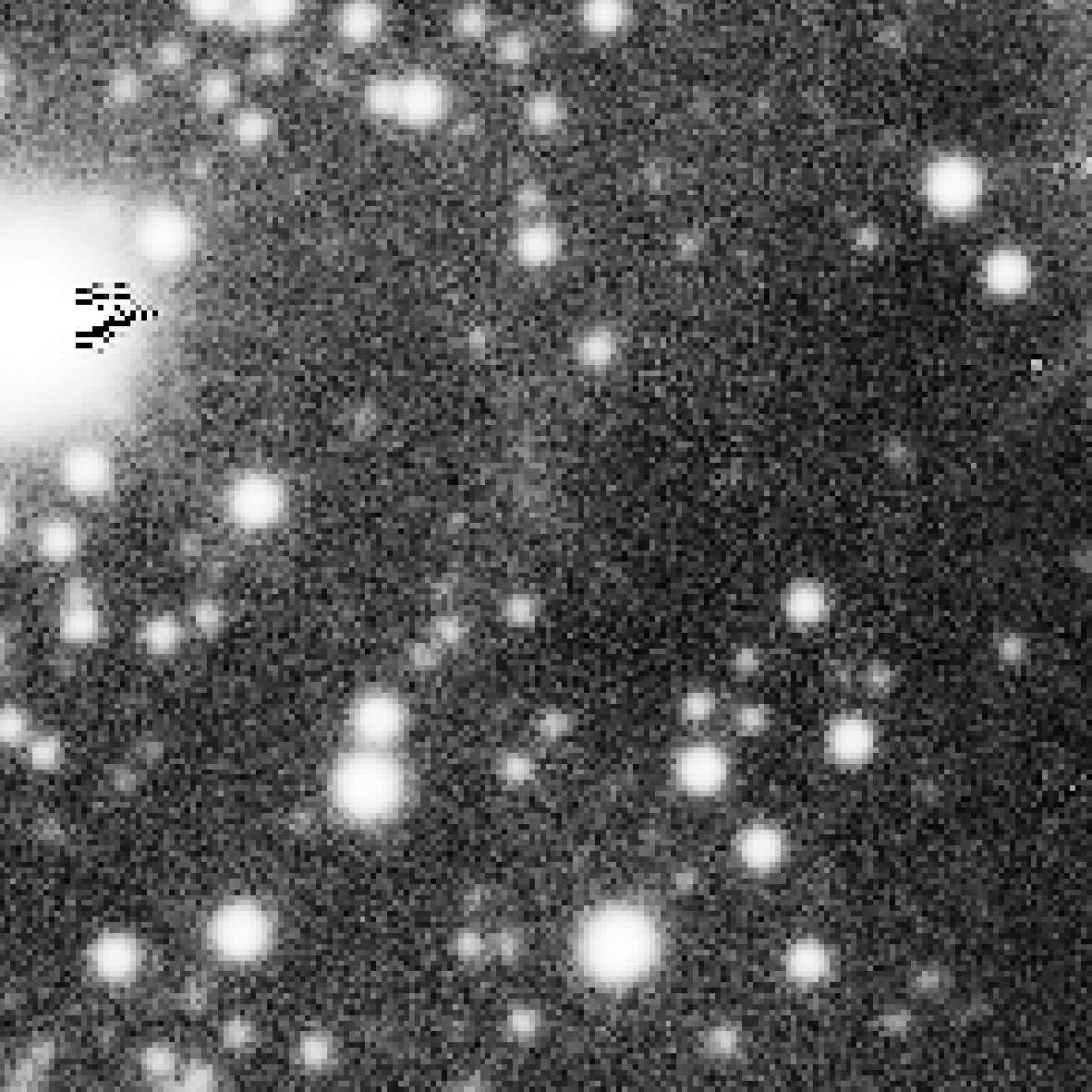}{0.25\textwidth}{HH~1258a}
          \fig{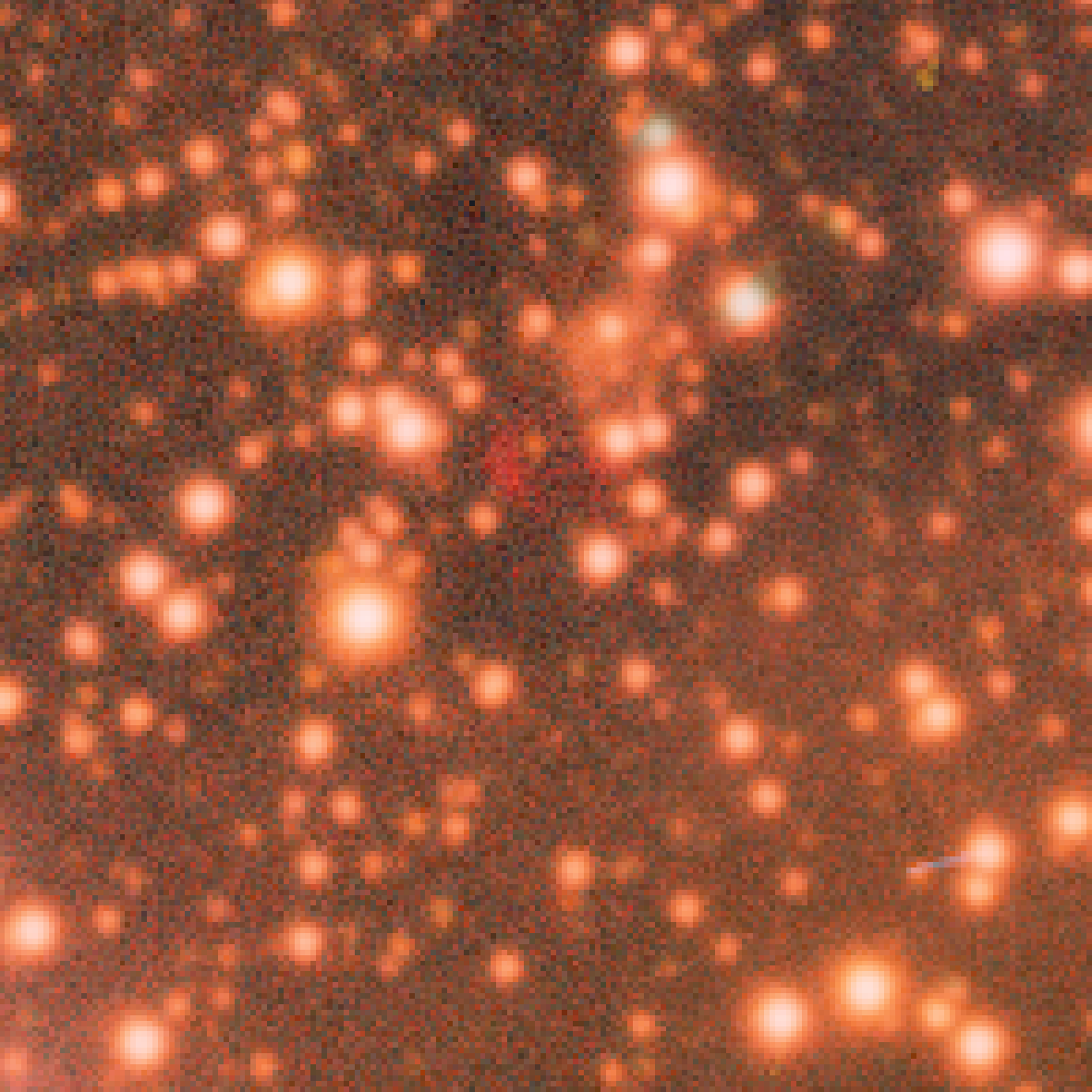}{0.25\textwidth}{HH~1258b}
          \fig{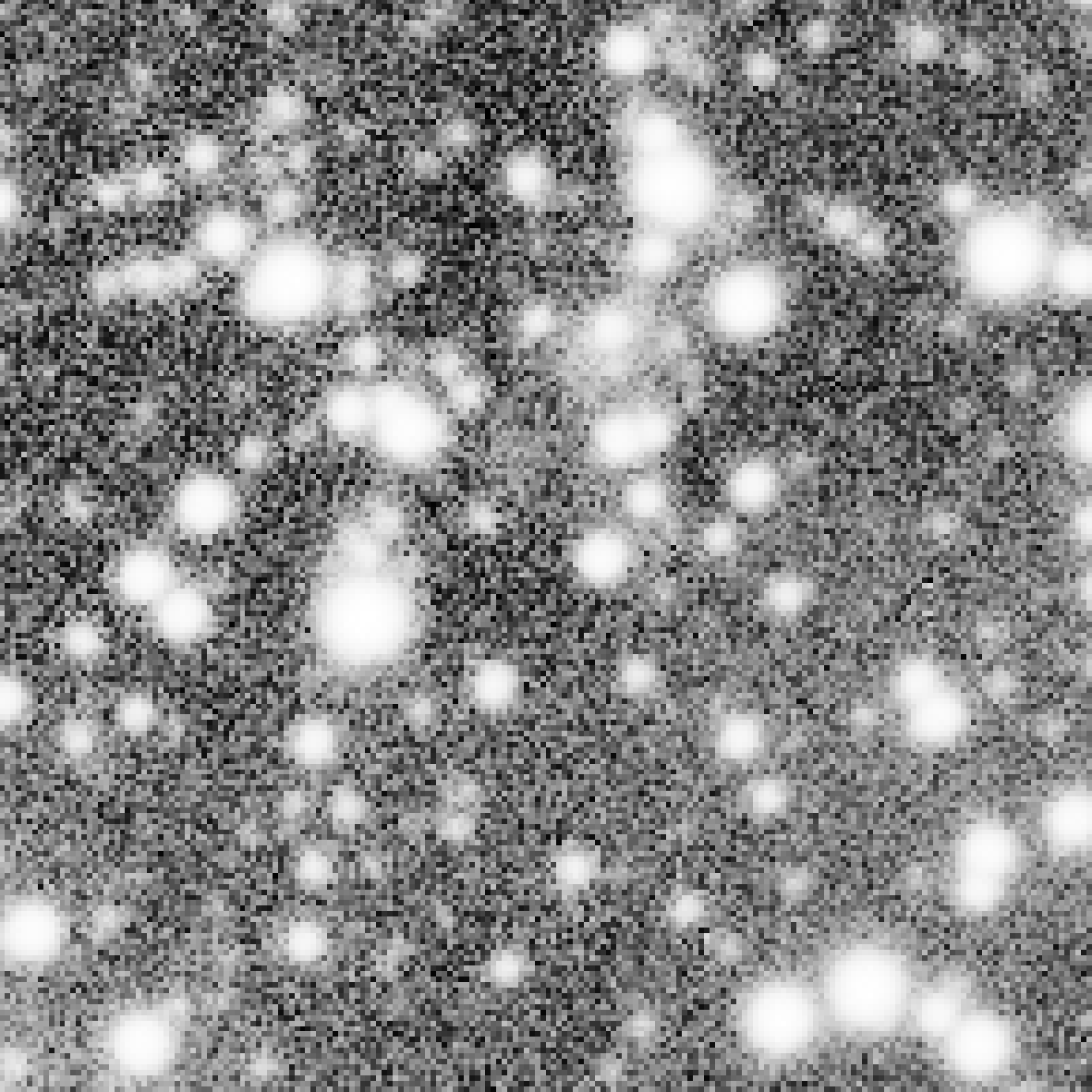}{0.25\textwidth}{HH~1258b}
          }
\gridline{\fig{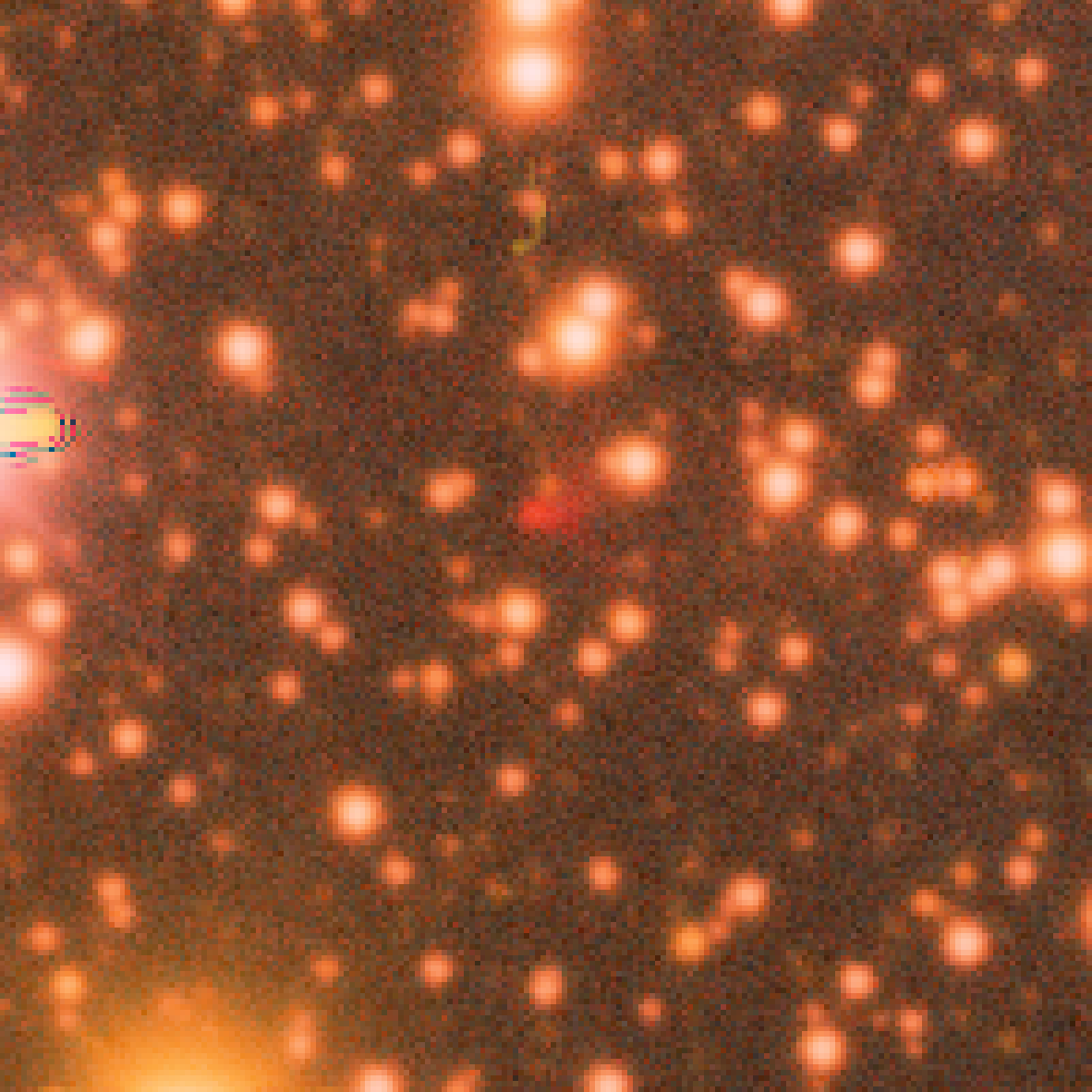}{0.25\textwidth}{HH~1258c}
          \fig{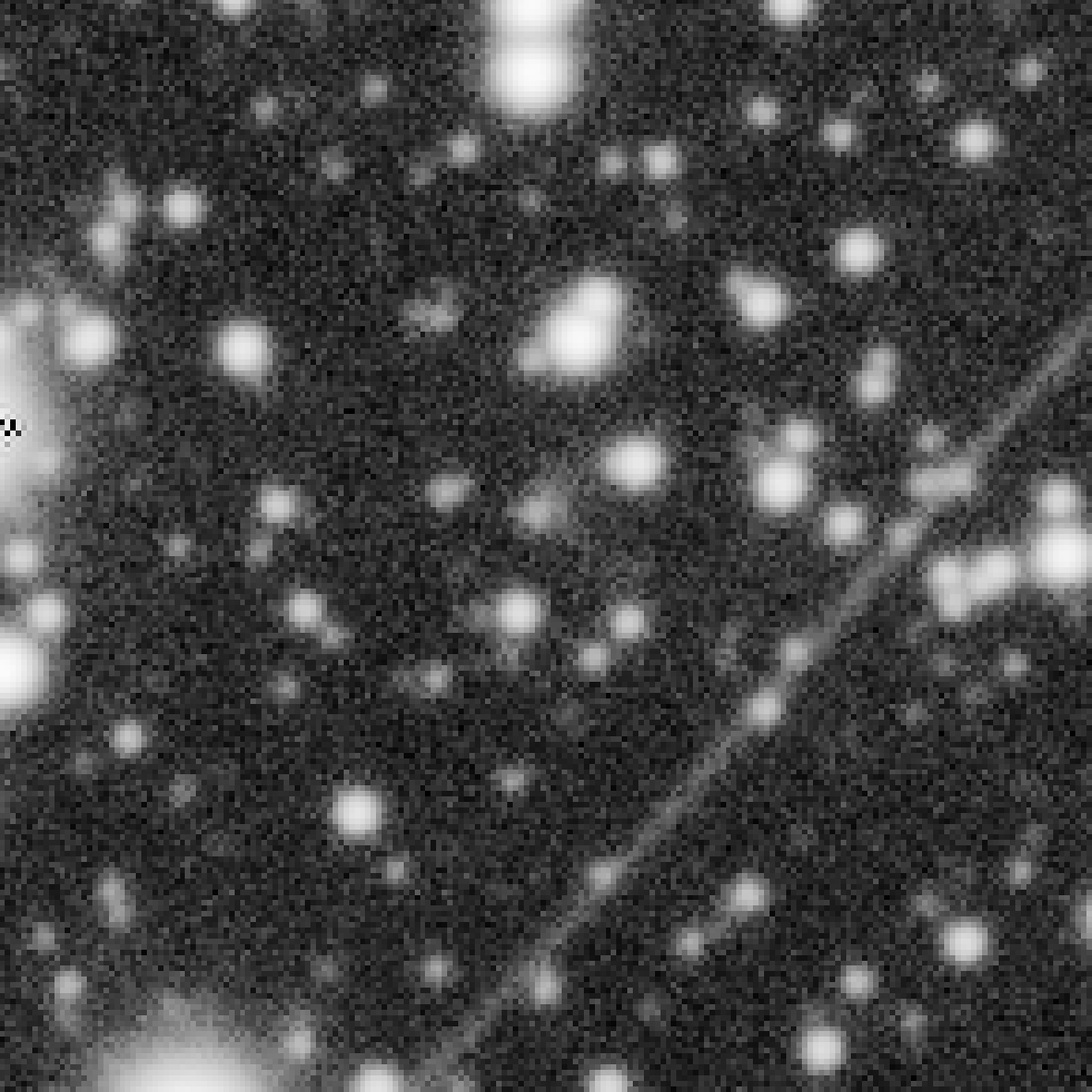}{0.25\textwidth}{HH~1258c}
          \fig{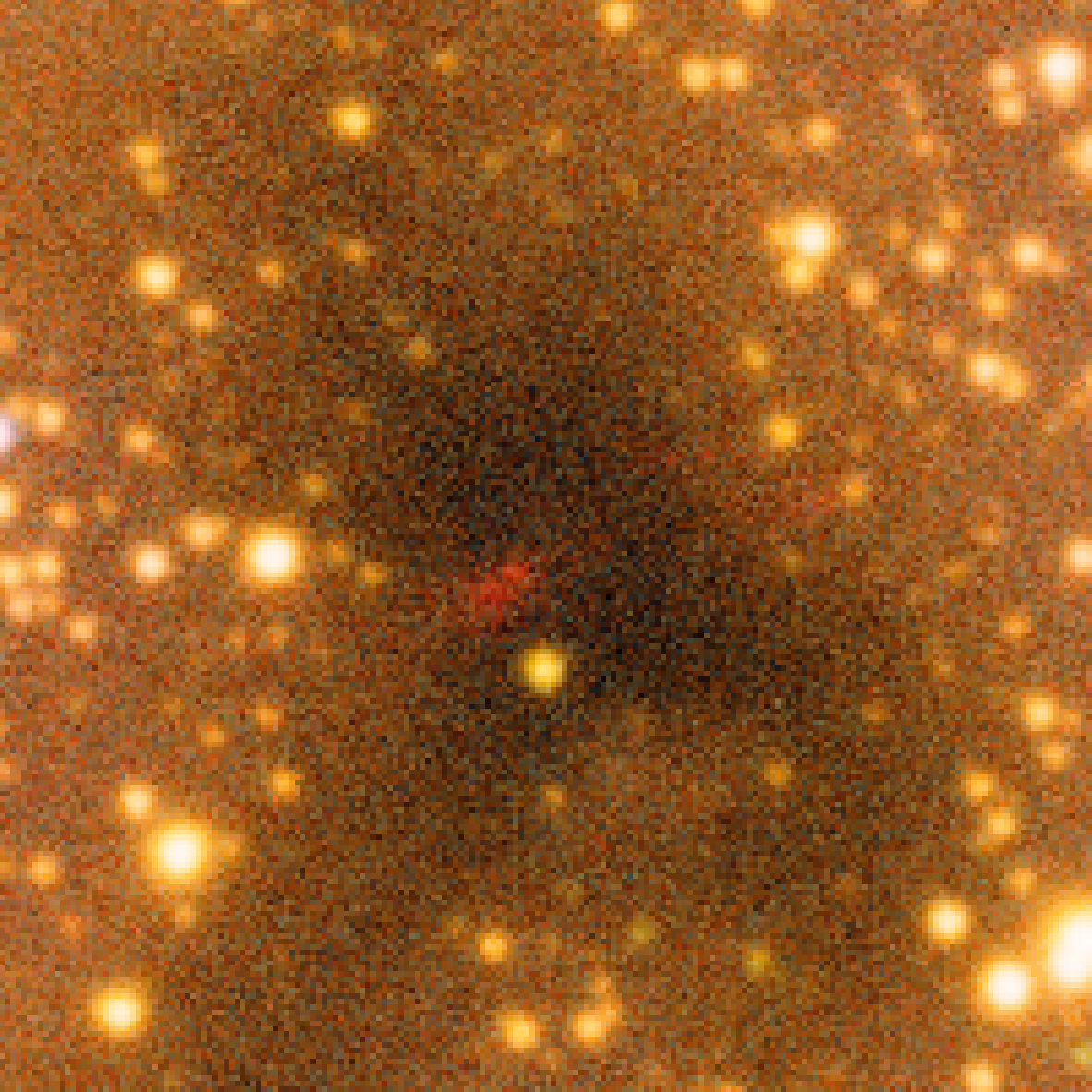}{0.25\textwidth}{HH~1259}
          \fig{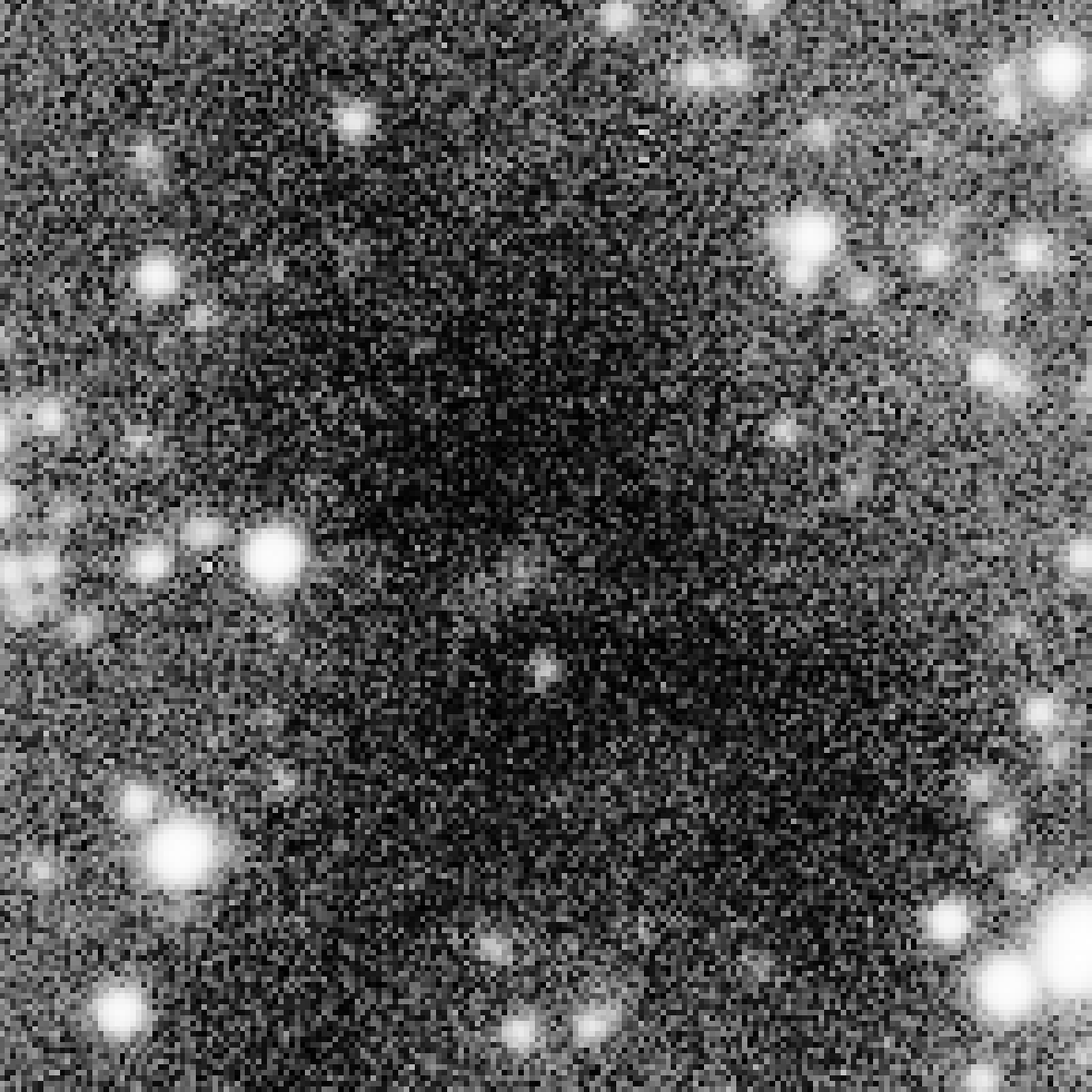}{0.25\textwidth}{HH~1259}
          }
\caption{Cutouts of four new HH objects in the southeastern region of Cir-W.  Cutout size and color schemes are the same as for Figure~\ref{fig:cutouts_AABB}
\label{fig:cutouts_AD}}
\end{figure}

HH~1258a-c:  These three HH objects (Figure~\ref{fig:cutouts_AD}) are colinear and appear to be part of an outflow originating at \object{WISEP J150353.05-632554.8}, which is identified as a Class~I YSO candidate by \citet{2011ApJ...733L...2L}.  It is the only red IR source in the region and therefore the most plausible progenitor.  HH~1258b is only weakly detected in \stwo.

HH~1259:  Figure~\ref{fig:cutouts_AD} shows that it is emerging from a dark cloud that contains the embedded red IR source \object{WISEA J150431.01-633803.2}, the likely progenitor.

\subsection{Circinus East}


\begin{deluxetable}{lll}
\tablecaption{Newly Discovered HH objects in Cir-E\label{tbl:newHH_e}}
\tablewidth{0pt}
\tablehead{\colhead{HH} & \colhead{RA(2000)} & \colhead{DEC}}
\startdata
1260a & 15:13:16.5 & -62:34:50 \\
1260b & 15:13:22.8 & -62:35:52 \\
1261a & 15:13:50.5 & -62:25:14 \\
1261b & 15:13:52.6 & -62:24:27 \\
1262 & 15:13:52.2 & -62:25:24 \\
1263a & 15:14:40.2 & -62:45:06 \\
1263b & 15:14:41.3 & -62:44:07 \\
1264 & 15:15:35.1 & -62:35:27 \\
1265 & 15:18:03.0 & -62:27:08 \\
\enddata
\end{deluxetable}

\begin{figure}
\gridline{\fig{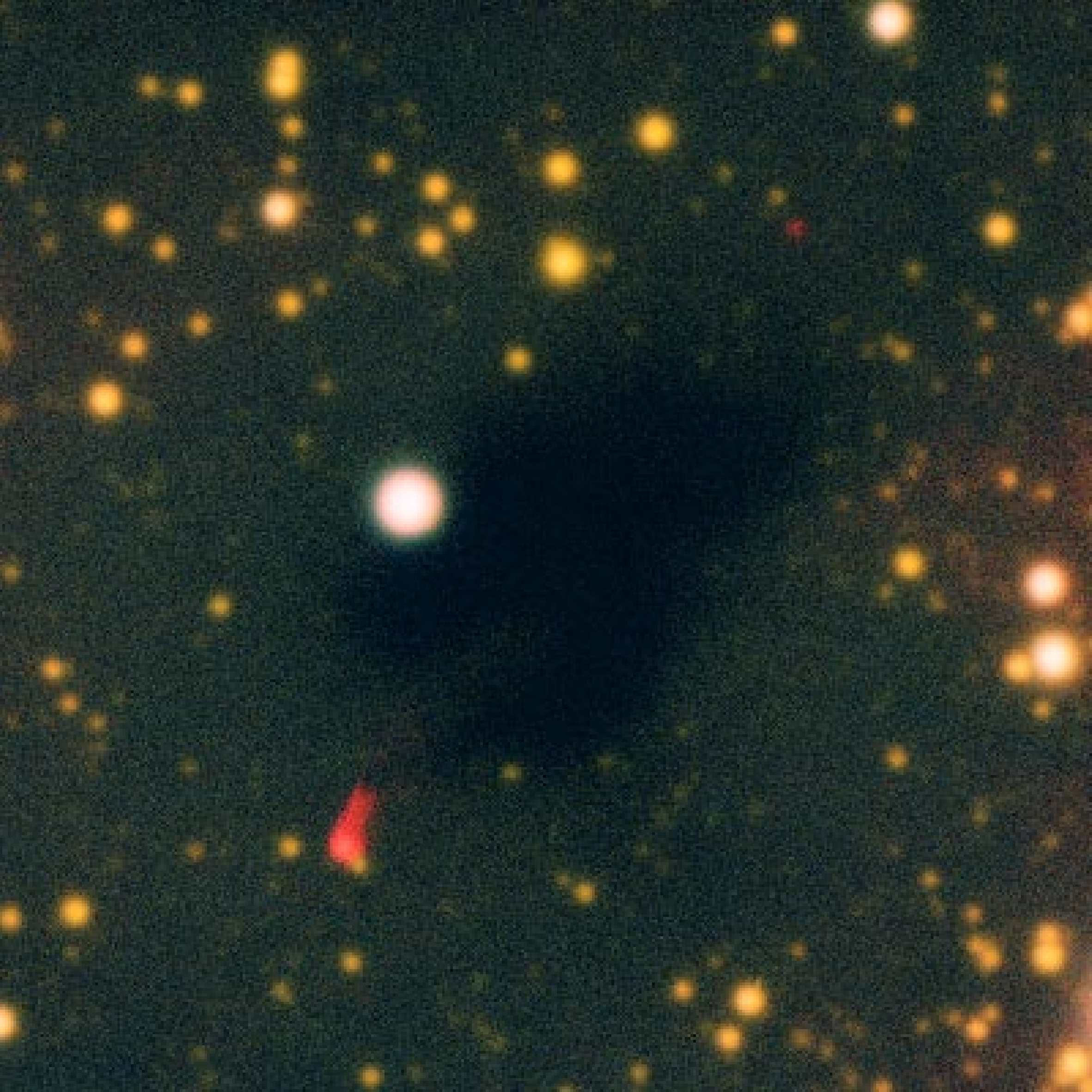}{0.5\textwidth}{HH~1260b \& HH~1260a}
          \fig{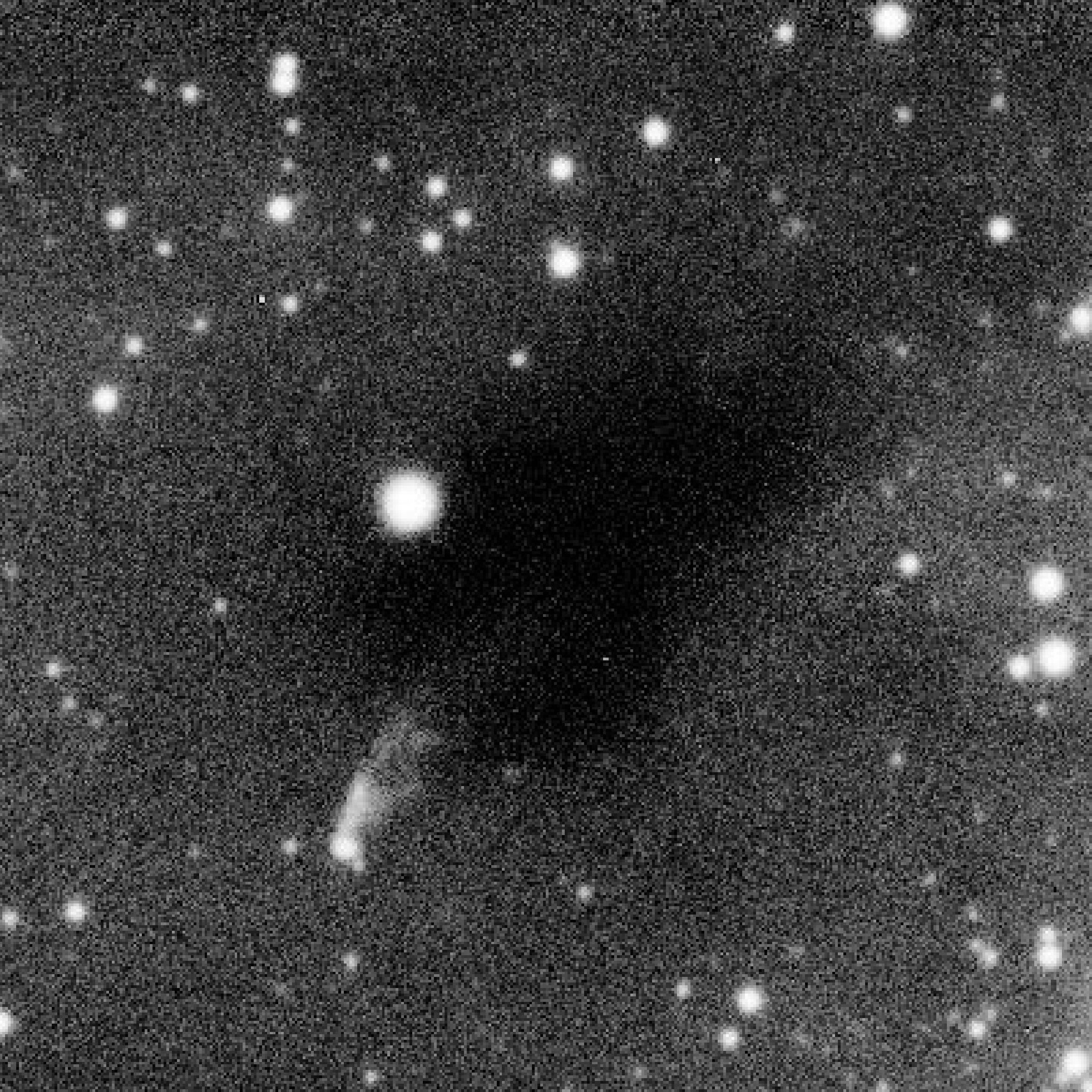}{0.5\textwidth}{HH~1260b \& HH~1260a}
          }
\caption{Cutouts of the field around HH~1260b (lower-left) and HH~1260a (upper-right) within Cir-E.  WISEA~J151319.86-623522.3 is located in the cloud between them.  Each cutout is 400 pixels (104\arcsec) on a side.  Color schemes are the same as for Figure~\ref{fig:cutouts_AABB}.  
\label{fig:ce_cutouts_CD}}
\end{figure}

HH~1260a \& 1260b: These HH objects (Figure~\ref{fig:ce_cutouts_CD}) lie on either side of a globule that contains the red IR source \object{WISEA J151319.86-623522.3}, which is assumed to be the progenitor.

\begin{figure}
\gridline{\fig{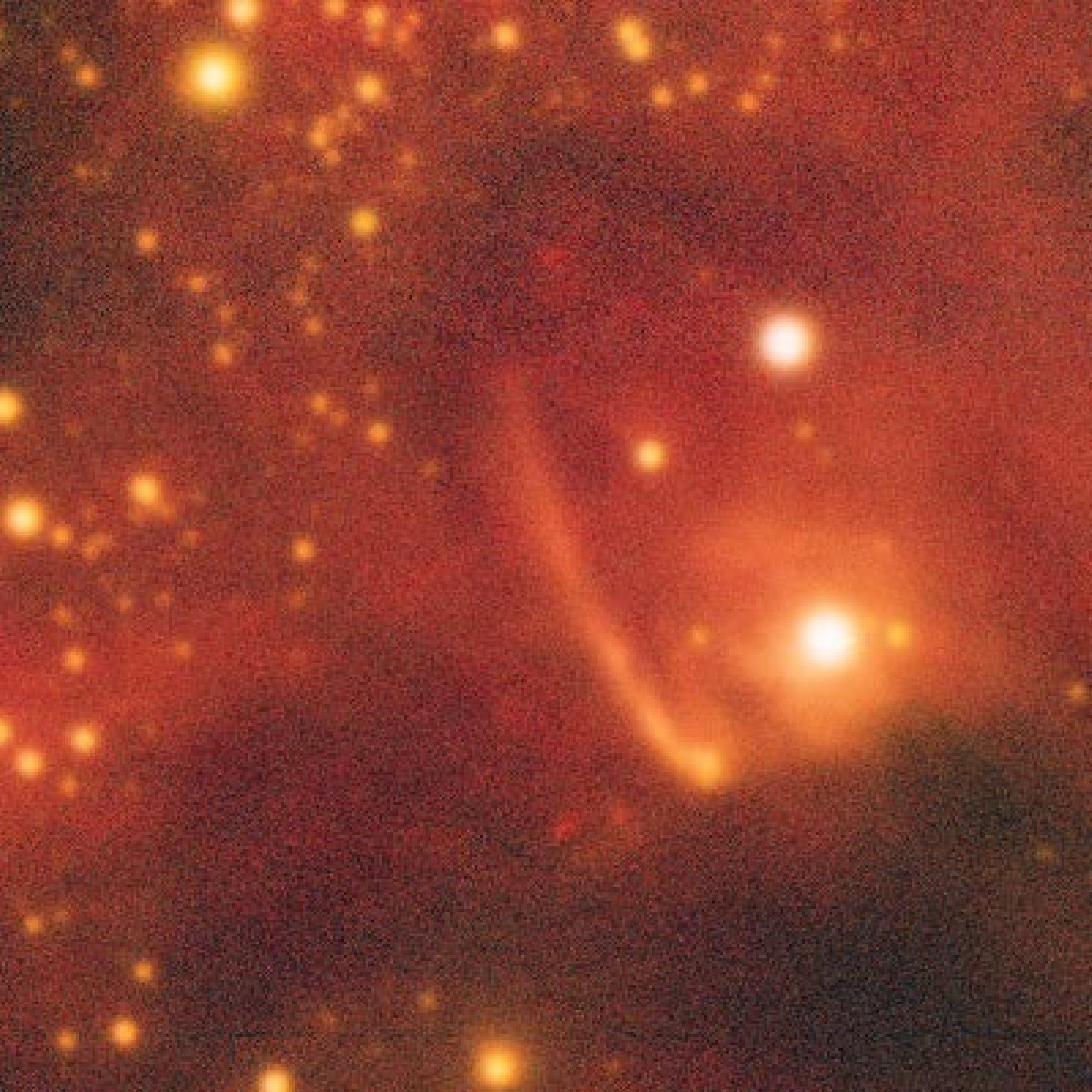}{0.5\textwidth}{HH~1261 \& HH~1262}
          \fig{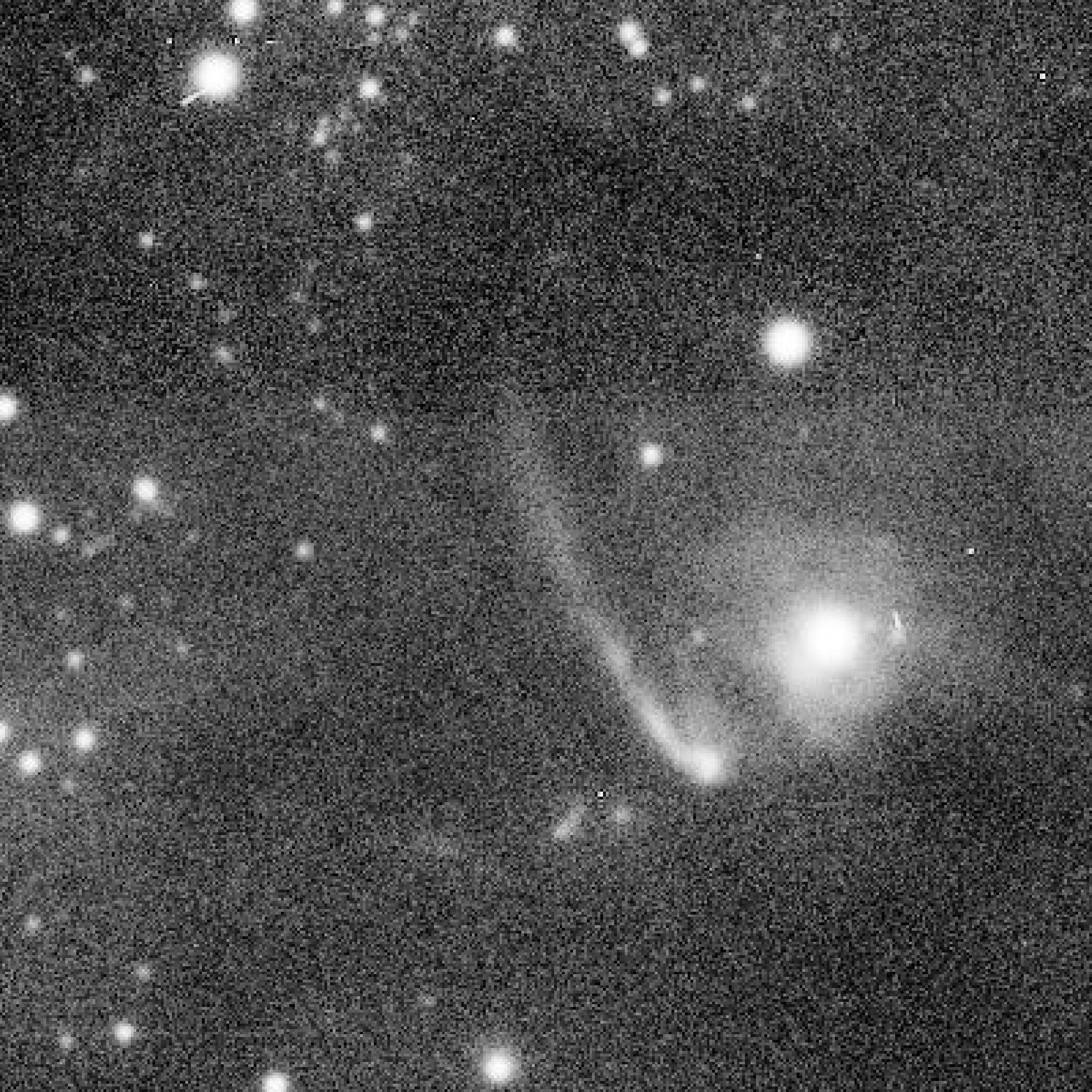}{0.5\textwidth}{HH~1261 \& HH~1262}
          }
\caption{Cutouts of the field around HH~1261 (center-top) and HH~1262 (center-bottom) within Cir-E.  HH~1261a is only faintly detected in \stwo. Cutout size and color schemes are the same as for Figure~\ref{fig:ce_cutouts_CD}.
\label{fig:ce_cutouts_AB}}
\end{figure}

HH~1261-1262: These HH objects (Figure~\ref{fig:ce_cutouts_AB}) reside in the complex, dusty environment around the emission-line star \object{[MO94] 20} \citep{1994MNRAS.270..199M}.  It is not clear what is the progenitor.  The red IR source \object{WISEA J151350.26-622518.8}, which is $\sim$20\arcsec\ to the southeast of [MO94]~20, appears to be the progenitor for HH~1261a and 1261b. This source could be a binary system and responsible for HH~1262 as well.  There is also a redder IR source \object{WISEA J151342.78-622806.2}, which is $\sim$3\arcmin\ to the southwest.

\begin{figure}
\gridline{\fig{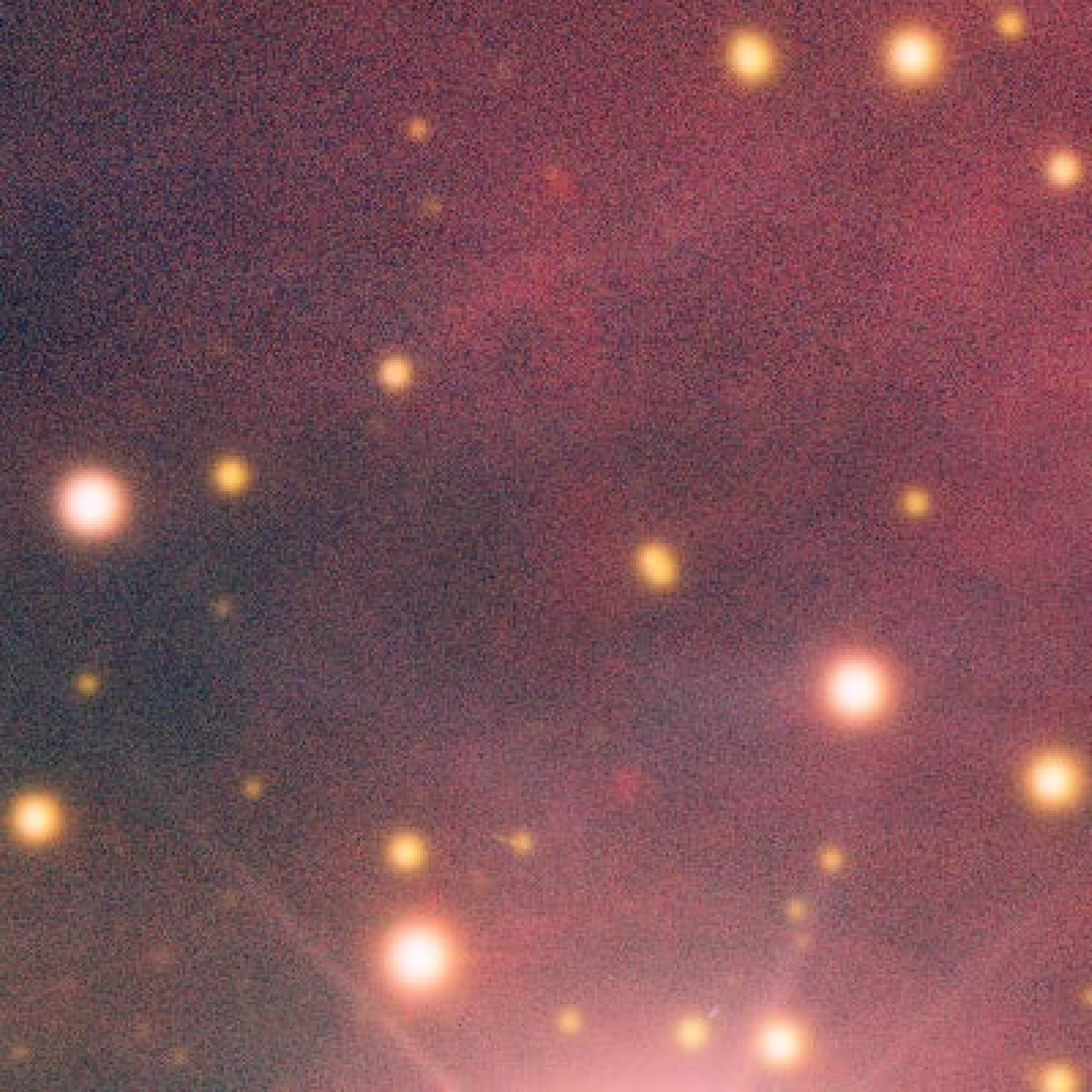}{0.5\textwidth}{HH~1263}
          \fig{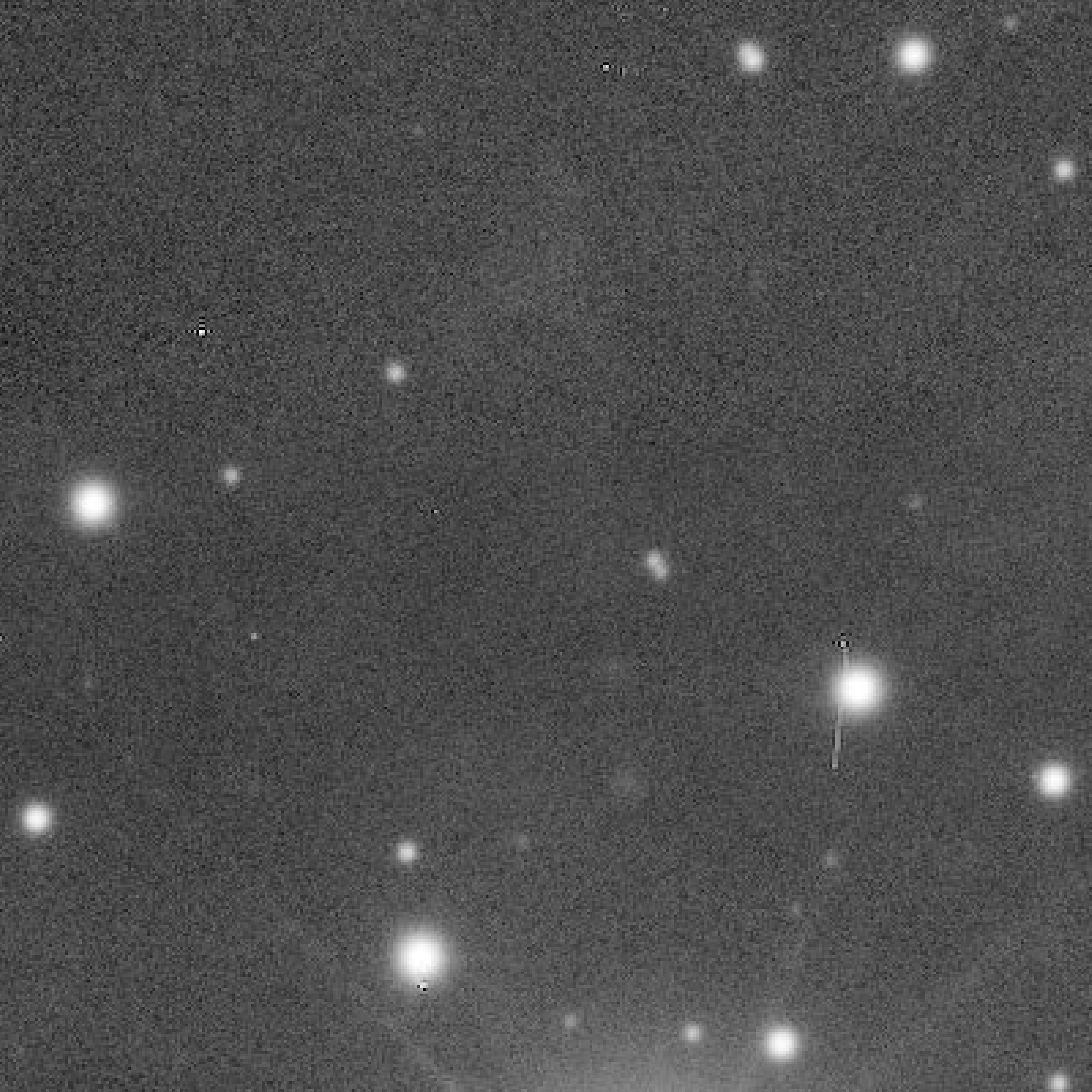}{0.5\textwidth}{HH~1263}
          }
\caption{Cutouts of the field around HH~1263a (center-top) and HH~1263b (center-bottom) within Cir-E.  A faint stream with one or more knots appears to connect the two.  Cutout size and color schemes are the same as for Figure~\ref{fig:ce_cutouts_CD}.
\label{fig:ce_cutouts_EF}}.
\end{figure}

HH~1263: There are no red IR sources in the region, however the alignment and morphology of this outflow (Figure~\ref{fig:ce_cutouts_EF}) suggests that \object{HD 134733}, a B3 III star \citep{1975mcts.book.....H}, may be the progenitor. Alternatively, the red IR sources \object{WISEA J151429.96-624409.6} and \object{WISEA J151429.39-624652.1} lie about 1.5\arcmin and 2\arcmin\ away respectively.

\begin{figure}
\gridline{\fig{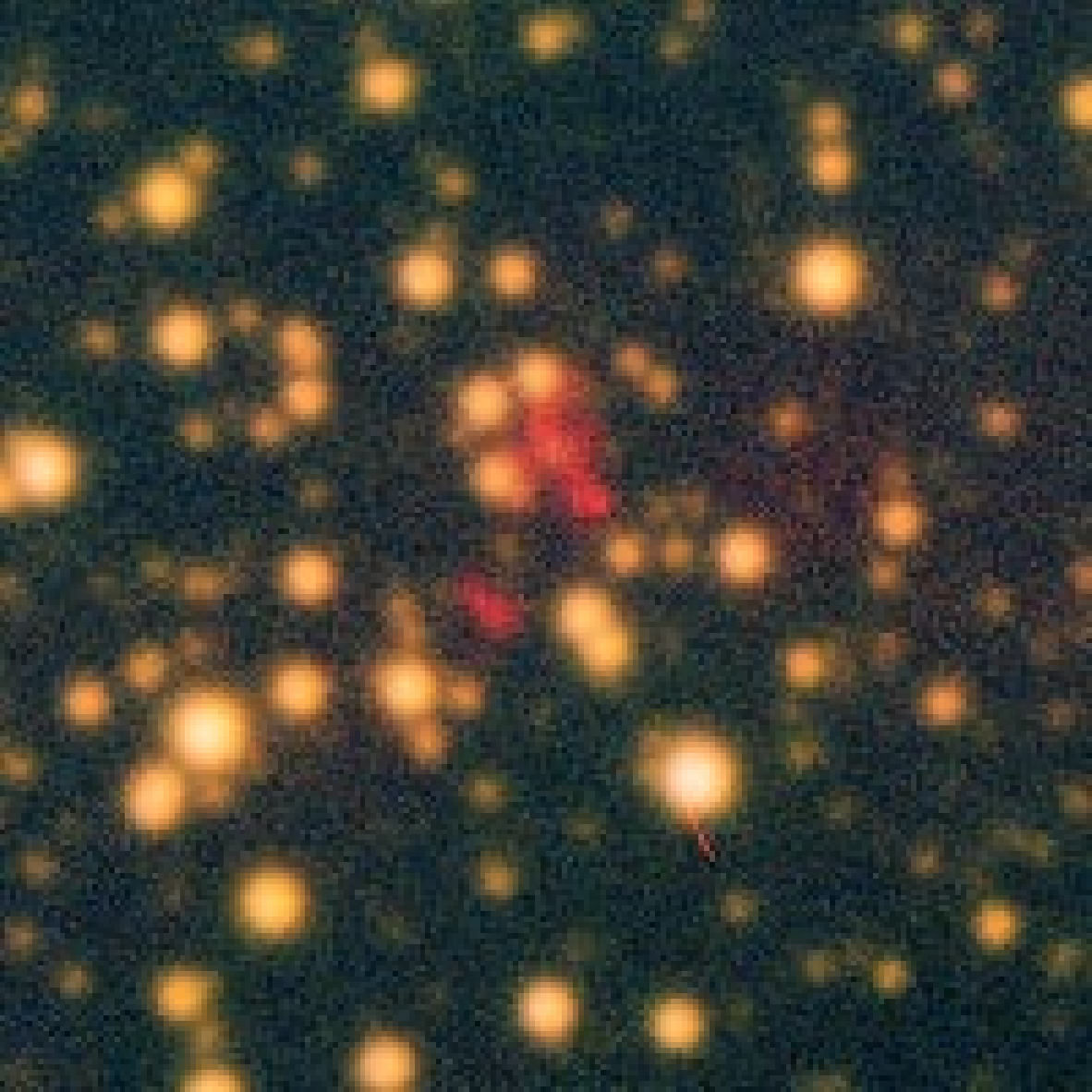}{0.25\textwidth}{HH~1264}
          \fig{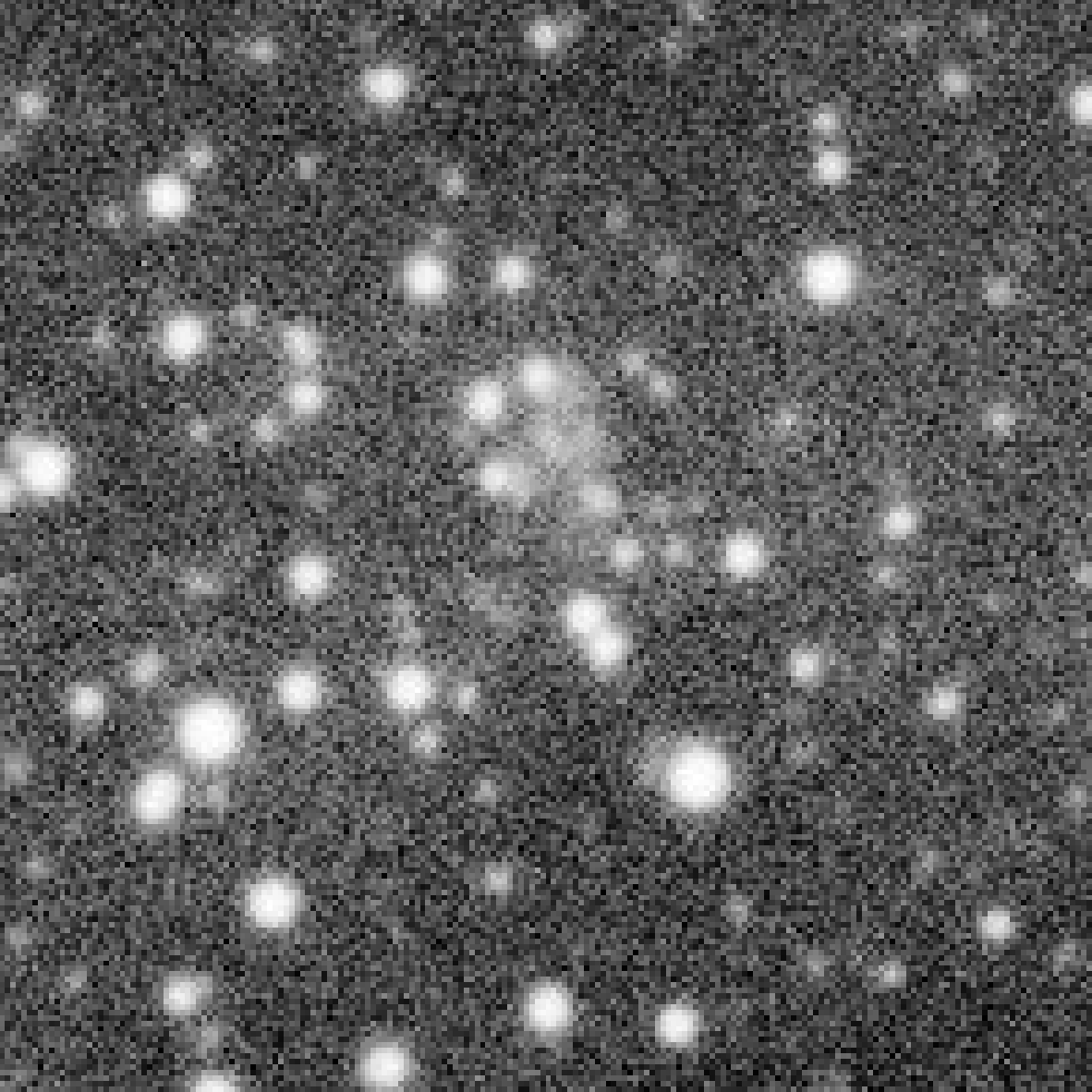}{0.25\textwidth}{HH~1264}
          \fig{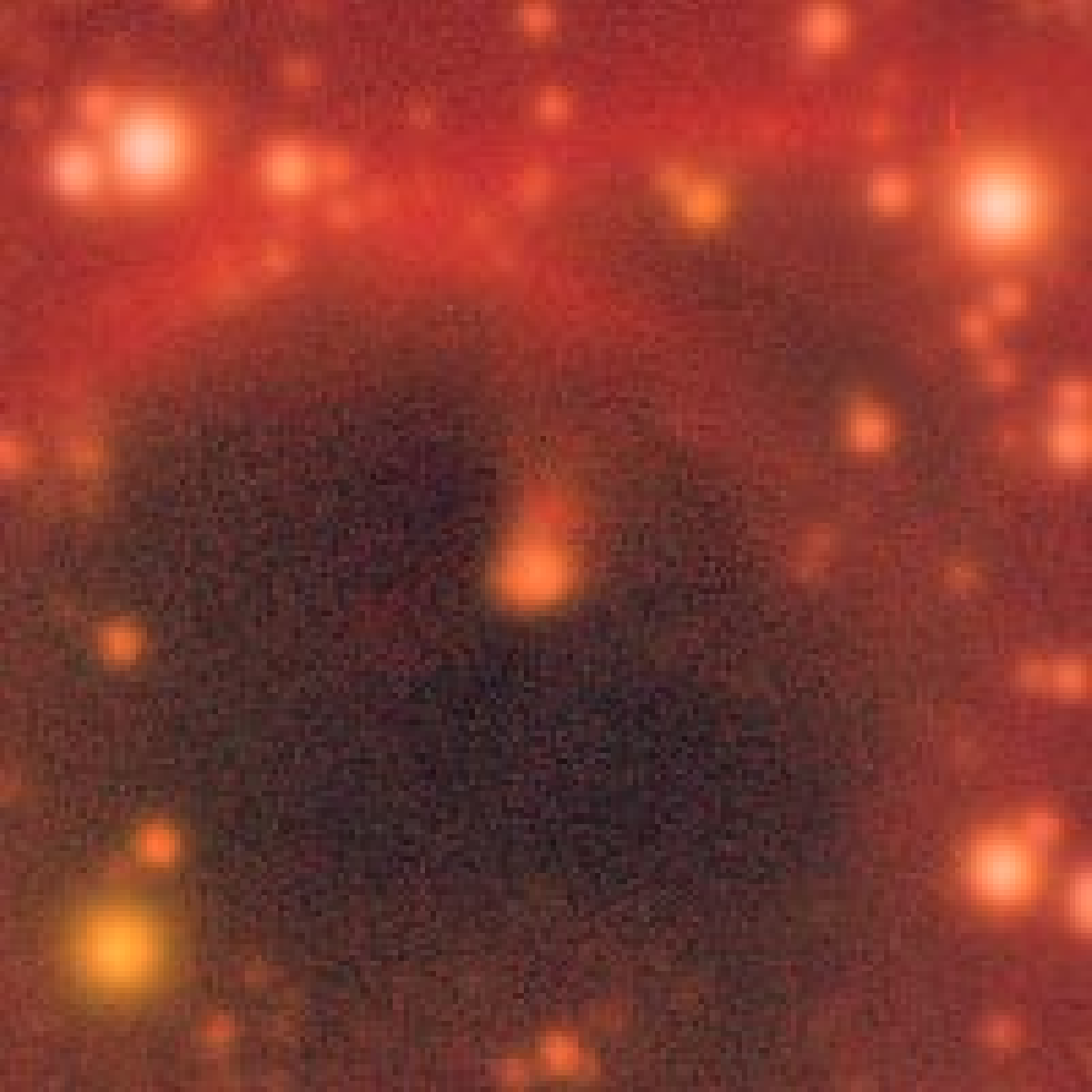}{0.25\textwidth}{HH~1265}
          \fig{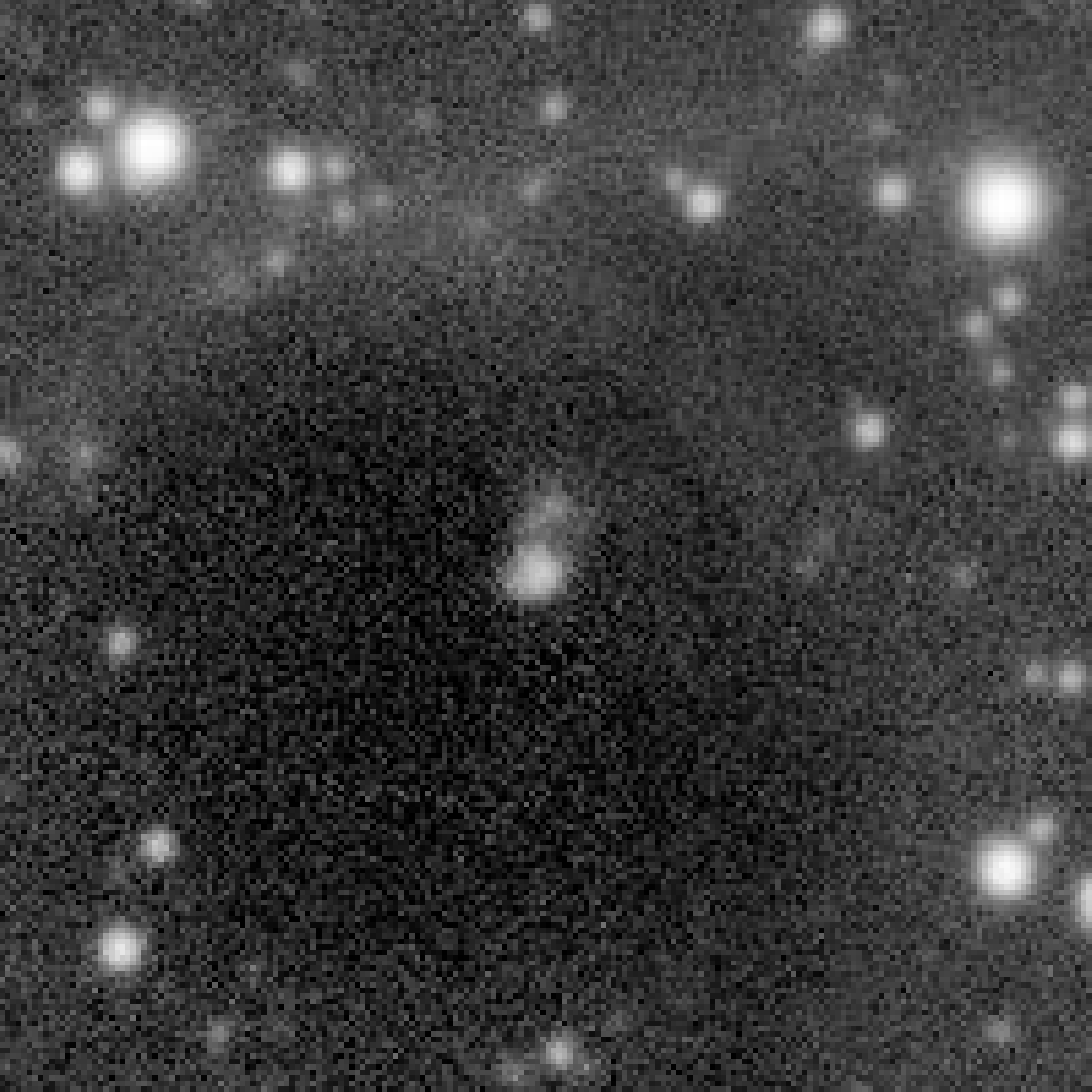}{0.25\textwidth}{HH~1265}
          }
\caption{Cutouts of new HH components in Cir-E. Each cutout is 200 pixels (52\arcsec) on side.  Cutout size and color schemes are the same as for Figure~\ref{fig:cutouts_AABB}.
\label{fig:ce_cutouts_GH}}
\end{figure}

HH~1264: There are no red IR sources in the region.  This HH object  (Figure~\ref{fig:ce_cutouts_GH}) is 12\arcmin\ to the northeast of HD~134733 and could share a common progenitor with HH~1263.

HH~1265: This object (Figure~\ref{fig:ce_cutouts_GH}) appears to be embedded in a globule that contains the red IR source \object{WISEA J151803.04-622710.9}, which is assumed to be the progenitor.

\section{Circinus in the Local Environmental Context} \label{sec:context}

\begin{figure}
\centering
\includegraphics[width=8.1cm]{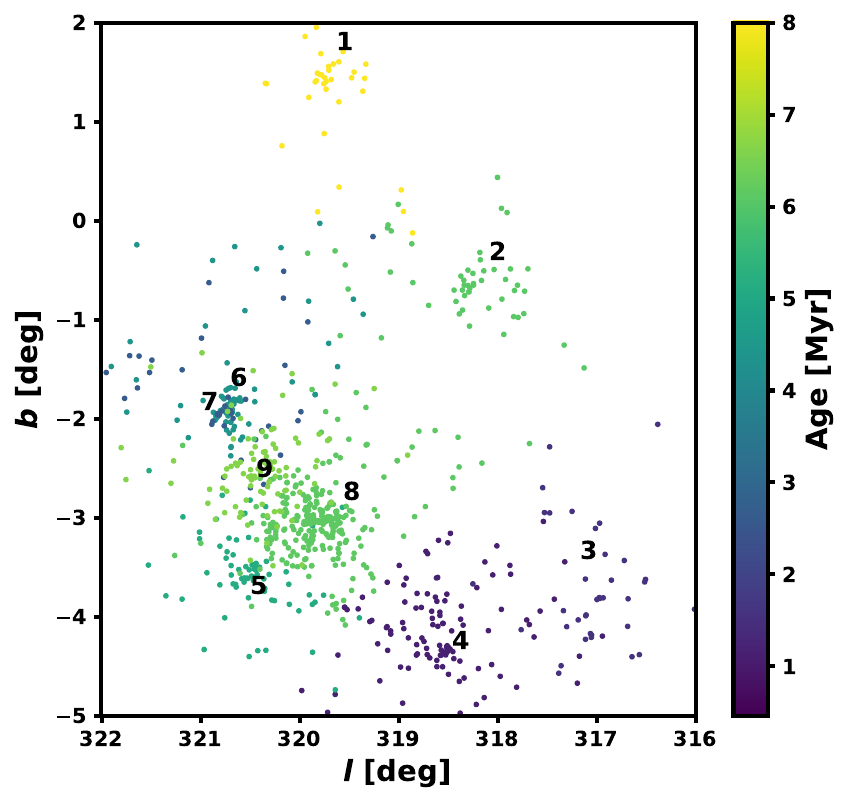}
\includegraphics[width=7cm]{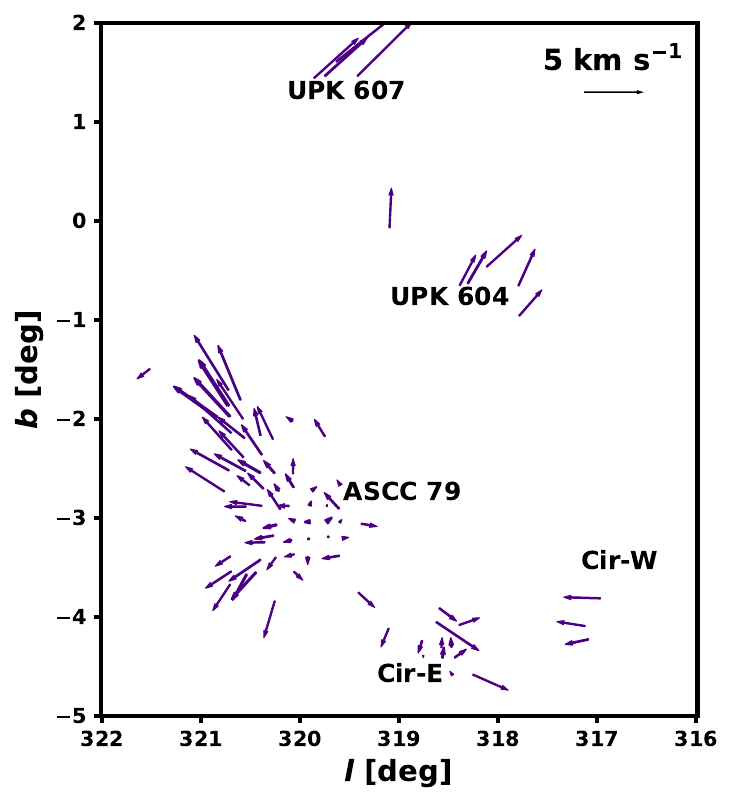}\hfill
\caption{In the left panel, a sky-plane map of the Circinus Complex subgroups defined by \citet{KerrFarias25}, colored by their ages and with subgroup IDs annotated in black. Subgroups 3 and 4 contain the populations associated with the Cir-W and Cir-E clouds, respectively. In the right panel, the mean stellar velocities by on-sky bin relative to the velocity of ASCC~79 with the locations of notable structures labelled, as previously presented in Figure 9 of \citet{KerrFarias25}. }
\label{fig:circinusages}
\end{figure}

In the left panel of Figure \ref{fig:circinusages}, we show the on-sky distribution of stars in the \citet{KerrFarias25} high-quality member sample, colored by the isochronal age of their parent subgroup defined in Table 2 of \citet{KerrFarias25}. There we show that with the exception of the largely discontiguous groups to the galactic north, older populations in the Circinus Complex are concentrated around ASCC~79 (group 8); and subgroups are progressively younger moving away from that cluster. In the right panel of Figure \ref{fig:circinusages}\footnote{previously presented in Figure 9 in \citet{KerrFarias25}}, we show the mean transverse velocities by on-sky bin relative to ASCC~79 for stars in Circinus, which are computed by combining \textit{Gaia} distances and proper motions and correcting the result for virtual expansion \citep[e.g., see][]{Kuhn19}. There we show that subgroups 5-9, which are collectively referred to as the Circinus Central (CirCe) Region in \citet{KerrFarias25}, share a common expansion trend, with outlying populations having higher velocities in proportion to their distance from ASCC~79. By comparing the CirCe region to an analogous STARFORGE simulation \citep{Grudic21, Guszejnov22}, \citet{KerrFarias25} finds that the dynamics and age distribution there can be explained by an initial cluster-building stage that forms ASCC~79, followed by a stage of triggered star formation in a shell that forms the later generations. While the velocities of the stars associated with the Circinus Molecular Cloud do not generally follow the expansion trend present in the CirCe region, they do adhere to the age trend, indicating that the presence of this shell may drive gas morphology in the Circinus Molecular Cloud. 

\begin{figure}[ht]
\plottwo{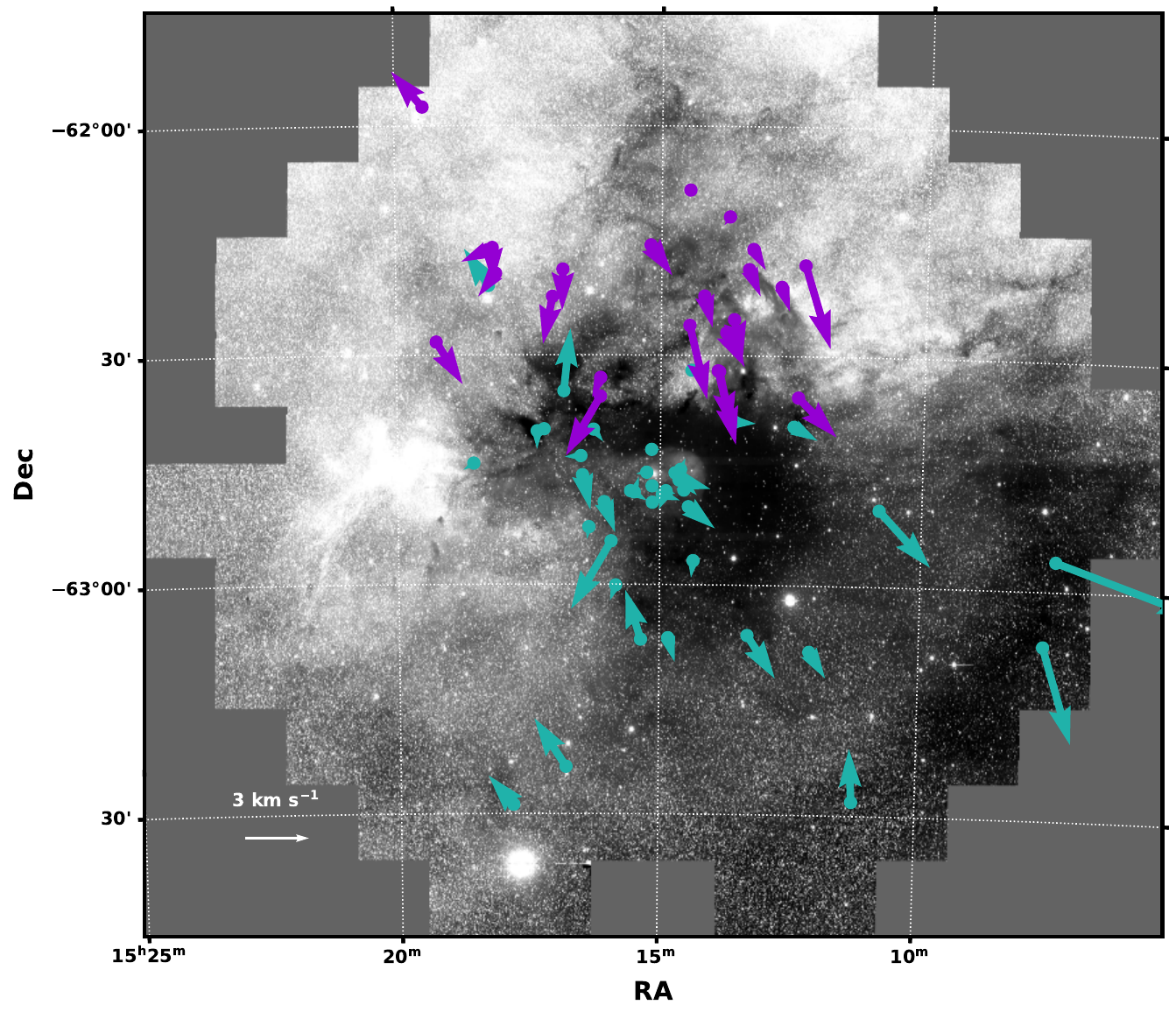}{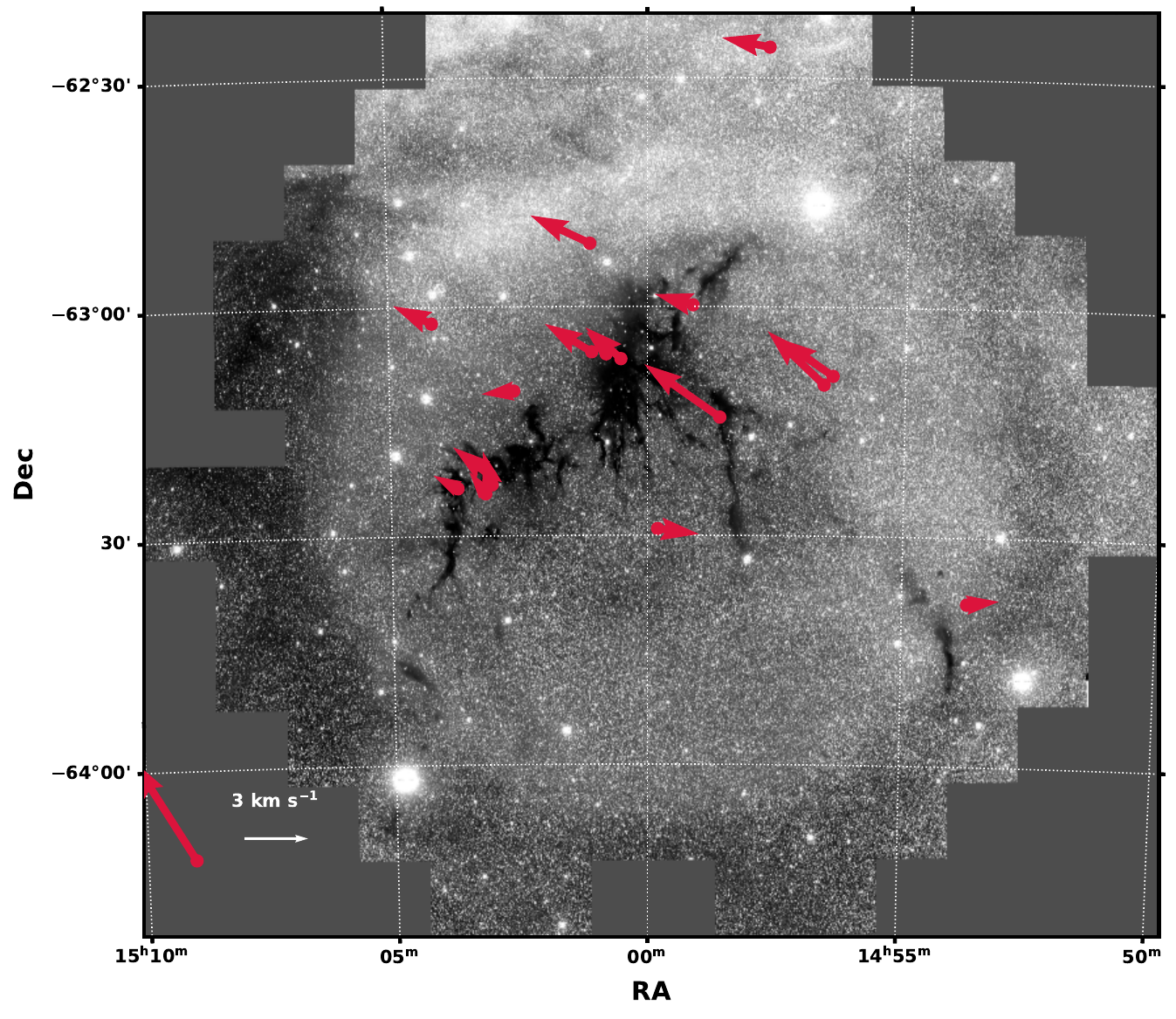}
\caption{Motions of Circinus members with $P_{Age<50 Myr}>0.95$ in \citet{KerrFarias25}, overlaid on the $H\alpha$ DECam frames. Velocities are shown relative to ASCC~79, the central cluster in the Circinus Complex, and corrected for virtual expansion. The left panel shows Cir-E, and the right panel shows Cir-W. We split up the membership of Cir-E according to its two components defined in \citep{KerrFarias25}, which correspond roughly to the Cir-Ea (south) and Cir-Eb (north) subcomponents from \citet{2011ApJ...731...23S}.
\label{fig:cmc_starmot}}
\end{figure}

\begin{figure}[ht]
\plotone{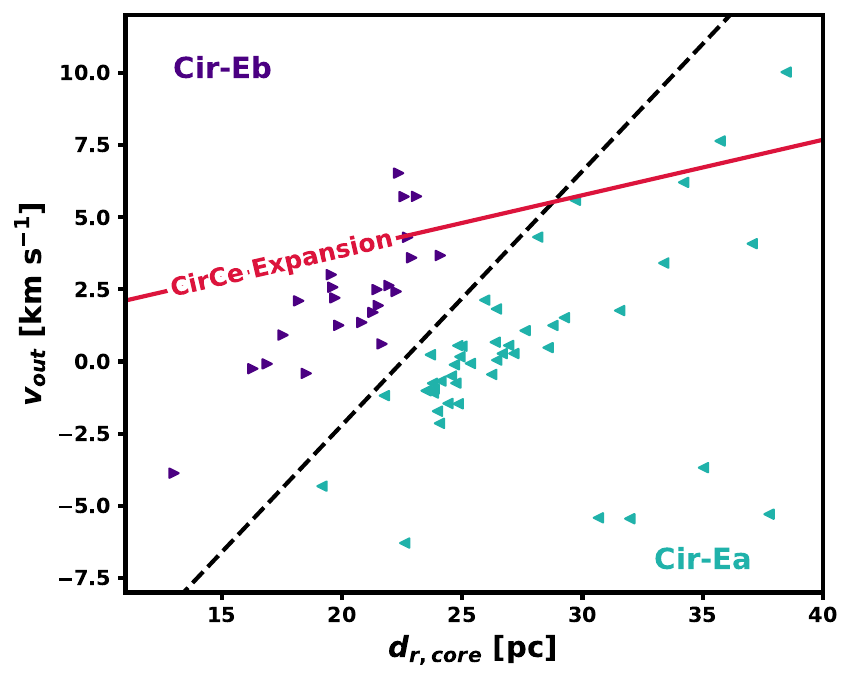}
\caption{Distance from ASCC 79 in the plane of the sky against velocity along that separation vector for Cir-E members. The dashed line separates the two velocity components, Cir-Eb and Cir-Ea. We include the fit to the expansion trend in the CirCe Region from Figure 10 of \citet{KerrFarias25} for reference.
\label{fig:c4velocities}}
\end{figure}

In Figure \ref{fig:cmc_starmot}, we show the transverse velocities of stars in the \textit{Gaia} Circinus Molecular Cloud sample, measuring velocities relative to ASCC~79 as in \citet{KerrFarias25}. In Cir-E, we show the two dynamically distinct velocity components that \citet{KerrFarias25} identifies in the region: one centered on the core of Cir-E that has velocities nearly identical to ASCC~79 (Cir-Ea), and another offset to the north with velocities pointing away from ASCC~79 and towards the core of the Cir-E cloud (Cir-Eb). This appears to reproduce the multi-component nature of Cir-E proposed by \citet{2011ApJ...731...23S}, indicating that the bimodal velocity signature presented there is visible in both the stellar transverse motions and the radial motions of the gas. It also indicates that the velocities of the two components are convergent in the plane of the sky. In Figure \ref{fig:c4velocities}, we plot the radial component of the Cir-E transverse velocities relative to ASCC~79 in the plane of the sky against the on-sky distance from ASCC~79, showing how the velocities vary with distance from the CirCe region. 

The convergent velocities of this northern component and the core of Cir-E support the suggestion by \citet{2011ApJ...731...23S} that a cloud-cloud collision is underway. Figure \ref{fig:c4velocities} shows a velocity gradient across Cir-Eb, with stars closer to ASCC~79 having lower velocities, and stars closer to the core of Cir-E having velocities similar to the expansion gradient of the CirCe Region. This signature is consistent with material ejected by feedback from the CirCe region colliding with an adjacent cloud, forming stars as they compress one another. Stars formed primarily from the ejected material have higher velocities relative to ASCC~79 in the plane of the sky, causing them to move away from ASCC~79 and towards the core of the Cir-E. This leaves the low-velocity stars in Cir-Eb that formed primarily of the interloping cloud further from the core of Cir-E, while higher-velocity stars are found closer to the core.

Sky-plane velocities in the dense core of Cir-Ea are low relative to ASCC~79, suggesting that the stars there formed primarily from the low-velocity interloping cloud, not from ejected material. The uniform velocities in the core of Cir-Ea imply that its gas was not mixed with or substantially accelerated by this proposed collision. However, the compression caused by the collision in the north of Cir-E may have nonetheless had a role in triggering star formation there. Figure \ref{fig:c4velocities} shows that velocities in Cir-Ea outside of the core may host a similar gradient to the one seen in Cir-Eb, potentially indicating substructure. However, the spatial view in Figure \ref{fig:cmc_starmot} shows that this trend largely lacks organization in the plane of the sky, so we conclude that this signature is likely consistent with contamination, not another velocity component. 

This narrative is also supported by the gas dynamics of the region presented in \citet{2011ApJ...731...23S}. \citet{KerrFarias25} provide a bulk radial velocity measurement for the Circinus Complex of -3 km s$^{-1}$, which is dominated by stars in and around ASCC~79. \citet{2011ApJ...731...23S} revealed that the gas associated with Cir-Eb has a slightly more blue-shifted velocity of -4 km s$^{-1}$. Given that the 821 pc distance to ASCC~79 from \citet{KerrFarias25} is more distant than any of our distance estimates to Cir-E (see Appendix \ref{app:distance}), ejected material from ASCC~79 associated with the Cir-E cloud would need slightly blue-shifted velocities. The Cir-Ea component has velocities that are even more blue-shifted at $\sim-6$ km s$^{-1}$, indicating that if a cloud-cloud collision is underway, it would occur through the main Cir-E cloud overtaking the slower-moving ejected gas in the radial direction. The distances to Cir-Ea and Cir-Eb are consistent with one another, so the positions of these two components also support the collision hypothesis. 

In Cir-W, Figure \ref{fig:cmc_starmot} shows relatively uniform velocities across the entire cloud, demonstrating a lack of clear convergence signatures similar to those that we see in Cir-E . This could explain the morphological differences between the two clouds, with Cir-E being sculpted a combination of a cloud-cloud collision and feedback from the older stellar populations in Circinus, while the evolution of Cir-W is dominated by self-gravity. The relative positions of Cir-W and Cir-E may contribute to these different morphologies, as Cir-E's location may partially shield Cir-W from the feedback emanating from ASCC 79. 

\section{Discussion}\label{sec:disc}

\subsection{Relationship Between Cir-W and Cir-E}

\citet{2020AJ....160..189R} detailed the detection of fourteen new HH objects in Cir-W and suggested that the complex velocity field in Cir-E, as described by \citet{2011ApJ...731...23S}, may be the result of external impacts to the region which inhibited star formation, as corroborated by the lack of HH objects  in Cir-E \citep{2020AJ....160..189R}. In \S~\ref{sec:context}, we demonstrated that feedback from sequential waves of star formation, emanating from the ASCC 79 open cluster, appear to now be acting on Cir-E, given the turbid dynamics of both its stellar (Figure \ref{fig:cmc_starmot}) and gas components \citep{2011ApJ...731...23S}. If the evolution of the amorphous Cir-E cloud has been shaped by feedback in its extended environment, the filamentary structure of Cir-W, rife with star formation and HH objects, may be representative of what clouds do in the absence of external perturbations. Thus Cir-E and Cir-W may be of similar age and evolutionary state, but the relatively unperturbed Cir-W, possibly shielded in part by Cir-E from the disruptive radiative force of the high-mass stars in ASCC~79 \citep{2022MNRAS.515.4929G}, may be a site of more active and efficient star formation.

In this current work, we have identified nine HH objects in Cir-E, the first ever detected in this region, but we also found an additional twenty-eight in Cir-W, bringing the total there to 57. Although these two clouds are spatially adjacent, of similar apparent age \citep{KerrFarias25}, at comparable distances (Appendix \ref{app:distance}), and may well have formed simultaneously, their current morphologies and apparently their dynamical histories are strikingly distinct (Figure \ref{fig:cmc_starmot}). If a relationship exists between these clouds beyond co-location in the broader Circinus environment, it is not obviously apparent. 

\subsection{Future Work}

The greater Circinus region provides a rich and diverse laboratory for the study of star formation on a variety of scales, from Galactic to local. The range of properties and active processes ongoing in this cloud complex motivates a number of new studies to illuminate the detailed sequence of star formation and stellar plus gas kinematics. The presence of the relatively isolated Cir-W acts as a control sample, demonstrating the evolution of an apparently externally unperturbed cloud.

\citet{KerrFarias25} demonstrate the presence of a young population in the CIR-7 region, the northeastern-most extension of the CirCe region. This area may represent the local active star formation front, an interface between the stellar populations in the CirCe region and a previously unnamed gas cloud to the north of ASCC 79, which we call Circinus North (Cir-N) for future reference. This area is therefore ripe for infrared and optical IFU and photometric studies to characterize the stellar properties (e.g., T$_{eff}$, blue and infrared excess, etc), permitting analyses of ages, circumstellar disk properties, and mass distribution. Similar DECam observations as those highlighted in this current work to identify HH objects will help to characterize the evolutionary stage of this young population. Furthermore, focused millimeter observations of the molecular gas in the region could reveal potential coupling between the dynamical state of the cloud and the newly-formed stars.

High-spectral resolution (e.g., $>$40,000) observations of selected targets across the entire Circinus complex will yield radial velocities (RVs) and a precise three-dimensional map of the stellar space motions. Such observations also are useful to determine the individual radial velocity dispersions of the sub-components of the complex and to flag tight binaries that distort the kinematics.

Understanding the detailed distribution, physical characteristics, isotopic composition, and kinematics of distributed CO gas across the complex as well as other molecular tracers of dense cores will provide a complete picture of Circinus' gas properties. The detailed CO maps of \citet{2011ApJ...731...23S} provide invaluable data for understanding the distinct gas kinematics in the cores of Cir-W and Cir-E, but these results did not extend above a Galactic latitude of $\sim$-3.5 degrees and cover only Galactic longitudes of $\sim$315--319 degrees. The CO maps of \citet{1987ApJ...322..706D} include the extended Circinus environment, but at too low angular resolution to study the detailed properties of the region. Millimeter mapping of Circinus, and in particular virial analyses of dense cores \citep[e.g.,][]{2017ApJ...846..144K, Kerr19} will be required to supplement the stellar observations and dust extinction maps \citep{2005PASJ...57S...1D} and illuminate the global relationships between the gas, dust, and stellar content of the region.

Finally, our DECam observations of HH objects in Circinus and other regions indicate that cloud morphology is linked to the detection of optically visible HH objects. A logical expansion of this work would be to target dense, dusty cores in Cir-W and Cir-E with narrow-band forbidden and molecular line imaging in the infrared to search for more deeply embedded HH objects not visible at shorter wavelengths. \citet{2001A&A...380L...1G} demonstrated this approach with deep K-band imaging of extinguished HH objects in the Ophiuchus region.

%
%
%
%

\section{Summary}\label{sec:concl}

Between the work described here and in \citet{2020AJ....160..189R}, we have discovered 57 new HH objects in Cir-W and nine in Cir-E. HH objects imply stellar ages of only $\sim$1 Myr and are thus excellent tracers of ongoing early star formation. The discrepancy between the numbers of HH objects in Cir-W and Cir-E, combined with their distinct morphologies and gas properties as described by \citet{2011ApJ...731...23S}, suggest different star formation histories for the two co-located clouds. The results of \citet{Kerr23} demonstrated the presence of an extended region of young stars to the northeast.

In this current work, we have characterized the transverse velocity and age distribution of this population, described in detail in a sister paper \citep{KerrFarias25}, and find that Cir-E likely formed through a collision between an interloping cloud and ejected material associated with the ASCC 79 open cluster. Cir-W appears to be unrelated to this broader environment and thus provides a control sample of an externally unperturbed star forming region at similar age and distance to Cir-E. The discrepancy in the number of HH objects we have identified across the two regions may reflect not a difference in age but a disruption in star formation efficiency in Cir-E resulting from the complex stellar and gas dynamics and feedback from prior waves of star formation impacting the Cir-E cloud. These results open up numerous possible pathways for future exploration of this rich region, including deep infrared imaging for embedded HH objects, the determination of stellar space motions with the addition of RV measurements of stars across the complexes sub-components, and high-angular resolution millimeter mapping of the CO molecular distribution across the complex and of denser molecular tracers of star forming cores, such as NH$_3$, associated with the younger stellar populations. At a distance of only $\sim$800 pc, Circinus provides largely unexplored new terrain for understanding the processes of triggered and sequential star formation and its impact on the efficiency of the star formation, and thus HH object formation, processes.


\begin{acknowledgments}
We are grateful to the National Science Foundation for support through award AST-2206703 to TR and LP. This research has made use of ``Aladin sky atlas" developed at CDS, Strasbourg Observatory, France.  This publication makes use of data products from the Wide-field Infrared Survey Explorer, which is a joint project of the University of California, Los Angeles, and the Jet Propulsion Laboratory/California Institute of Technology, funded by the National Aeronautics and Space Administration.  This work has made use of data from the European Space Agency (ESA) mission Gaia.  We thank B. Reipurth and J. Bally for sharing their infrared data for this region, as well as B. Reipurth for the numbering of the HH objects and helpful comments. We also wish to thank Cerro Tololo Interamerican Observatory and its excellent support staff.  The figures in this paper were created with the help of the ESA/ESO/NASA FITS Liberator. The Dunlap Institute is funded through an endowment established by the David Dunlap family and the University of Toronto.
\end{acknowledgments}

%

\vspace{5mm}
\facilities{Blanco, WISE, Gaia}





\appendix

\section{Measured Distances} \label{app:distance}

\citet{KerrFarias25} suggests that Cir-W and the two subcomponents of Cir-E all lie at similar distances, with mean distances of 777 and 809 pc to Cir-W and Cir-E, respectively, and subcomponent-specific distances of 810 and 808 pc to Cir-Ea and Cir-Eb, respectively. However, those measurements include all stars attributed to those populations and may be contaminated by background sources, ejecta from ASCC~79, or other young components along the same sight line that may not relate directly to the cloud. Therefore, to confirm the coherence of these measurements, we conduct our own distance measurements that include additional restrictions that limit the distance measurements to populations directly associated with the clouds. 

To more precisely determine and compare the distances to Cir-E and Cir-W, we searched for stars that appear to be embedded in these clouds, based upon the presence of reflection nebula adjacent to or surrounding these stars. We used the most probable geometric distances from \textit{Gaia} Data Release 3, as determined by \citet{2021AJ....161..147B}.  The properties of these stars are given in Table~\ref{tbl:gedr3}. Of the five stars identified, four are in Cir-W; we used the weighted mean of their distances to compute an average distance of 773.3 $\pm$ 7.5 pc for Cir-W. This matches closely with the distance of 777 pc from \citep{KerrFarias25}. 

Cir-E has only one member connected to a reflection nebula, so additional data are necessary to facilitate reliable distance measurements to each of the two components in this region \citep{2011ApJ...731...23S, KerrFarias25}. We therefore base our distance measurements in Cir-E on the \textit{Gaia} Circinus Molecular Cloud Sample introduced in Section \ref{sec:gaiadata}. This sample is a subset of the stars used to compute distances to Cir-E in \citet{KerrFarias25}, with the additional youth probability cut $P_{Age<50 Myr} > 0.95$, which limits the sample to near-certain members of the Circinus Complex. The distance to the two components of Cir-E with this restriction are $814.3 \pm 3.9$ pc in Cir-Ea and $807.3 \pm 4.9$ pc in Cir-Eb, indicating distances broadly consistent with one another. The distance to Gaia DR3 5875245541334410624, which has a position and velocity consistent with Cir-Eb, has a \citet{2021AJ....161..147B} distance of 813$^{+30}_{-21}$ parsecs, which is consistent with both components. Because the core of Cir-Ea has much more consistent velocities compared to the rest of Cir-Ea (see Fig. \ref{fig:cmc_starmot}), we compute a separate distance measurement that restricts the \textit{Gaia} Circinus Molecular Cloud sample further to include only objects within 5 arcminutes of $(l, b) = (318.53, -4.34)$. This produces a distance of $780.4 \pm 8.5$ for Cir-Ea, putting it in the foreground of Cir-Eb, with a distance about equal to Cir-W. This revision suggests that contaminants left behind after the $P_{Age<50 Myr}$ restriction are preferentially located in the background, so the distance for Cir-Eb may also be an overestimate. However, without a major concentration of members in Cir-Eb, there are no clear ways to refine this sample further. We therefore conclude that the distances to Cir-W and the two components of Cir-E are consistent with each other, and that follow-up membership studies will be necessary to confirm a distance offset between Cir-Ea and Cir-Eb. 

\begin{deluxetable}{llllccc}
\tablecaption{Distances to Embedded Stars\label{tbl:gedr3}}
\tablewidth{0pt}
\tablehead{\colhead{GAIA DR3 ID} & \colhead{Alt ID} & \colhead{RA(2000)} & \colhead{DEC} & \colhead{Med\tablenotemark{a}} & \colhead{Lo} & \colhead{Hi}}
\startdata
\multicolumn7c{Circinus West}\\
\hline
5874396271662479872 & 2MASS J15003078-6306521 & 15:00:30.78 & -63:06:52.2 & 773 & 762 & 785	\\
5874396580900132224 & 2MASS J15004868-6305373 & 15:00:48.69 & -63:05:37.3 & 775 & 765 & 787	\\
5873636642237123328 & 2MASS J15021535-6320284 & 15:02:15.34 & -63:20:28.3 & 747 & 595 & 1011 \\
5873624414492498944 & 						  & 15:03:45.72 & -63:23:41.2 & 767 & 744 & 792 \\
\hline
\multicolumn7c{Circinus East}\\
\hline
5875245541334410624	& 						  & 15:13:37.47 & -62:25:16.4 & 813 & 792 & 843	\\
\enddata
\tablenotetext{a}{Med, Lo, and Hi distances (in pc) \citet{2021AJ....161..147B} }

\end{deluxetable}

\bibliography{rector}{}
\bibliographystyle{aasjournal}



\end{document}